\documentclass[11pt]{article}
\usepackage{amsmath, amssymb}
\usepackage{graphicx}
\usepackage{caption}
\usepackage{subcaption}
\usepackage{geometry}
\usepackage{authblk}
\usepackage{cite}
\usepackage{float}
\usepackage{hyperref}
\usepackage{multirow}

\usepackage[utf8]{inputenc}
\usepackage[T1]{fontenc}
\usepackage{amsmath, amsfonts, amssymb, amsthm}
\usepackage{geometry}
\newtheorem{theorem}{Theorem}[section]
\newtheorem{lemma}[theorem]{Lemma}
\newtheorem{definition}[theorem]{Definition}

\newtheorem{claim}[theorem]{Claim}
\title{A Coupled Fourth-Order Telegraph-Diffusion Framework Using Grayscale Indicators for Image despeckling}

  \author{
    \textbf{Manish Kumar$^{1*}$,Rajendra K. Ray $^{2}$} \\
    \textsuperscript{1,2} School of Mathematical and Statistical Sciences, \\
    Indian Institute of Technology Mandi, Mandi (H.P), 175075, India \\
    \texttt{manishkumarkitlana@gmail.com}, \texttt{rajendra@iitmandi.ac.in} \\
    \textsuperscript{*}\footnote{Corresponding author: Rajendra K. Ray, \texttt{rajendra@iitmandi.ac.in}}}
\date{}

\usepackage{float}
\begin{document}

\maketitle

\begin{abstract}

Speckle noise severely limits the quality of images acquired from coherent imaging systems such as Synthetic Aperture Radar (SAR) and medical ultrasound. Traditional second-order PDE-based despeckling approaches, although popular, often introduce staircase artifacts and blur fine details. To overcome these limitations, we present a nonlinear, fourth-order coupled hyperbolic-parabolic PDE model that effectively reduces noise while preserving the structure. The framework consists of two evolution equations: one governing fourth-order diffusion for effective speckle reduction and smooth intensity transitions, and another refining an edge indicator to protect textures and structural features. The diffusion coefficient is adaptively constructed using both the image intensity variable u and a grayscale-based indicator function, ensuring structure-aware denoising while avoiding blocky artifacts and preserving fine details. We also prove the existence of a weak solution to the proposed model by applying Schauder’s fixed-point theorem. A finite-difference scheme with Gauss–Seidel iteration is employed for efficient implementation. We compare the proposed model with the existing coupled second-order PDE model (HPCPDE) and the fourth-order telegraph diffusion model (TDFM). The results show that our model consistently outperforms these approaches. Experiments on standard grayscale images, real SAR and ultrasound data, as well as speckle-corrupted color images, demonstrate that the proposed method achieves superior performance over conventional PDE-based techniques in terms of PSNR, MSSIM, and Speckle Index.

\end{abstract}

\section{Introduction}
In practical imaging scenarios, the quality of images is compromised by various types of noise, making the extraction of meaningful content a challenging task. Among these, multiplicative speckle noise is particularly problematic because of its multiplicative nature, which significantly impacts the visual quality of images obtained from systems such as synthetic aperture radar (SAR), ultrasound, and laser imaging. Removing such noise is critical not only for better visual quality but also for improving the performance of subsequent tasks such as segmentation, object detection, and scene interpretation.

This study focuses specifically on the suppression of multiplicative speckle noise~\cite{1}. The degraded observation $J$ is modeled as
\begin{equation}
    J = I \cdot \eta,
\end{equation}
where $I$ represents the original clean image and $\eta$ denotes the multiplicative speckle noise, typically modeled using a gamma distribution with shape and scale parameters $(L, L)$, where $L \in \mathbb{N}$ is called the number of looks~\cite{2,5,6}. The principal challenge in despeckling lies in achieving a balance between noise suppression and the preservation of essential image features such as edges and textures. The statistical properties of speckle noise have been extensively studied~\cite{7,8,9,10}, leading to the development of a broad spectrum of denoising techniques. These include Bayesian estimation in both spatial and transform domains~\cite{8,9,10,11,12}, filters based on order statistics and morphology~\cite{13,14}, simulated annealing~\cite{15}, non-local means (NLM) approaches~\cite{16,17,18}, homomorphic filtering~\cite{19,20}, wavelet shrinkage~\cite{21,25}, and non-linear diffusion in both spatial and pyramid-based representations~\cite{26,26,28,29,30,31,33}. Variational formulations~\cite{34,35,36,37,38,39,40} and deep learning methods~\cite{41,42,43,44,45} have also shown promising results. Broadly, despeckling methods fall into three categories: (i) spatial and wavelet-based methods, (ii) non-local patch-based techniques, and (iii) variational and PDE-driven models. Early spatial domain techniques such as multi-look averaging~\cite{47} offer noise reduction at the cost of spatial resolution. Adaptive filters aim to preserve edges more effectively~\cite{8}, while MMSE and MAP estimators~\cite{10,48} based on statistical priors can fail to preserve fine structures when assumptions deviate from reality. Wavelet-based MAP estimators~\cite{36} provide detail preservation but often introduce ringing artifacts. Homomorphic filtering, which transforms the multiplicative model into an additive one via logarithmic scaling, can suffer from bias due to the non-linearity of the logarithmic transformation~\cite{2}. Nonlocal filtering methods, such as NLM~\cite{49}, PPB~\cite{18}, and BM3D~\cite{50}, utilize patch-based self-similarity for high-quality denoising. SAR-specific adaptations like SAR-BM3D~\cite{16} perform well, but are computationally demanding and can blur low-contrast regions. Total variation (TV) models~\cite{39} preserve edges by minimizing a functional energy with regularization, but they can cause stair-casing effects in smooth regions. Deep learning approaches~\cite{46} have set new benchmarks by learning mappings from noisy to clean images, although they require large annotated datasets and significant computational resources.
An emerging and effective direction involves the use of non-linear partial differential equations (PDEs), which provide a unified framework for noise reduction and feature preservation~\cite{31,32,34}. These models are grounded in solid mathematical principles and can be efficiently implemented with iterative solvers. Despite their advantages, classical second-order PDE-based methods often suffer from blocky artifacts. To counteract this, recent research has proposed higher-order and hybrid models combining diffusion and wavelike behavior to better capture textures and fine structures. For example, Majee et al.~\cite{29} proposed a telegraph model for speckle noise that helps to handle the variation. Other notable works include Roy et al.'s Bayesian MAP-TV approach~\cite{67}, Li et al.'s hybrid anisotropic-TV framework~\cite{66}, and Cuomo et al.'s deep variational model for ultrasound and SAR despeckling~\cite{65}.

Let us discuss some anisotropic second-order models, both parabolic and hyperbolic, with particular focus on their diffusion coefficients. The general form of the second-order parabolic model is given by

\begin{equation}
   \frac{\partial I}{\partial t} = \nabla \cdot \left( C(|\nabla I|) \nabla I \right) + \lambda \, A(I, J), 
   \quad \text{for } (x, t) \in \Omega \times (0, T),
\end{equation}

\begin{equation}
   \frac{\partial u}{\partial \mathbf{n}} = 0, 
   \quad \text{on } \partial \Omega \times (0, T),
\end{equation}

\begin{equation}
   u(x, 0) = f(x), \quad \text{for } x \in \Omega.
\end{equation}

with boundary conditions and initial conditions. Here, \(\Omega\) denotes the spatial domain, \(T\) represents the total evolution time, and \(\lambda\) is a regularization parameter. The operators \(\nabla\) and \(\operatorname{div}\) represent the gradient and the divergence, respectively. The function \(A(J, I)\), often originating from variational principles~\cite{34,39}, acts as a fidelity term. The coefficient C controls the degree of diffusion and plays a key role in maintaining image structures while denoising.

In 2014, Prasath and Vorotnikov~\cite{51} proposed a coupled PDE system for image restoration. The model is described by:
\begin{align}
    \frac{\partial I}{\partial t} - \nabla \cdot \left( g(u) \, \nabla I \right) &= 0, \quad \text{for } (x,t) \in \Omega \times (0,T), \\[10pt]
    \frac{\partial u}{\partial t} - \lambda \, \Delta u - (1 - \lambda) \left( \lvert \nabla I \rvert - u \right) &= 0, \quad \text{for } (x,t) \in \Omega \times (0,T),
\end{align}

where \(\lambda > 0\) balances the influence of smoothing and edge fidelity. The diffusion function \(g(s)\) is defined by:
\[
    g(s) = \frac{1}{1 + (\frac{s}{K}) ^2}, 
\]
with \(K > 0\) as a sensitivity parameter.

Later, Jain et al.~\cite{28} proposed an enhanced despeckling framework known as the Modified Coupled Partial Differential Equation (MCPDE) model, which comprises the following system of coupled equations:

\begin{align}
    \frac{\partial I}{\partial t} &= \nabla \cdot \left( \psi(I) \, g(u) \, \nabla I \right) - 2\lambda v, \quad \text{for } (x,t) \in \Omega \times (0,T), \\
    \frac{\partial u}{\partial t} &= \kappa_1 \left( \left| \nabla I_\xi \right|^2 - u + \frac{\kappa_2^2}{2} \, \Delta u \right), \quad \text{for } (x,t) \in \Omega \times (0,T), \\
    \frac{\partial v}{\partial t} &= v - \frac{I - J}{I}, \quad \text{for } (x,t) \in \Omega \times (0,T).
\end{align}

These equations are solved with the Neumann boundary conditions:

\[
    \frac{\partial I}{\partial \mathbf{n}} = 0, \quad 
    \frac{\partial u}{\partial \mathbf{n}} = 0, \quad 
    \frac{\partial v}{\partial \mathbf{n}} = 0, \quad 
    \text{on } \partial \Omega \times (0,T),
\]

and the following initial conditions:

\[
    I(x, 0) = J(x), \quad 
    v(x, 0) = 0, \quad 
    u(x, 0) = G_\xi * \left| \nabla J \right|^2, \quad 
    \text{for } x \in \Omega.
\]

In this formulation, \(\Delta\) denotes the Laplacian, \(g(s) = 1 / (1 + (s/k)^2)\), and \(\psi(I) = \frac{2 |I|^\alpha}{|I|^\alpha + M^\alpha}\), where \(M = \sup_{x \in \Omega} (A_\xi * J)(x)\) and \(I_\xi = A_\xi * I\). The constants \(\alpha, \xi, k, \kappa_1, \kappa_2\) are all positive, and \(A_\xi\) is a Gaussian kernel used for convolution in the spatial domain. The outward normal derivative is denoted by \(\partial_n\). The model separates the computation of the edge indicator \(u\) and the fidelity variable \(v\) to improve the quality of denoising. We now turn to grayscale indicator–based diffusion coefficients and discuss some related parabolic and hyperbolic models.
Shan et al. \cite{30} introduced a regularized version of the DD model\cite{70} to ensure well-posedness.
\begin{equation}\frac{\partial I}{\partial t} = \nabla \cdot \left( C(I_\xi, |\nabla I_\xi|)\nabla I \right),
\end{equation}
with boundary and initial conditions:
\begin{equation}
\frac{\partial I}{\partial n} = 0, \quad \text{on } \partial \Omega_T,
\end{equation}
\begin{equation}
I(a, 0) = f(a), \quad \text{in } \Omega.
\end{equation}
where $I_\xi$ is a Gaussian-smoothed image, and the diffusion coefficient is defined as:
\begin{equation}
C(I_\xi, |\nabla I_\xi|) = \left(\frac{I_\xi}{M_\xi }\right)^\alpha \cdot \frac{1}{1 + |\nabla I_\xi|^\nu}.
\end{equation}
To further enhance the performance of despeckling, Majee et al. \cite{29} proposed the TDE model, which combines diffusion and wave equations. The governing equation for the TDE model is:
\begin{equation}
\frac{\partial^2 I}{\partial t^2} + \gamma \frac{\partial I}{\partial t} = \nabla \cdot \left( C(I_\xi, |\nabla I_\xi|) \nabla I \right),
\end{equation}
where $\gamma$ represents the damping coefficient. The initial condition and the boundary conditions are as follows:
\begin{equation}
I(a, 0) = J(a), \quad \frac{\partial I}{\partial t}(a, 0) = 0, \quad \text{in } \Omega,
\end{equation}
\begin{equation}
\frac{\partial I}{\partial n} = 0, \quad \text{on } \partial \Omega_T.
\end{equation}
The diffusion coefficient in the TDE model is expressed as:
\begin{equation}
C(I_\xi, |\nabla I_\xi|) = \frac{2 |I_\xi|^\alpha}{M_\xi^\alpha + |I_\xi|^\alpha} \cdot \frac{1}{1 + (|\nabla I_\xi| / k)^2},
\end{equation}
where $M_\xi = \max_{x \in \Omega} | I_\xi(x,t) |$, $\alpha$ is a parameter that regulates intensity adaptation and $k$ controls gradient-based diffusion. The inclusion of wave dynamics in the TDE model helps in maintaining fine details and providing improved image restoration results. Higher-order PDEs have been explored to overcome the limitations of second-order models. You and Kaveh \cite{68} replaced the gradient with the Laplacian, resulting in the fourth-order equation:\begin{equation}\frac{\partial I}{\partial t} = -\Delta \left( C(\Delta I) \Delta I \right),\end{equation}with boundary and initial conditions:\begin{equation}\frac{\partial I}{\partial n} = 0, \quad \text{on } \partial \Omega_T,\end{equation}\begin{equation}I(a, 0) = f(a), \quad \text{in } \Omega.\end{equation}where the diffusion coefficient is:\begin{equation}C(s) = \frac{1}{1 + (s/k)^2}.\end{equation}This approach improves noise reduction while preserving edges. 
Rajendra Ray\cite{71} proposed a  telegraph diffusion fourth-order PDE model(TDFM) for effective speckle noise reduction:

\begin{equation}
    \frac{\partial^2 I}{\partial t^2} + \gamma \frac{\partial I}{\partial t} 
    = -\Delta \left( C(I_\xi, |\Delta I_\xi|)\, \Delta I \right) 
    - \lambda \left( \frac{I - f}{I} \right)^2, 
    \quad (x, t) \in \Omega \times (0, T),
    \label{eq:ray_model}
\end{equation}

with the initial conditions
\begin{equation}
    I(x, 0) = J(x), \quad I_t(x, 0) = 0, \quad \forall x \in \Omega,
    \label{eq:ray_ic}
\end{equation}

and Neumann-type boundary conditions
\begin{equation}
    \frac{\partial I}{\partial n} = 0, 
    \quad \frac{\partial (\Delta I)}{\partial n} = 0, 
    \quad (x, t) \in \partial \Omega \times (0, T).
\end{equation}

Here, $f$ denotes the observed noisy image and $t$ represents the artificial time parameter.
The nonlinear diffusion coefficient is defined as
\begin{equation}
    C(I_\xi, |\Delta I_\xi|) 
    = \frac{2 |I_\xi|^\alpha}{M_\xi^\alpha + |I_\xi|^\alpha} 
      \cdot \frac{1}{1 + \left(\frac{|\Delta I_\xi|}{k}\right)^2},
    \label{eq:ray_diffusion}
\end{equation}
where $I_\xi$ is the grayscale indicator, $M_\xi$ is the maximum value of $I_\xi$, 
$\alpha$ controls sensitivity to intensity variations, and $k$ tunes the Laplacian-based smoothing strength.

Also majee.\cite{69} proposes a second-order coupled system of PDE (HPCPDE) that is 

\begin{equation}
    \frac{\partial^2 I}{\partial t^2} + \gamma \frac{\partial I}{\partial t} 
    - \nabla \cdot \left( \frac{1}{(1 + s^\alpha)(1 + \iota \lvert u_\xi \rvert^\beta)} \nabla I \right) = 0, 
    \quad \text{for } (x,t) \in \mathcal{T},
    \label{eq:extisting3a}
\end{equation}

\begin{equation}
    \frac{\partial u}{\partial t} - h\left( \lvert \nabla I_\xi \rvert \right) + u - \frac{\nu^2}{2} \Delta u = 0, 
    \quad \text{for } (x,t) \in \mathcal{T},
    \label{eq:extisting3b}
\end{equation}

\noindent with the associated initial and boundary conditions:

\begin{align}
    I(x,0) &= J(x), \quad \frac{\partial I}{\partial t}(x,0) = 0, \quad u(x,0) = A_\xi * \lvert \nabla J(x) \rvert^2, \quad x \in \Omega, \nonumber \\
    \frac{\partial I}{\partial \mathbf{n}} &= 0, \quad \frac{\partial u}{\partial \mathbf{n}} = 0, \quad \text{on } \partial \Omega \times (0,T).
    \label{eq:extisting_boundary}
\end{align}

Second-order PDE models have shown promise in image denoising but often generate blocky artifacts and fail to preserve fine structures. To overcome these drawbacks, we propose a coupled PDE framework that merges a fourth-order diffusion equation with a hyperbolic telegraph equation. The diffusion term enables enhanced smoothing and structure preservation, while the hyperbolic component introduces wave-like dynamics that maintain sharp features and reduce staircase effects. This approach works especially well for removing speckle noise, offering a good balance between noise reduction and detail preservation.

\section{Proposed Method}
Inspired by the ideas of hyperbolic-parabolic PDEs in~\cite{29} and the coupled PDE in~\cite{69}, we propose the following improved nonlinear and coupled hyperbolic-parabolic model:

\begin{equation}\label{eq:main_system}
\begin{cases}
I_{tt} + \gamma I_t + \Delta \left(\mathcal{D}(I, u_\xi) \Delta I \right) + \lambda (I - J)^2 = 0, & \text{in } \mathcal{Q}_T, \\
u_t - \frac{\nu^2}{2} \Delta u + u = h(|\Delta I_\xi|), & \text{in } \mathcal{Q}_T.
\end{cases}
\end{equation}

The system is subject to the following initial conditions:
\begin{equation}\label{eq:initial_cond}
\begin{aligned}
    I(x, 0) &= J(x), \quad I_t(x, 0) = 0, && x \in \Omega, \\
    u(x, 0) &= u_0(x) = G_\xi * |\nabla J(x)|^2, && x \in \Omega,
\end{aligned}
\end{equation}
and the Neumann-type boundary conditions:
\begin{equation}\label{eq:boundary_cond}
   \frac{\partial I}{\partial n} = 0, \quad \frac{\partial(\Delta I)}{\partial n} = 0 , \quad \frac{\partial u}{\partial n} = 0 \quad \text{on } \partial\mathcal{Q}_T.
\end{equation}
The diffusion coefficient is given by:
\[
\mathcal{D}(I, u_\xi) =  \frac{2 |I_\xi|^\alpha}{|I_\xi|^\alpha + M_\xi^\alpha}. \frac{1}{1 + \iota |u_\xi|^\beta}.
\]

The given noisy input image is represented by \( I_0 \). The model incorporates parameters \( \alpha \geq 1 \), \( \beta \geq 1 \), \( \gamma > 0 \), \( \nu > 0 \), and \( \iota > 0 \), all of which are fixed and positive. Furthermore, the function \( h: \mathbb{R}^+ \rightarrow \mathbb{R}^+ \) is required to be both bounded and Lipschitz continuous. The proposed model adopts a structure similar to the partial differential equation (PDE) described in Equation~\ref{eq:extisting3a}, but it introduces several important enhancements. Unlike the existing model, which employs a separable second-order diffusion weight, the proposed model incorporates a fourth-order adaptive diffusion operator in place of the conventional divergence gradient formulation, along with a fidelity term that enforces closeness to the observed data, thus improving robustness. These enhancements allow the model to achieve a better trade-off between noise suppression and edge or texture preservation, leading to superior restoration performance compared to conventional telegraph second-order PDE-based methods. To the best of our knowledge, this work presents the first despeckling framework based on a nonlinear coupled hyperbolic–parabolic fourth-order PDE formulation. The proposed model integrates wave propagation dynamics with higher-order diffusion processes, yielding a rigorous and unified mathematical approach to enhanced speckle noise suppression while preserving structure of the images.

\section{Well-posedness of the weak solution}
In this section, we prove the existence and uniqueness of the weak solution of the proposed model \eqref{eq:main_system}--\eqref{eq:boundary_cond}. Since the problem \eqref{eq:main_system}--\eqref{eq:boundary_cond} is nonlinear, we first consider the linearized problem and then use the Schauder fixed-point theorem to show the existence of a weak solution.
\subsection{Technical framework and statement of the main result}
Throughout the article, we denote by $C > 0$ a generic constant that may change from line to line. Let $L^p(\Omega)$ with $1 \le p \le \infty$ denote the standard Lebesgue spaces. For $r \in \mathbb{N}$, $(H^r(\Omega), \|\cdot\|_{H^r})$ represents the usual Hilbert-Sobolev spaces. Specifically, we denote $V = H^2(\Omega)$ for the fourth-order diffusion and $W = H^1(\Omega)$ for the edge indicator. We denote $(H^1(\Omega))'$ as the dual space of $H^1(\Omega)$. For brevity, we write $L^p, H^r, V, W$ instead of $L^p(\Omega), H^r(\Omega), V(\Omega), W(\Omega)$, respectively. The solution space $\mathcal{W}(0,T)$ is defined as $\mathcal{W}(0,T) = \mathcal{W}_1(0,T) \times \mathcal{W}_2(0,T)$, where:
\begin{align*}
    \mathcal{W}_1(0,T) &= \left\{ I \in L^\infty(0,T; V) : \partial_t I \in L^\infty(0,T; L^2), \partial_{tt} I \in L^2(0,T; V') \right\}, \\
    \mathcal{W}_2(0,T) &= \left\{ u \in L^\infty(0,T; W) : \partial_t u \in L^\infty(0,T;L^2) \right\}.
\end{align*}
Equipped with the norm $\|(I, u)\|_{\mathcal{W}} = \|I\|_{\mathcal{W}_1} + \|u\|_{\mathcal{W}_2}$, the space $\mathcal{W}(0,T)$ is a Banach space.
\begin{definition}[Weak Solution]\label{def:weak_solution}
A pair $(I, u)$ is said to be a weak solution of \eqref{eq:main_system}--\eqref{eq:boundary_cond}, if:
\begin{enumerate}
    \item[a)] $I \in \mathcal{W}_1(0, T)$, $u \in \mathcal{W}_2(0, T)$ and satisfies the initial conditions \eqref{eq:initial_cond} and \eqref{eq:boundary_cond}.
    \item[b)] For all $\phi \in H^2(\Omega)$, $\psi \in H^1(\Omega)$ and a.e. $t \in (0, T)$, there hold
\end{enumerate}
\begin{equation} \label{eq:weak_I_new}
    \langle I_{tt}, \phi \rangle + \gamma \int_{\Omega} I_t \phi \, dx + \int_{\Omega} \mathcal{D}(I, u_\xi) \Delta I \Delta \phi \, dx + \lambda \int_{\Omega} (I-J)^2 \phi \, dx = 0,
\end{equation}
\begin{equation} \label{eq:weak_u_new}
    \int_{\Omega} u_t \psi \, dx + \frac{\nu^2}{2} \int_{\Omega} \nabla u \cdot \nabla \psi \, dx + \int_{\Omega} u \psi \, dx = \int_{\Omega} h(|\Delta G_\xi * I|) \psi \, dx.
\end{equation}
\end{definition}
\begin{theorem}[Existence and Uniqueness] \label{theo:existence_uniqueness}
The coupled telegraph-diffusion system \eqref{eq:main_system}--\eqref{eq:boundary_cond} admits a unique weak solution $(I, u) \in \mathcal{W}(0, T)$ in the sense of Definition \ref{def:weak_solution}, provided the following two conditions hold:
\begin{enumerate}\label{ass:A1}
    \item[\textbf{A.1}]  $J \in H^2(\Omega)$ satisfying $0 < \rho := \inf_{x \in \Omega} J(x)$.
    \item[\textbf{A.2}]  $h : \mathbb{R}^+ \to \mathbb{R}^+$ is a bounded, Lipschitz continuous function with Lipschitz constant $C_h$ such that $0 \le h(s) \le 1$ for all $s \in \mathbb{R}^+$.
\end{enumerate}
\end{theorem}
\subsection{Linearized Problem}
As mentioned earlier, due to the nonlinear nature of the problem, we first consider linear auxiliary problem and study its wellposedness result. To proceed, we first consider the following space: 
For positive constants $M_1, M_2 > 0$, define the convex set of admissible functions:
\begin{equation*}
\mathcal{B}_{M_1, M_2} = \left\{
\begin{aligned}
    (\bar{I}, \bar{u}) : \ & \bar{I} \in \mathcal{W}_1(0,T), \ \|\bar{I}\|_{L^\infty(0,T; H^2)} + \|\bar{I}_t\|_{L^\infty(0,T; L^2)} \le M_1 \|J\|_{H^2}, \\
    & \bar{u} \in \mathcal{W}_2(0,T), \ \|\bar{u}\|_{L^\infty(0,T; H^1)} + \|\bar{u}_t\|_{L^\infty(0,T; L^2)} \le M_2 \|J\|_{H^2}, \\
    & \bar{I}(x,t) \ge \rho > 0, \ \text{for a.e. } (t, x) \in \mathcal{Q}_T
\end{aligned} \right\}
\end{equation*}

For a fixed pair $(\bar{I}, \bar{u}) \in \mathcal{B}_{M_1, M_2}$, we consider the linearized system:
\begin{align}
    I_{tt} + \gamma I_t + \Delta \left( \bar{\mathcal{D}}(x,t) \Delta I \right) + \mathcal{S}(x,t)&=0, \label{eq:linI} \\
    u_t - \frac{\nu^2}{2} \Delta u + u - h(|\Delta \bar{I}_\xi|)&=0, \label{eq:linu}
\end{align}
where 
\begin{equation} 
\bar{\mathcal{D}}(x,t) =\frac{2 |\bar{I}_\xi|^\alpha}{|\bar{I}_\xi|^\alpha + M_\xi^\alpha}.\frac{1}{1 + \iota |\bar{u}_\xi|^\beta},\quad \text{and} \quad \mathcal{S}(x,t):= \lambda \left({\bar{I}(x,t)-J(x)}\right)^2.
\end{equation}

\begin{claim}\label{claim:diffusion_bounds}
Let $(\bar{I}, \bar{u}) \in \mathcal{B}_{M_1, M_2}$. Then there exist positive constants $\kappa, C_t$, and $C_{\mathcal{S}},$depending only on the initial data $J$, parameters  $M_1, M_2, G_\xi, \alpha, \beta, \rho$, and the domain $\Omega$ such that for any $\overline{I} \in \mathcal{B}_{M_1, M_2}$:
\begin{enumerate}
    \item [(a)] $0 < \kappa \le \bar{\mathcal{D}}(x,t) \le 1$, 
    \item [(b)]$|\partial_t \bar{\mathcal{D}}(x,t)| \le C_t$, 
    \item [(c)]$\|\mathcal{S}(\bar{I})\|_{L^2(\Omega)}=\lambda (\bar{I} - J)^2 \le C_{\mathcal{S}}$,
\end{enumerate}

\end{claim} 
\begin{proof}
\textbf{ (a,b):} Since $(\bar{I}, \bar{u}) \in \mathcal{B}_{M_1, M_2}$, a similar argument as in the proof of \cite[Claim 3.1]{69} yields the desired bound.

\textbf{Proof of (c):} Since $(\bar{I}, \bar{u}) \in \mathcal{B}_{M_1, M_2}$, we have the bound $\bar{I}(x,t) \ge \rho > 0$. Thus:
\[
|\mathcal{S}(x,t)| = \lambda \left| {\bar{I}(x,t)} - J(x) \right|^2 \le \lambda \left({\rho}+|J(x)| \right)^2\le \lambda \left({\rho}+|J(x)|_{H^2} \right).
\]
Therefore, there exists a constant $C_{\mathcal{S}}>0$ such that
\[
\|\mathcal{S}\|_{L^2(\Omega_T)} \le C_{\mathcal{S}}.
\]
Thus, (c) holds.
\end{proof}

\begin{lemma}\label{lem:energy}

For any fixed $(\bar{I},\bar{u}) \in \mathcal{B}_{M_1, M_2}$, the linearized problem 
\begin{align}
    I_{tt} + \gamma I_t + \Delta \left( \bar{\mathcal{D}}(x,t) \Delta I \right) &= -\lambda (\bar{I} - J)^2, \label{eq:linI} \\
    u_t - \frac{\nu^2}{2} \Delta u + u &= h(|\Delta \bar{I}_\xi|), \label{eq:linu2}
\end{align}
subject to the initial and boundary conditions, admits a unique weak solution $(I, u) \in \mathcal{W}(0,T)$. Furthermore, this solution satisfies the following energy estimates:
\begin{enumerate}
    \item[a)] $\| I \|_{L^\infty(0,T; H^2)} + \| I_t \|_{L^\infty(0,T; L^2)} \le C \| J \|_{H^2}$,
    \item[b)] $\| I_{tt} \|_{L^2(0,T; (H^2)')} \le C \| J \|_{H^2}$,
    \item[c)] $\| u \|_{L^\infty(0,T; H^1)} + \| u_t \|_{L^\infty(0,T; L^2)} \le C \| u_0 \|_{H^1}$,
\end{enumerate}
where $C > 0$ is a constant depending only on $G_\xi$, $J$, $h$, $M_1, M_2$, $\xi$, $\alpha$, $\beta$, and $\rho$.
\end{lemma}
\begin{proof}Thanks to Claim \ref{claim:diffusion_bounds}, one can apply the classical Galerkin method to show that there exists a unique weak solution $I \in \mathcal{W}(0, T)$ of the linearized problem \eqref{eq:linI} with the condition \eqref{eq:initial_cond} and \eqref{eq:boundary_cond}.
Next we prove that the solution of the linearized problem satisfy the above regularity:

\textbf{Proof of (a):} We multiply equation \eqref{eq:linI} by $I_t$ and integrate by parts to have 
\begin{equation} \label{eq:energy_I_1}
\frac{1}{2} \frac{d}{dt} | I_t |^2_{L^2} + \gamma | I_t |^2_{L^2} + \int_\Omega \bar{\mathcal{D}} \Delta I \Delta I_t , dx = -\lambda \int_\Omega \mathcal{S} I_t , dx.
\end{equation}
Thanks to green's identity and the neumann boundary conditions, we can re-write the third term of \eqref{eq:energy_I_1} as
\begin{align}\label{eq:energy_I_2}
  \int_\Omega\bar{\mathcal{D}} \Delta I \Delta I_t = \frac{1}{2} \int_\Omega\frac{d}{dt} (\bar{\mathcal{D}} |\Delta I|^2) - \frac{1}{2} \int_\Omega\bar{D}_t |\Delta I|^2  
\end{align}
and using Claim \ref{claim:diffusion_bounds} we observe that
\begin{equation} \label{eq:B.3}
\frac{1}{2} \left| \int_{\Omega} \bar{\mathcal{D}}_t (\Delta I)^2 dx \right| \le \frac{C_{t}}{2} \|\Delta I\|_{L^2}^2 \le \frac{C_{t}}{2\kappa} \int_{\Omega} \bar{\mathcal{D}} (\Delta I)^2 dx.
\end{equation}
For the source term, using the Cauchy-Schwarz and Young's inequalities:
\begin{equation}  \label{eq:B.4}
\left| \int_{\Omega} \mathcal{S} I_t \, dx \right| \le \frac{1}{2} \|\mathcal{S}\|_{L^2}^2 + \frac{1}{2} \|I_t\|_{L^2}^2.
\end{equation}
Now using \eqref{eq:energy_I_2},\eqref{eq:B.3}, and \eqref{eq:B.4} in \eqref{eq:energy_I_1}, and dropping the non-negative term $\gamma \|I_t\|_{L^2}^2$, we have

\begin{equation}  \label{eq:B.5}
\frac{d}{dt} \left[ \|I_t\|_{L^2}^2 + \int_{\Omega} \bar{\mathcal{D}} (\Delta I)^2 dx \right] \le C \left( \|I_t\|_{L^2}^2 + \int_{\Omega} \bar{\mathcal{D}} (\Delta I)^2 dx \right) + \|\mathcal{S}\|_{L^2}^2.
\end{equation}
The application of Gronwall's lemma gives the following: for a.e. $t \in (0, T]$,
\begin{equation} \label{eq:B.6}
\|I_t(t)\|_{L^2}^2 + \|\Delta I(t)\|_{L^2}^2 \le C \left( \|J\|_{H^2}^2 + \int_0^T \|\mathcal{S}(s)\|_{L^2}^2 ds \right) \le C.
\end{equation}
Since $I(x, t) = J(x) + \int_0^t I_t(s) ds$, we have by way of Young's inequality and \eqref{eq:B.6}
\begin{equation} \label{eq:B.7}
\|I(t)\|_{L^2}^2 \le 2 \|J\|_{L^2}^2 + 2T \int_0^t \big\| I_t(s) \big\|_{L^2}^2 ds \le C.
\end{equation}
By elliptic regularity for the Laplacian with Neumann boundary conditions, we know $\|I\|_{H^2} \le C_\Omega (\|I\|_{L^2} + \|\Delta I\|_{L^2})$. We combine this with \eqref{eq:B.6} and \eqref{eq:B.7} to conclude
\begin{equation}  \label{eq:B.8}
\|I\|_{L^\infty(0,T; H^2)} + \|I_t\|_{L^\infty(0,T; L^2)} \le C \|J\|_{H^2}.
\end{equation}
Hence (a) of the lemma follows.

\textbf{Proof of b):} Multiplying the linearized equation by a test function $\phi \in H^2$ with $\|\phi\|_{H^2} \le 1$ and integrating over $\Omega$, we have
\begin{equation} \label{eq:weak_form_itt}
    \langle I_{tt}, \phi \rangle_{(H^2)', H^2} + \int_{\Omega} \left( \gamma I_t \phi + \bar{\mathcal{D}} \Delta I \Delta \phi + \mathcal{S} \phi \right) dx = 0.
\end{equation}

We use the Cauchy-Schwarz inequality, along with the energy estimates from part (a) and the boundedness of $\bar{\mathcal{D}}$, to obtain
\begin{align*}
    |\langle I_{tt}, \phi \rangle| &\le \left( \gamma \|I_t\|_{L^2} + \|\bar{D}\|_{L^\infty} \|\Delta I\|_{L^2} + \|\mathcal{S}\|_{L^2} \right) \|\phi\|_{H^2} \\
    &\le C \left( \|J\|_{H^2} + \|\mathcal{S}(t)\|_{L^2} \right) \|\phi\|_{H^2}.
\end{align*}

Hence, by the definition of the norm in the dual space $(H^2)'$, we get
\begin{equation} \label{eq:dual_norm_pointwise}
    \|I_{tt}(t)\|_{(H^2)'} \le C \left( \|J\|_{H^2} + \|\mathcal{S}(t)\|_{L^2} \right).
\end{equation}

Moreover, squaring both sides of \eqref{eq:dual_norm_pointwise} and integrating over $(0, T)$, and using the fact that the source term $\mathcal{S} \in L^2(0,T; L^2(\Omega))$, one arrives at the assertion that $\|I_{tt}\|_{L^2(0,T; (H^2)')} \le C\left( \|J\|_{H^2}\right)$. Hence (b) of the lemma follows.

\textit{Proof of c):} 
Multiply  \eqref{eq:linu} by $u_t$, integrate by parts over $\Omega$, use Cauchy-Schwarz and Young's inequalities, and then integrate w.r.t. time between $0$ to $t$. We have, for a.e. $t \in (0, T)$,
\begin{equation*}
    \|u\|_{H^1}^2 + \int_0^t \|u_t\|_{L^2}^2 \, ds \leq C \left( 1 + t |\Omega| \right).
\end{equation*}
Moreover, since $u_0 \in H^2$ and $\bar{h}_t \in L^\infty(0, T; L^2)$, by regularity theory, $u_t \in L^\infty(0, T; L^2)$ with
\begin{equation}
    \|u\|_{H^1}^2 + \|u_t\|_{L^2}^2 \leq C e^t \left( 1 + t |\Omega| \right). \label{eq:19}
\end{equation}
This completes the proof.
\end{proof}

\begin{lemma}[Maximum Principle] \label{lemma:L_infinity_bounds}
Let $(I, u)$ be a weak solution of the system \eqref{eq:main_system}--\eqref{eq:boundary_cond}, and $\theta := \sup_{x \in \Omega} J(x) < \infty$. Then, the following bounds hold:
\begin{equation}
    0 < \rho \le I(x, t) \le \theta, \quad \text{for a.e. } (x, t) \in \Omega_T.
\end{equation}
\end{lemma}
\begin{proof}
Integrating the fourth-order equation \eqref{eq:main_system} and utilizing the initial condition $I_t(x,0) = 0$, we obtain the following integral-differential form:
\begin{equation} \label{eq:integrated_I}
    I_t + \gamma(I - J) + \int_0^t \Delta \left( \mathcal{D}(I, u_\xi) \Delta I \right) ds + \lambda \int_0^t (I - J)^2 ds = 0, \quad \forall (t, x) \in \mathcal{Q}_T.
\end{equation}
We define the truncated function $(I - \theta)_+ = \max\{0, I - \theta\}$. Since $I \in L^2(0,T; H^2(\Omega))$, it follows that $(I - \theta)_+ \in H^2(\Omega)$. Multiplying \eqref{eq:integrated_I} by $(I - \theta)_+$ and integrating over the spatial domain $\Omega$, we get:
\begin{align}
    \frac{1}{2} \frac{d}{dt} \int_\Omega |(I - \theta)_+|^2 dx &+ \gamma \int_\Omega (I - J)(I - \theta)_+ dx \nonumber \\
    &+ \int_0^t \int_{\{I \ge \theta\}} \mathcal{D}(I, u_\xi) |\Delta I|^2 dx \, ds + \lambda \int_0^t \int_\Omega (I - J)^2 (I - \theta)_+ dx \, ds = 0.
\end{align}
Consequently, we have the differential inequality:
\begin{equation}
    \frac{d}{dt} \int_\Omega |(I - \theta)_+|^2 dx \le 0.
\end{equation}
Integrating with respect to time and noting that at $t=0$, $I(0) = J \le \theta$, we conclude:
\begin{equation}
    \int_\Omega |(I - \theta)_+|^2 dx \le 0, \quad \text{for a.e. } t \in [0, T].
\end{equation}
This implies $I(t, x) \le \theta$ for a.e. $(t, x) \in \mathcal{Q}_T$. By a similar argument using the truncated function $(I - \rho)_- = \min\{0, I - \rho\}$, one can establish the lower bound $I(x, t) \ge \rho$. Thus, lemma \eqref{lemma:L_infinity_bounds} holds.
\end{proof}

\begin{lemma} \label{lemma:coupled_coeff_prop}
Let $(I_1, u_1), (I_2, u_2) \in L^2(\Omega)$. For the diffusion coefficient $\mathcal{D}(I, u_\xi)$, the following properties hold:

\begin{enumerate}
    \item[(i)] There exists a constant $C_0 > 0$, depending only on $\xi$ and $\Omega$, such that:
    \begin{equation} \label{eq:coupled_bound_lap}
    \|\Delta (G_\xi * w)\|_{L^\infty(\Omega)} \le C_0 \|w\|_{L^2(\Omega)}.
    \end{equation}
    
    \item[(ii)] The mapping $(I, u) \mapsto \mathcal{D}(I, u_\xi)$ is Lipschitz continuous. And, there exists a constant $C(\alpha, \beta, \xi, J, \rho) > 0$ such that for all $t \in [0,T]$:
    \begin{equation} \label{eq:coupled_lipschitz_D}
    \|\mathcal{D}(I_1, u_1)(t) - \mathcal{D}(I_2, u_2)(t)\|_{L^\infty(\Omega)} \le C(\alpha, \xi, J, \rho) \left( \|I_1(t) - I_2(t)\|_{L^2(\Omega)} + \|u_1(t) - u_2(t)\|_{L^2(\Omega)} \right).
    \end{equation}
\end{enumerate}
\end{lemma}

\begin{proof}
\textit{Proof of (i):} We utilize the properties of the Gaussian convolution and Young's inequality for convolutions with $p=2, q=2$:
\begin{equation}
\|\Delta (G_\xi * w)\|_{L^\infty(\Omega)} = \|(\Delta G_\xi) * w\|_{L^\infty(\Omega)} \le \|\Delta G_\xi\|_{L^2(\Omega)} \|w\|_{L^2(\Omega)}.
\end{equation}
Setting $C_0 = \|\Delta G_\xi\|_{L^2(\Omega)}$ yields the required bound.

\paragraph{Proof of (ii).} We decompose the difference $\mathcal{D}_1 - \mathcal{D}_2$ into two terms:
\begin{align}
\mathcal{D}(I_1, u_1) - \mathcal{D}(I_2, u_2) &= \frac{2|I_{1,\xi}|^\alpha}{(M_{I_1}^\xi)^\alpha + |I_{1,\xi}|^\alpha} 
\left( \frac{1}{1 + \iota |u_{1,\xi}|^\beta} - \frac{1}{1 + \iota |u_{2,\xi}|^\beta} \right) \nonumber \\
&\quad + \frac{1}{1 + \iota |u_{2,\xi}|^\beta} 
\left( \frac{2|I_{1,\xi}|^\alpha}{(M_{I_1}^\xi)^\alpha + |I_{1,\xi}|^\alpha} - \frac{2|I_{2,\xi}|^\alpha}{(M_{I_2}^\xi)^\alpha + |I_{2,\xi}|^\alpha} \right). \label{eq:diff_decomp}
\end{align}
By way of the convolution properties and the Lipschitz continuity of the function $f(s) = (1+\iota|s|^\beta)^{-1}$ on bounded sets, we have
\begin{align}
    \| \mathcal{D}_1 - \mathcal{D}_2 \|_{L^\infty} &\le C(\xi, \beta, \iota, J) \|u_1 - u_2\|_{L^2} \nonumber \\
    &\quad + \left\| \frac{2|I_{1,\xi}|^\alpha (M_{I_2}^\xi)^\alpha - 2|I_{2,\xi}|^\alpha (M_{I_1}^\xi)^\alpha}{\bigl( (M_{I_1}^\xi)^\alpha + |I_{1,\xi}|^\alpha \bigr) \bigl( (M_{I_2}^\xi)^\alpha + |I_{2,\xi}|^\alpha \bigr)} \right\|_{L^\infty} \nonumber \\
    &\equiv C(\xi, \beta, \iota, J) \|u_1 - u_2\|_{L^2} + \mathcal{B}. \label{eq:g_diff}
\end{align}
Since the initial data $J \ge \rho > 0$ implies the uniform lower bound $(M_{I_i}^\xi)^\alpha \ge (\rho \|G_\xi\|_{L^1})^\alpha$ for $i=1, 2$, the denominator is strictly separated from zero. We can bound the term $\mathcal{B}$ as follows:
\begin{align*}
    \mathcal{B} &\le \frac{2}{(\rho \|G_\xi\|_{L^1})^{2\alpha}} \left\| |I_{1,\xi}|^\alpha (M_{I_2}^\xi)^\alpha - |I_{2,\xi}|^\alpha (M_{I_1}^\xi)^\alpha \right\|_{L^\infty} \\
    &\le C(\rho, \xi, \alpha) \left\{ \left\| |I_{1,\xi}|^\alpha \left( (M_{I_2}^\xi)^\alpha - (M_{I_1}^\xi)^\alpha \right) \right\|_{L^\infty} + \left\| (M_{I_1}^\xi)^\alpha \left( |I_{1,\xi}|^\alpha - |I_{2,\xi}|^\alpha \right) \right\|_{L^\infty} \right\} \\
    &\le C(\rho, \xi, \alpha, J) \|I_1 - I_2\|_{L^2}.
\end{align*}
Combining the above relation with \eqref{eq:g_diff}, we obtain the required result:
\begin{equation}
\|\mathcal{D}(I_1, u_1)(t) - \mathcal{D}(I_2, u_2)(t)\|_{L^\infty(\Omega)} \le C(\alpha, \beta, \xi, J, \rho) \left( \|I_1(t) - I_2(t)\|_{L^2} + \|u_1(t) - u_2(t)\|_{L^2} \right).
\end{equation}
\end{proof}

\section{Proof of Theorem \ref{theo:existence_uniqueness}} In this section, we prove the well-posedness of the weak solution via the Generalized Schauder fixed-point theorem.
\subsection{ Existence of weak solution} We introduce a non empty, weakly compact, and convex subspace $W_0$ of $W(0, T)$, defined by
\begin{equation} \label{eq:W0_definition}
W_0 = \left\{ (w, v) \in W(0, T) : 
\begin{aligned}
& \|w\|_{L^\infty(0,T;H^2)} + \|w_t\|_{L^\infty(0,T;L^2)} + \|w_{tt}\|_{L^2(0,T;(H^2)')}\le M \|J\|_{H^2}, \\
& \|v\|_{L^\infty(0,T;H^1)} + \|v_t\|_{L^2(0,T;L^2)}\le M \|u_0\|_{H^1}, \\
& 0 < \alpha \le w(x,t) \text{ for a.e. } (x,t) \in \Omega_T, \\
& \text{and } (w, v) \text{ satisfies inital condition (\ref{eq:initial_cond})}
\end{aligned} 
\right\}.
\end{equation}

Consider the mapping:
\begin{align*}
\mathcal{P} : W_0 &\longrightarrow W_0 \\
(w, v) &\longmapsto (I_w, u_v),
\end{align*}
If we can show that the mapping $\mathcal{P} : (w, v) \mapsto (I_w, u_v)$ is weakly continuous from $W_0$ into itself, then by the Schauder fixed-point theorem, there exists at least one fixed point $(w, v) \in W_0$ such that $(w, v) = \mathcal{P}(w, v)$.
In order to prove the weak continuity of $\mathcal{P}$, let $\{(w_k, v_k)\}$ be a sequence that converges weakly to some $(w, v)$ in $W_0$, and let $(I_k, u_k) = \mathcal{P}(w_k, v_k)$ be the corresponding sequence of solutions. We aim to show that $\mathcal{P}(w_k, v_k) := (I_k, u_k)$ converges weakly to $\mathcal{P}(w, v) := (I, u)$.

Using the Lemma \ref{lem:energy} and the classical theorem of compact inclusion in the Sobolev space, we can extract subsequences $\{w_{k_n}\}$, $\{v_{k_n}\}$, $\{I_{k_n}\}$, and $\{u_{k_n}\}$, still denoted by the same indices $\{k\}$, such that for some $(I, u) \in W_0$, the following convergences hold as $k \to \infty$:

\begin{equation} \label{eq:convergences_full}
\left\{
\begin{aligned}
    & w_k \to w, \quad v_k \to v && \text{in } L^2(0, T; L^2(\Omega)) \text{ and a.e. on } \Omega_T, \\
    & G_\xi \ast w_k \to G_\xi \ast w && \text{in } L^2(0, T; L^2(\Omega)) \text{ and a.e. on } \Omega_T, \\
     & |G_\sigma \ast w_k|^\nu \rightarrow |G_\sigma \ast w|^\nu && \text{in } L^2(0, T; L^2(\Omega)) \text{ and a.e. on } \Omega_T, \\
    & \frac{2|G_\xi \ast w_k|^\alpha}{|G_\xi \ast w_k|^\alpha + (M_{w_k}^\xi)^\alpha} \to \frac{2|G_\xi \ast w|^\alpha}{|G_\xi \ast w|^\alpha + (M_w^\xi)^\alpha} && \text{in } L^2(0, T; L^2(\Omega)) \text{ and a.e. on } \Omega_T, \\
    & h(|\Delta G_\xi \ast w_k|) \to h(|\Delta G_\xi \ast w|)  && \text{in } L^2(0, T; L^2(\Omega)) \text{ and a.e. on } \Omega_T, \\
     & |G_\sigma \ast v_k| \rightarrow |G_\sigma \ast v| && \text{in } L^2(0, T; L^2(\Omega)) \text{ and a.e. on } \Omega_T, \\
    & \frac{1}{1 + |G_\xi \ast v_k|^\beta} \to \frac{1}{1 + |G_\xi \ast v|^\beta}  && \text{in } L^2(0, T; L^2(\Omega)) \text{ and a.e. on } \Omega_T, \\
    & \mathcal{S}(w_k) \to \mathcal{S}(w) && \text{strongly in } L^2(\Omega_T) \text{ and a.e. on } \Omega_T, \\
     & I_k \rightharpoonup^* I  && \text{weakly-* in } L^\infty(0, T; H^2(\Omega)), \\
    & I_k \to I && \text{in } L^2(0, T; L^2(\Omega)), \\
    & \partial_t I_k \rightharpoonup^* \partial_t I && \text{weakly-* in } L^\infty(0, T; L^2(\Omega)), \\
    & \partial_{tt} I_k \rightharpoonup \partial_{tt} I && \text{weakly in } L^2(0, T; (H^2(\Omega))'), \\
    & u_k \rightharpoonup^* u && \text{weakly-* in } L^\infty(0, T; H^1(\Omega)), \\
    & u_k \to u && \text{in }  L^2(0, T; L^2(\Omega)), \\
    & \partial_t u_k \rightharpoonup \partial_t u && \text{weakly in } L^2(0, T; L^2(\Omega)).
\end{aligned}
\right.
\end{equation}

The convergences established in the above section allow us to pass to the limit $k \to \infty$ in the weak formulation of the linearized coupled system \eqref{eq:linu}.
Let $\{(I_k, u_k)\}$ be the sequence of weak solutions to linearized pde corresponding to $(w_k, v_k)$, i.e., $\{(I_k, u_k)\} = \mathcal{P}(w_k, v_k)$. For any test function $\phi \in H^2$ and $\psi \in H^1$ , we have:
\begin{equation} \label{eq:coupled_weak_k}
\begin{cases}
\displaystyle \int_0^T \int_{\Omega} \left( \partial_{tt}I_{k} \phi + \gamma \partial_{t}I_k \phi + \bar{\mathcal{D}}(w_k, v_k) \Delta I_k \Delta \phi + \mathcal{S}(w_k) \phi \right) \, dx dt = 0, \\
\displaystyle \int_0^T \int_{\Omega} \left( \partial_{t}u_k \psi + \frac{\nu^2}{2} \nabla u_k \cdot \nabla \psi + u_k \psi - h(|\Delta G_\xi \ast w_k|) \psi \right) \, dx dt = 0.
\end{cases}
\end{equation}

From the weak convergences $\partial_{tt}I_{k} \rightharpoonup \partial_{tt}I$, $\partial_{t}I_{k} \rightharpoonup \partial_{t}I$, and $\partial_{t}u_{k} \rightharpoonup \partial_{t}u$ in respective spaces, We have:
\begin{align}
    \int_0^T \int_{\Omega} (\partial_{tt}I_{k} \phi + \gamma \partial_{t}I_{k} \phi) \, dx dt &\longrightarrow \int_0^T \int_{\Omega} (\partial_{tt}I \phi + \gamma \partial_{t}I \phi) \, dx dt, \\
    \int_0^T \int_{\Omega} \left(\partial_{t}u_k \psi + \frac{\nu^2}{2} \nabla u_k \cdot \nabla \psi + u_k \psi \right) dx dt &\longrightarrow \int_0^T \int_{\Omega} \left( \partial_{t}u \psi + \frac{\nu^2}{2} \nabla u \cdot \nabla \psi + u \psi \right) dx dt. \label{eq:limit_u}
\end{align}
Similarly, the weak convergence $u_k \rightharpoonup u$ in $H^1$ ensures that we can pass to the limit in the linear spatial terms of the edge equation, yielding:
\begin{equation*}
    \int_0^T \int_\Omega \left( \frac{\nu^2}{2} \nabla u_k \cdot \nabla \psi + u_k \psi \right) dx dt \longrightarrow \int_0^T \int_\Omega \left( \frac{\nu^2}{2} \nabla u \cdot \nabla \psi + u \psi \right) dx dt,
\end{equation*}

Considering the nonlinear diffusion term, we add and subtract $\bar{g}(w) \Delta I_k \Delta \phi$ and apply the triangle inequality to obtain:
\begin{align} \label{eq:weak_derivative}
    &\left| \int_0^T \int_{\Omega} \left( \bar{\mathcal{D}}(w_k, v_k) \Delta I_k - \bar{\mathcal{D}}(w, v) \Delta I \right) \Delta \phi \, dx dt \right| \nonumber \\
    &\quad \le \int_0^T \int_{\Omega} |\bar{\mathcal{D}}(w_k, v_k) - \bar{\mathcal{D}}(w, v) |\Delta I_k| |\Delta \phi| \, dx dt + \left| \int_0^T \int_{\Omega}\bar{\mathcal{D}}(w,v) (\Delta I_k - \Delta I) \Delta \phi \, dx dt \right|.
\end{align}

For the first term on the right-hand side of \eqref{eq:weak_derivative}, we apply the Cauchy-Schwarz inequality, the Lipschitz continuity of $\bar{\mathcal{D}}$ in lemma \ref{lemma:coupled_coeff_prop}, and the uniform $L^2$-bound of $\Delta I_k$:
\begin{equation*}
    \int_0^T \int_{\Omega} |\bar{\mathcal{D}}(w_k, v_k) - \bar{\mathcal{D}}(w, v)| |\Delta I_k| |\Delta \phi| \, dx dt \le C \|\Delta \phi\|_{L^\infty(\Omega_T)} \|\Delta I_k\|_{L^2(\Omega_T)} \|(w_k, v_k) - (w, v)\|_{L^2(\Omega_T)},
\end{equation*}
which converges to $0$ as $k \to \infty$ because $(w_k, v_k) \to \mathcal{D}(w, v)$ strongly in $L^2(\Omega_T)$. 

For the second term, $\Delta I_k \rightharpoonup \Delta I$ weakly in $L^2(\Omega_T)$, the second term also naturally converges to $0$ as $k \to \infty$. Thus, the entire difference in \eqref{eq:weak_derivative} vanishes in the limit.

For the source and coupling terms, and By the Lebesgue Dominated Convergence Theorem, we obtain:
\begin{align}
    \int_0^T \int_{\Omega} \mathcal{S}(w_k) \phi \, dx dt &\longrightarrow \int_0^T \int_{\Omega} \mathcal{S}(w) \phi \, dx dt, \\
    \int_0^T \int_{\Omega} h(|\Delta G_\xi \ast w_k|) \psi \, dx dt &\longrightarrow \int_0^T \int_{\Omega} h(|\Delta G_\xi \ast w|) \psi \, dx dt.
\end{align}
By substituting the limit results of the individual terms into the approximate weak formulation \eqref{eq:coupled_weak_k}, the system of equations becomes:
\begin{equation} \label{eq:weak_limit_coupled}
\begin{cases}
\displaystyle \int_0^T \int_{\Omega} \left( I_{tt} \phi + \gamma I_t \phi + \bar{\mathcal{D}}(w, v) \Delta I \Delta \phi + \mathcal{S}(w) \phi \right) \, dx dt = 0, \\
\displaystyle \int_0^T \int_{\Omega} \left( u_t \psi + \frac{\nu^2}{2} \nabla u \cdot \nabla \psi + u \psi - h(|\Delta G_\xi \ast w|) \psi \right) \, dx dt = 0.
\end{cases}
\end{equation}
which implies $(I, u) = \mathcal{P}(w, v)$. By the uniqueness of the solution to the linearized problem, the entire sequence $(I_k, u_k) = \mathcal{P}(w_k, v_k)$ converges weakly in $\mathcal{W}(0, T)$ to $(I, u) = \mathcal{P}(w, v)$. Hence, the mapping $\mathcal{P}$ is weakly continuous from $W_o$ into $W_0$. This weak continuity, combined with the compactness of the operator, allows us to apply the Schauder fixed-point theorem. We conclude that there exists a fixed point $(w, v) \in W_0$ such that $(w, v) = \mathcal{P}(w, v) = (I, u)$, which solves the original nonlinear problem \eqref{eq:main_system}--\eqref{eq:boundary_cond}.

\subsection*{Uniqueness of the Weak Solution}
Under Assumption \ref{ass:A1}, we prove the uniqueness of the weak solution for the coupled fourth-order telegraph-diffusion problem \eqref{eq:main_system}--\eqref{eq:boundary_cond}. Let $(I_1, u_1)$ and $(I_2, u_2)$ be two weak solutions of \eqref{eq:main_system}--\eqref{eq:boundary_cond}. Then, we have
\begin{align}
    &I_{tt} + \gamma I_t + \Delta(\mathcal{D}_1 \Delta I) = - \Delta\left((\mathcal{D}_1 - \mathcal{D}_2)\Delta I_2\right) - \left( \mathcal{S}(I_1) - \mathcal{S}(I_2) \right), \label{eq:diff_I} \\
    &u_t - \frac{\nu^2}{2} \Delta u + u = h(|\Delta G_\xi \ast I_1|) - h(|\Delta G_\xi \ast I_2|), \label{eq:diff_u}
\end{align}
with initial conditions $I(0,x) = 0$, $I_t(0,x) = 0$, $u(0,x) = 0$, and boundary conditions $\partial_{\mathbf{n}} I = \partial_{\mathbf{n}} (\Delta I) = \partial_{\mathbf{n}} u = 0$. Here, the composite diffusion coefficients are $\mathcal{D}_i = \mathcal{D}(I_i, u_i)$ and the source terms are $\mathcal{S}_i = \mathcal{S}(I_i)$ for $i=1,2$.
To show $I=0$
Fix $0 < s < T$, and define the backward-integrated variable:
\begin{equation} \label{eq:backward_v}
    v_i(t, x) = 
    \begin{cases} 
    \int_t^s I_i(\tau, x) \, d\tau, & 0 < t \le s, \\
    0, & s \le t < T.
    \end{cases}
\end{equation}
Note that, for $t \in (0, T)$,
\begin{equation} \label{eq:v_dist}
    \begin{cases}
    \partial_t v_i(t, x) = -I_i(t, x), \quad i = 1, 2, \\
    v_i(t, \cdot) =0 
    \end{cases}
\end{equation}
in the sense of distributions.
Setting $v = v_1 - v_2$, we have $v_t = -I$ on $(0,s)$, and $v(s,x) = 0$. 
Multiplying \eqref{eq:diff_I} by $v$, integrating over $\Omega \times (0, s)$, and applying integration by parts along with Cauchy-Schwarz inequality, we analyze the terms. For the diffusion term, since $\Delta I = -\partial_t \Delta v$, we utilize the following identity: 
\begin{equation}
\mathcal{D}_1 (-\partial_t \Delta v) \Delta v = -\frac{1}{2} \partial_t \left( \mathcal{D}_1 |\Delta v|^2 \right) + \frac{1}{2} \partial_t \mathcal{D}_1 |\Delta v|^2.
\end{equation}
Integrating this expression over the space-time domain $\Omega \times (0, s)$ yields:\begin{align*}
\int_0^s \int_\Omega \Delta(\mathcal{D}_1 \Delta I) v \, dx dt &= \int_0^s \int_\Omega \mathcal{D}_1 \Delta I \Delta v \, dx dt = -\int_0^s \int_\Omega \mathcal{D}_1 (\partial_t \Delta v) \Delta v \, dx dt \\
&= -\frac{1}{2} \int_\Omega \int_0^s \partial_t (\mathcal{D}_1 |\Delta v|^2) \, dt dx + \frac{1}{2} \int_0^s \int_\Omega \partial_t \mathcal{D}_1 |\Delta v|^2 \, dx dt \\
&= -\frac{1}{2} \int_\Omega \left[ \mathcal{D}_1 |\Delta v|^2 \right]_0^s dx + \frac{1}{2} \int_0^s \int_\Omega \partial_t \mathcal{D}_1 |\Delta v|^2 \, dx dt.
\end{align*}
We have $v(x, s) = 0$, which implies $\Delta v(x, s) = 0$. Consequently, the boundary term at $t=s$ vanishes, leaving:
\begin{equation*}
-\frac{1}{2} \int_\Omega \left[ \mathcal{D}_1(x, s) |\Delta v(x, s)|^2 - \mathcal{D}_1(x, 0) |\Delta v(x, 0)|^2 \right] dx = \frac{1}{2} \int_\Omega \mathcal{D}_1(x, 0) |\Delta v(x, 0)|^2 dx.
\end{equation*}
Since $v(x,s) = 0$, the boundary term yields $\frac{1}{2} \int_\Omega D_1(x,0) |\Delta v(x,0)|^2 dx$. Combining this with the time derivative terms, we obtain the energy inequality:
\begin{align}
\frac{1}{2} \| I(s) \|^2_{L^2} &+ \gamma \int_0^s \| I(t) \|^2_{L^2} dt + \frac{1}{2} \int_\Omega \mathcal{D}_1(x, 0) |\Delta v(x, 0)|^2 dx \nonumber \\
&\leq \frac{1}{2} \left| \int_0^s \int_\Omega |\Delta v|^2 \partial_t \mathcal{D}_1 \, dx dt \right| + \int_0^s \| \mathcal{D}_1 - D_2 \|_{L^\infty} \| \Delta I_2 \|_{L^2} \| \Delta v \|_{L^2} dt \nonumber \\
&\quad + \left| \int_0^s \int_\Omega (\mathcal{S}(I_1) - \mathcal{S}(I_2)) v \, dx dt \right|. \label{eq:v_energy}
\end{align}
Since $(I_i, u_i)$  belong to the admissible set $\mathcal{B}_{M_1, M_2}$. They satisfy the point-wise lower bound $I_i(x,t) \ge \rho > 0$. Therefore, the source term is locally Lipschitz continuous with an adaptive constant $L_{\mathcal{S}}$. Using Cauchy-Schwarz and Young's inequalities, we obtain:
\begin{align}
    \|\mathcal{S}(I_1) - \mathcal{S}(I_2)\|_{L^2(\Omega)} &= \lambda \|(I_1 - J)^2 - (I_2 - J)^2\|_{L^2(\Omega)} \nonumber \\
    &= \lambda \|(I_1 - I_2)(I_1 + I_2 - 2J)\|_{L^2(\Omega)} \nonumber \\
    &\le \lambda \|I_1 + I_2 - 2J\|_{L^\infty(\Omega)} \|I_1 - I_2\|_{L^2(\Omega)}. \label{eq:S_diff_step}
\end{align}
We obtain:
\begin{equation} \label{eq:lipschitz_source_uniqueness}
    \|\mathcal{S}(I_1) - \mathcal{S}(I_2)\|_{L^2(\Omega)} \le L_S \|I(t)\|_{L^2(\Omega)}.
\end{equation}
By claim \ref{claim:diffusion_bounds}, there exist positive constants $\kappa, C_t > 0$ such that:
\begin{equation}
    0 < \kappa \le \mathcal{D}_1(t,x) \le 1, \quad |\partial_t \mathcal{D}_1(t,x)| \le C_t.
\end{equation}

Moreover, by lemma \eqref{eq:weak_derivative} i.e. using the properties of convolution and the positive lower bound $\rho$ of the solutions $I_i$, we get:
\begin{equation}
\|\mathcal{D}(I_1, u_1)(t) - \mathcal{D}(I_2, u_2)(t)\|_{L^\infty(\Omega)} \le C(\alpha, \beta, \xi, J, \rho) \left( \|I_1(t) - I_2(t)\|_{L^2(\Omega)} + \|u_1(t) - u_2(t)\|_{L^2(\Omega)} \right).
\end{equation}
By applying Cauchy-Schwarz and Young's inequalities we obtain:
\begin{align}
    & \frac{1}{2} \|I(s)\|_{L^2}^2 + \int_0^s \gamma \|I(t)\|_{L^2}^2 \, dt + \frac{\kappa}{2} \|\Delta v(0)\|_{L^2}^2 \nonumber \\
    & \quad \le C \int_0^s \bigg( \|\Delta v(t)\|_{L^2}^2 + \|I(t)\|_{L^2}\|v(t)\|_{L^2} \nonumber \\
    & \quad \qquad \qquad + \big( \|I(t)\|_{L^2} + \|u(t)\|_{L^2} \big) \|\Delta I_2(t)\|_{L^2} \|\Delta v(t)\|_{L^2} \bigg) dt \nonumber \\
    & \quad \le C \int_0^s \Big( \|\Delta v(t)\|_{L^2}^2 + \|v(t)\|_{L^2}^2 + \|I(t)\|_{L^2}^2 + \|u(t)\|_{L^2}^2 \Big) dt. \label{eq:uniq_I_intermediate}
\end{align}
Notice that $\|v(t)\|_{L^2}^2 \le s \int_0^s \|I(t)\|_{L^2}^2 \, dt \le T \int_0^s \|I(t)\|_{L^2}^2 \, dt$. This allows us to absorb $\|v(t)\|_{L^2}^2$ into the integral of $I(t)$. 

Next, we define the forward-integrated variable:
\begin{equation}
    w_i(t, x) = \int_0^t I_i(\tau, x) \, d\tau, \quad w(t, x) = w_1(t, x) - w_2(t, x), \quad 0 \le t \le T.
\end{equation}
Observe that $v(0, x) = w(s, x)$ and $v(t, x) = w(s, x) - w(t, x)$ for $0 \le t \le s$. Substituting these relations into \eqref{eq:uniq_I_intermediate}, we arrive at:
\begin{equation}
\frac{1}{2} \|I(s)\|_{L^2}^2 + \int_0^s \|I(t)\|_{L^2}^2 dt + C \|w(s)\|_{H^2}^2 \le \tilde{C}s \|w(s)\|_{H^2}^2 + C \int_0^s \left( \|w(t)\|_{H^2}^2 + \|I(t)\|_{L^2}^2 \right) dt. \label{eq:energy_w}
\end{equation}
By choosing a time step $T_1 > 0$ sufficiently small such that $\tilde{\kappa} - C T_1 > 0$, we can absorb the term $C s \|w(s)\|_{H^2}^2$ into the left-hand side for all $0 < s \le T_1$. This yields:
\begin{equation} \label{eq:uniq_I_final}
    \|I(s)\|_{L^2}^2 + \|w(s)\|_{H^2}^2 \le C \int_0^s \left( \|w(t)\|_{H^2}^2 + \|I(t)\|_{L^2}^2 + \|u(t)\|_{L^2}^2 \right) dt.
\end{equation}

Next, multiplying the equation \eqref{eq:diff_u} by $u$ and integrating over the spatial domain $\Omega$, we obtain:
\begin{equation} \label{eq:edge_uniq_1}
    \frac{1}{2} \frac{d}{dt} \|u\|_{L^2}^2 + \|\nabla u\|_{L^2}^2 + \|u\|_{L^2}^2 = \int_\Omega \left( h(|\Delta G_\xi \ast I_1|) - h(|\Delta G_\xi \ast I_2|) \right) u \, dx.
\end{equation}
Applying the Cauchy-Schwarz and Young's inequalities on the right-hand side, we get:
\begin{equation}\label{eq:edge_uniq_2}
    \frac{d}{dt} \|u\|_{L^2}^2 + 2\|\nabla u\|_{L^2}^2 + \|u\|_{L^2}^2 \le \|h(|\Delta G_\xi \ast I_1|) - h(|\Delta G_\xi \ast I_2|)\|_{L^2}^2.
\end{equation}

The edge-indicator function $h$ is Lipschitz continuous with constant $C_h$. By Young’s convolution inequality, we observe:

\begin{equation} \label{eq:h_lip_uniqueness_final}
    \|h(|\Delta G_\xi * I_1|) - h(|\Delta G_\xi * I_2|)\|_{L^2}^2 \le C(C_h, \xi) \|I(t)\|_{L^2}^2.
\end{equation}
By putting the value back in equation \eqref{eq:edge_uniq_2} and Integrating it with respect to time from $0$ to $s$ (for $0 < s \le T_1$), we obtain:
\begin{equation} \label{eq:uniq_u_final}
    \|u(s)\|_{L^2}^2 \le C \int_0^s \left( \|I(t)\|_{L^2}^2 + \|u(t)\|_{L^2}^2 \right) dt.
\end{equation} and using the initial condition $u(0)=0$, we obtain:
\begin{equation} \label{eq:uniq_u_final2}
    \|u(s)\|_{L^2}^2 \le C \int_0^s \|I(t)\|_{L^2}^2 \, dt.
\end{equation}

Adding \eqref{eq:uniq_I_final} and \eqref{eq:uniq_u_final}, we obtain for $0 < s \le T_1$:
\begin{equation*}
    \|I(s)\|_{L^2}^2 + \|u(s)\|_{L^2}^2 + \|w(s)\|_{H^2}^2 \le C \int_0^s \left( \|I(t)\|_{L^2}^2 + \|u(t)\|_{L^2}^2 + \|w(t)\|_{H^2}^2 \right) dt.
\end{equation*}

By Grönwall's lemma, this implies $(I, u) \equiv (0, 0)$ on $[0, T_1]$. Repeating this argument on intervals $(T_1, 2T_1], (2T_1, 3T_1], \dots$, we conclude that $I_1 = I_2$ and $u_1 = u_2$ on $(0, T)$, completing the proof \ref{theo:existence_uniqueness}.

\section{Numerical Discretization}

Let $\tau$ and $\tilde{h}$ denote the time and space step sizes, respectively. Let \( I_{i,j}^n \) denote the discrete approximation of the solution at time \( t^n \) and spatial position \( (x_i, y_j) \), where \( x_i = i \tilde{h} \) for \( i = 0, 1, \ldots, M-1 \), \( y_j = j \tilde{h} \) for \( j = 0, 1, \ldots, N-1 \), and \( t^n = n \tau \) for \( n = 0, 1, 2, \ldots \). The discretized versions of equations ~\ref{eq:main_system} are derived using a weighted numerical scheme, as described in~\cite{60}.

\begin{align}
(1 + 0.5 \gamma \tau) I_{i,j}^{\,n+1} 
&- \tau \theta_1 \Bigl[ \Delta \cdot \mathcal{D}(I, u_\xi) \Delta I ) \Bigr]_{i,j}^{\,n+1} \nonumber \\
&= 2 I_{i,j}^n + \tau^2 (1 - \theta_1 - \theta_2) \Bigl[ \Delta \cdot \mathcal{D}(I, u_\xi) \Delta I ) \Bigr]_{i,j}^{\,n} \nonumber \\
&\quad+ \tau^2 \theta_2 \Bigl[ \Delta \cdot \mathcal{D}(I, u_\xi) \Bigr]_{i,j}^{\,n-1} + (0.5 \gamma \tau - 1) I_{i,j}^{\,n-1}.
\label{eq:32}
\end{align}

\begin{equation}
\frac{\partial^2 I}{\partial x^2}\Big|_{i,j} \approx \frac{I_{i+1,j} - 2I_{i,j} + I_{i-1,j}}{\tilde{h}^2}
\end{equation}

\begin{equation}
\frac{\partial^2 I}{\partial y^2}\Big|_{i,j} \approx \frac{I_{i,j+1} - 2I_{i,j} + I_{i,j-1}}{\tilde{h}^2}
\end{equation}

\begin{equation}
\Delta I_{i,j} \approx \frac{\partial^2 I}{\partial x^2}\Big|_{i,j} + \frac{\partial^2 I}{\partial y^2}\Big|_{i,j}
\end{equation}

Furthermore, the diffusion weight at each pixel is defined as
\[
 \mathcal{D}(I, u_\xi) =  \frac{2 |I_\xi|_{i,j}^\alpha}{|I_\xi|_{i,j}^\alpha + M_\xi^\alpha}. \frac{1}{1 + \iota |u_\xi|_{i,j}^\beta}.
\]
with \( \mathcal{A}_\xi \ast u_{i,j} \) denoting the convolution of the image \( u \) with a Gaussian kernel \( \mathcal{A}_\xi \), and \( \iota, \alpha, \beta \) being positive constants controlling the diffusion behavior.

The variable $u(t^n, x_i, y_j) = u_{i,j}^n$ is updated using:

\begin{align}
\begin{aligned}
(1 + \tau \theta) u_{i,j}^{n+1}
&- 0.5 \tau \nu^2 \left( \delta_x^2 u_{i,j}^{n+1} + \delta_y^2 u_{i,j}^{n+1} \right) = \\
&(1 - \tau (1 - \theta)) u_{i,j}^n 
+ 0.5 \tau \nu^2 (1 - \theta) \left( \delta_x^2 u_{i,j}^n + \delta_y^2 u_{i,j}^n \right) \\
&+ \tau h\left( |\Delta (G_\xi * I_{i,j}^n)| \right),
\end{aligned} \label{eq:34}
\end{align}

where $\theta$ is a weighting parameter, and $\delta_x^2$, $\delta_y^2$ are second-order central difference operators.

\begin{equation}
\Delta I^n_{i,j} = I^n_{i+1,j} + I^n_{i-1,j} + I^n_{i,j+1} + I^n_{i,j-1} - 4I^n_{i,j}.
\end{equation}
Symmetric boundary conditions are used:
\[
I^n_{-1,j} = I^n_{0,j}, \quad I^n_{I+1,j} = I^n_{I,j}, \quad I^n_{i,-1} = I^n_{i,0}, \quad I^n_{i,J+1} = I^n_{i,J}.
\]
 The nonlinearity $g$
\begin{equation}
g^n_{i,j} =  \mathcal{D}(I_{i,j}, (u_\xi)_{i,j})\cdot \nabla^2 I^n_{i,j}.
\end{equation}

\begin{equation}
\Delta g^n_{i,j} = \frac{g^n_{i+1,j} + g^n_{i-1,j} + g^n_{i,j+1} + g^n_{i,j-1} - 4g^n_{i,j}}{h^2}.
\end{equation}
Again, symmetric boundary conditions apply:
\[
g^n_{-1,j} =g^n_{0,j}, \quad g^n_{I+1,j} = g^n_{I,j}, \quad g^n_{i,-1} = g^n_{i,0}, \quad g^n_{i,J+1} = g^n_{i,J}.
\]

\subsection* {Extension to color image restoration}

To extend the proposed discretized fourth-order PDE model to color images, we treat each color channel (red (R), green (G) and blue (B) independently. Let the observed color image be denoted as \( I = [I^R, I^G, I^B] \), where each \( I^C \in \mathbb{R}^{M \times N} \) corresponds to the respective color channel for \( C \in \{R, G, B\} \).

For each channel \( I^C \), the coupled PDE formulation, as described in equations \eqref{eq:32} and \eqref{eq:34}, is applied separately. At every iteration \( n \), the discrete evolution equations are solved for \( I^{C,n} \) and the auxiliary variable \( u^{C,n} \), producing updated values \( I^{C,n+1} \) and \( u^{C,n+1} \).

After updating all three channels, the denoised color image at iteration \( n+1 \) is obtained by recombining the individual channels:
\[
I^{n+1}_{\text{RGB}} = \left[ I^{R,n+1},\ I^{G,n+1},\ I^{B,n+1} \right].
\]

To evaluate the performance of the restoration process, standard quality metrics are computed at each iteration. This approach ensures that the intrinsic structure of each color component is preserved while reducing speckle or multiplicative noise, leading to a high-quality reconstruction of the color image.

\section{Evaluation Metrics and Parameters}
When the ground truth image is available, the quality of the denoised image \( I^{(k)} \) can be quantified using the Peak Signal-to-Noise Ratio (PSNR). 
It measures the fidelity of \( I^{(k)} \) with respect to the reference image \( I \) and is computed as  
\begin{equation}
    \text{PSNR} = 10 \log_{10} \!\left( 
    \frac{ \max(I)^2 }{ 
    \frac{1}{MN} \sum_{i=1}^{M} \sum_{j=1}^{N} 
    \left( I(i,j) - I^{(k)}(i,j) \right)^2 } 
    \right),
\end{equation}
where \( M \times N \) represents the image size and \( \max(I) \) denotes the maximum intensity of the original image.  

To assess perceptual similarity, the Structural Similarity Index (SSIM) is employed, 
which compares luminance, contrast, and structure between \( I \) and \( I^{(k)} \). 
It is expressed as  
\begin{equation}
    \text{SSIM}(I, I^{(k)}) =
    \frac{(2 \mu_I \mu_{I^{(k)}} + c_1)(2 \sigma_{I I^{(k)}} + c_2)}
    {(\mu_I^2 + \mu_{I^{(k)}}^2 + c_1)(\sigma_I^2 + \sigma_{I^{(k)}}^2 + c_2)},
\end{equation}
where \( \mu_I \) and \( \mu_{I^{(k)}} \) are mean intensities, 
\( \sigma_I^2 \) and \( \sigma_{I^{(k)}}^2 \) are variances, 
\( \sigma_{I I^{(k)}} \) is the covariance between the two images, 
and \( c_1, c_2 \) are small constants to ensure numerical stability.

The Mean Structural Similarity Index (MSSIM) is the average of SSIM values of each iteration, defined as
\begin{equation}
    \text{MSSIM} = \frac{1}{K} \sum_{k=1}^{K} \text{SSIM}(I, I^{(k)}),
\end{equation}
where \( K \) is the total number of iterations. When ground truth is available, the stopping criterion is the best PSNR value. For real images without clean references, we use the relative change between consecutive restored images to determine convergence.
\begin{equation}         \frac{\| I^{(k+1)} - I^{(k)} \|_2^2}{\| I^{(k)} |_2^2} \leq \varepsilon.
    \end{equation}
 For the numerical simulations, we choose \(\varepsilon = 1 \times 10^{-4}\).
The \textbf{Speckle Index (SI)} is used to assess the level of speckle noise in SAR and ultrasound images. It is defined as the ratio of the standard deviation to the mean of the image:
\begin{equation}
    \text{SI} = \frac{\sigma}{\mu},
\end{equation}
Here, \( \sigma \) denotes the standard deviation, and \( \mu \) represents the mean intensity of the image pixels. All numerical evaluations are conducted on amplitude images degraded by speckle noise simulated using a Gamma distribution across varying numbers of looks \(\{1, 3, 5, 10\}\). 

The proposed method’s parameters are tuned based on thorough numerical experimentation. A fixed time step \( \tau = 0.25 \) is used throughout, as it provides stable and accurate results. The parameters \( \alpha, \beta, \gamma, \iota, \lambda \) are key to regulating performance, particularly in response to different image content and levels of noise.
 The iterative PDE-based denoising uses boundary and initial conditions applied consistently across coupled models and solves the discretized PDE system using the Gauss-Seidel iterative method. The forth order single pde model is discretized by the finite difference method. The same approach will be held for color images.
\begin{figure}[h!]
\centering
\begin{tabular}{cccc} 
\includegraphics[width=0.2\textwidth]{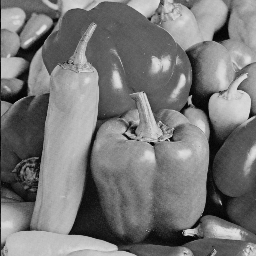} & 
\includegraphics[width=0.2\textwidth]{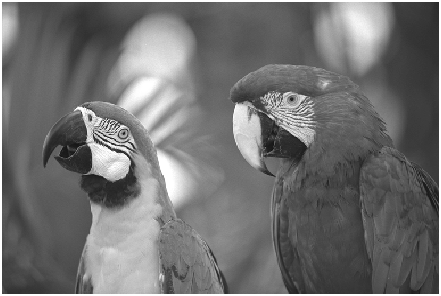} & 
\end{tabular}
\caption{Images: (a) Peppers, (b) Parrots}
\label{fig:images_0}
\end{figure}
\begin{figure}[h!]
\centering
\begin{tabular}{cccc} 
\includegraphics[width=0.25\textwidth]{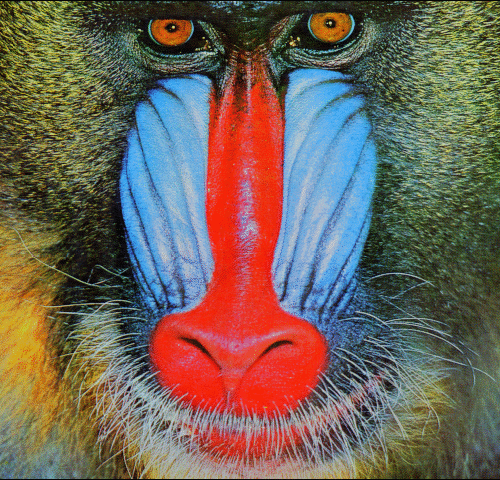} & 
\includegraphics[width=0.24\textwidth]{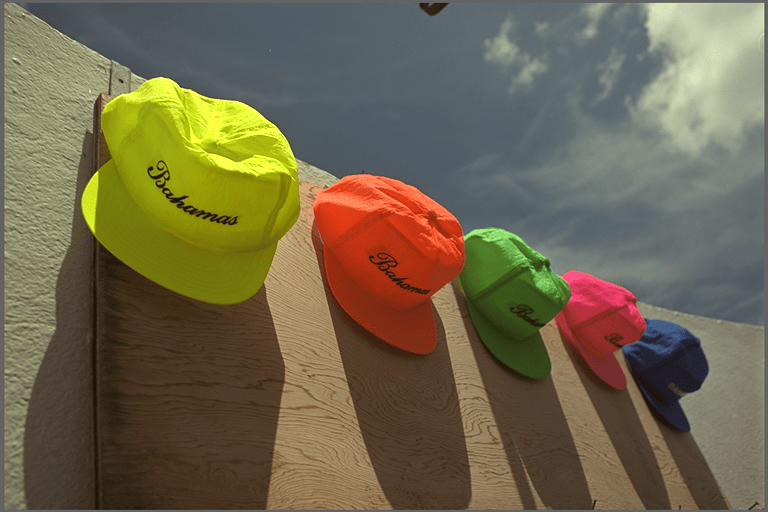}  
\end{tabular}
\caption{color Images: (a) Baboon, (b)caps.}
\label{fig:images_0c}
\end{figure}

\begin{figure}[H]
    \centering
    \begin{subfigure}[b]{0.24\textwidth}
        \centering
        \includegraphics[width=\textwidth]{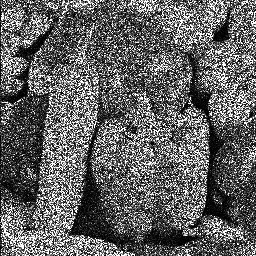}
        \caption{Look=1}
    \end{subfigure}
    \begin{subfigure}[b]{0.24\textwidth}
        \centering
        \includegraphics[width=\textwidth]{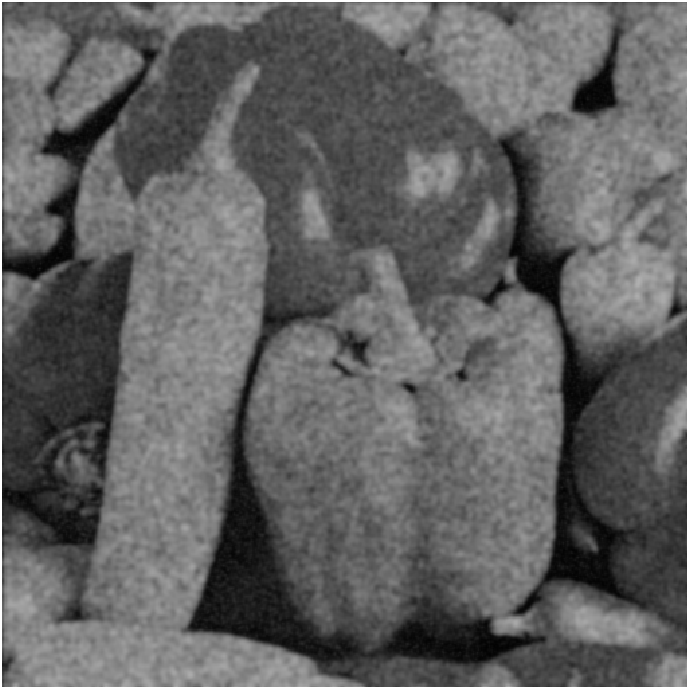}
        \caption{HPCPDE}
    \end{subfigure}
    \begin{subfigure}[b]{0.24\textwidth}
        \centering
        \includegraphics[width=\textwidth]{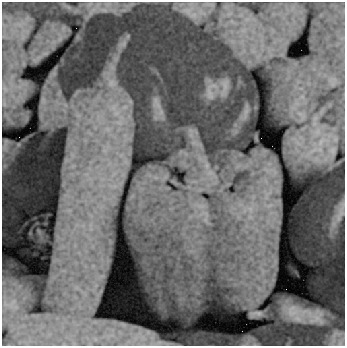}
        \caption{TDFM}
    \end{subfigure}
    \begin{subfigure}[b]{0.24\textwidth}
        \centering
        \includegraphics[width=\textwidth]{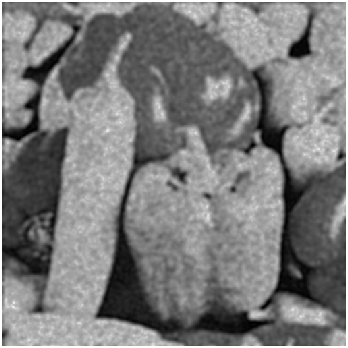}
        \caption{Proposed}
    \end{subfigure}

    \begin{subfigure}[b]{0.24\textwidth}
        \centering
        \includegraphics[width=\textwidth]{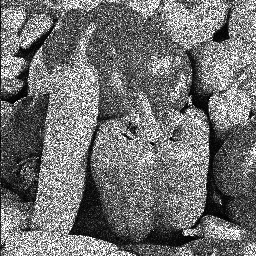}
        \caption{Look=3}
    \end{subfigure}
    \begin{subfigure}[b]{0.24\textwidth}
        \centering
        \includegraphics[width=\textwidth]{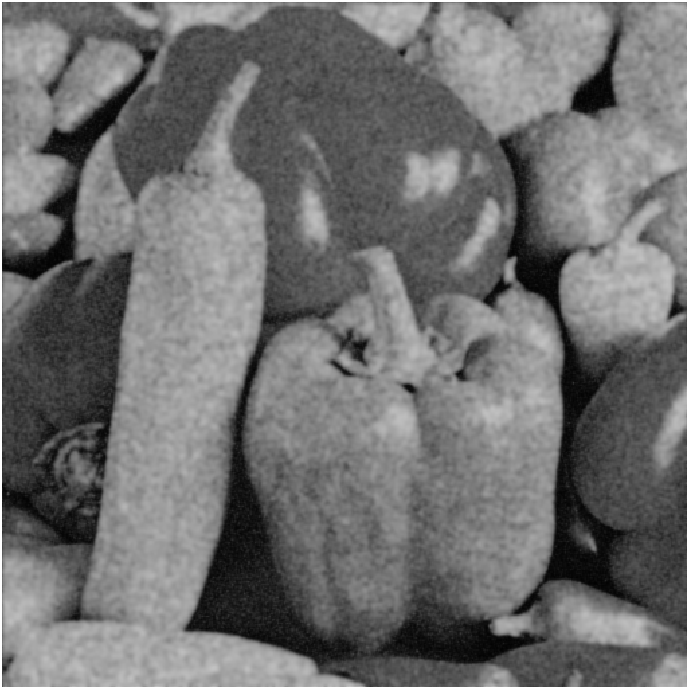}
        \caption{HPCPDE }
    \end{subfigure}
    \begin{subfigure}[b]{0.24\textwidth}
        \centering
        \includegraphics[width=\textwidth]{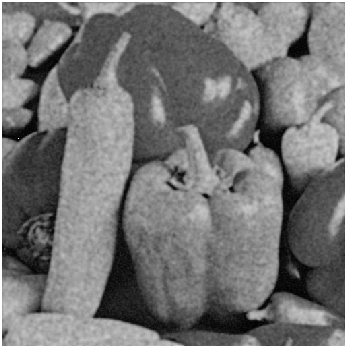}
        \caption{TDFM}
    \end{subfigure}
    \begin{subfigure}[b]{0.24\textwidth}
        \centering
        \includegraphics[width=\textwidth]{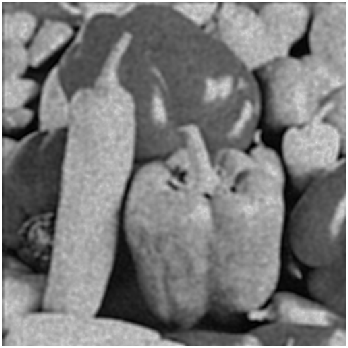}
        \caption{Proposed}
    \end{subfigure}

    \begin{subfigure}[b]{0.24\textwidth}
        \centering
        \includegraphics[width=\textwidth]{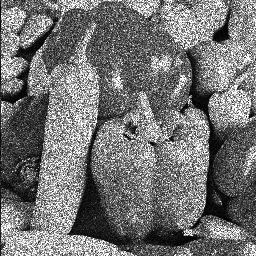}
        \caption{Look=5}
    \end{subfigure}
    \begin{subfigure}[b]{0.24\textwidth}
        \centering
        \includegraphics[width=\textwidth]{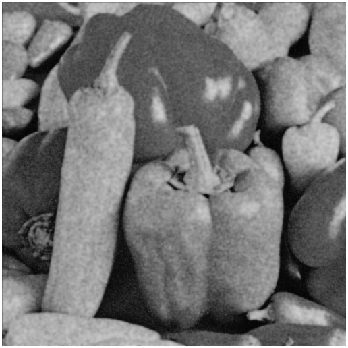 }
        \caption{HPCPDE}
    \end{subfigure}
    \begin{subfigure}[b]{0.24\textwidth}
        \centering
        \includegraphics[width=\textwidth]{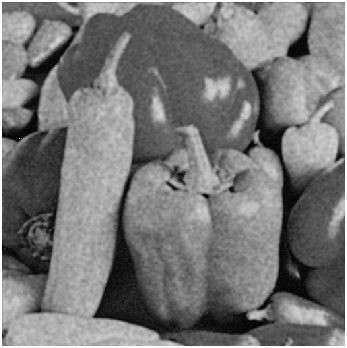}
        \caption{TDFM}
    \end{subfigure}
    \begin{subfigure}[b]{0.24\textwidth}
        \centering
        \includegraphics[width=\textwidth]{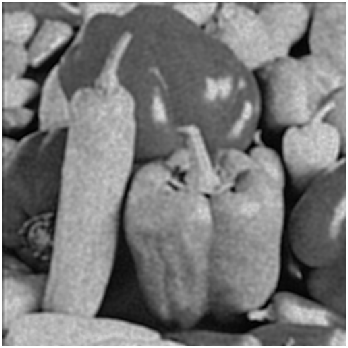}
        \caption{Proposed}
    \end{subfigure}

    \begin{subfigure}[b]{0.24\textwidth}
        \centering
        \includegraphics[width=\textwidth]{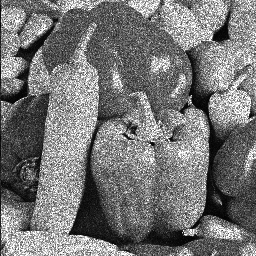}
        \caption{Look=10}
    \end{subfigure}
    \begin{subfigure}[b]{0.24\textwidth}
        \centering
        \includegraphics[width=\textwidth]{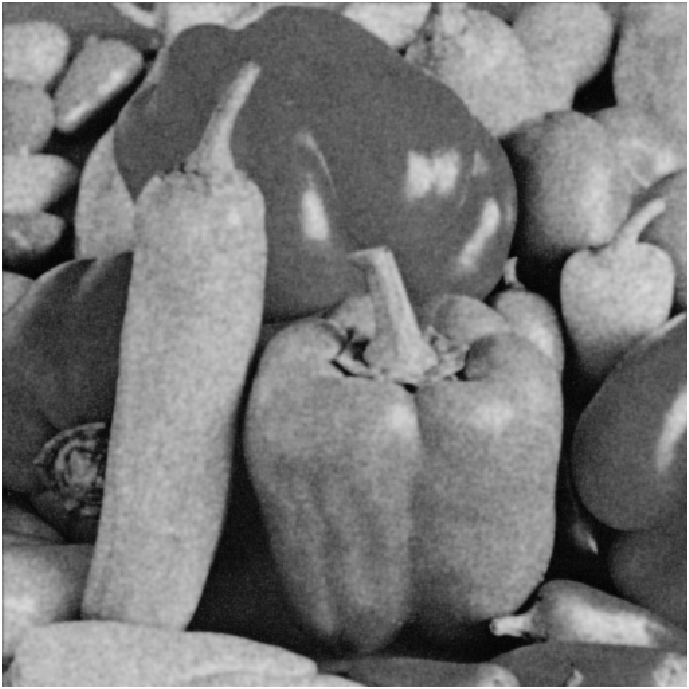}
        \caption{HPCPDE}
    \end{subfigure}
    \begin{subfigure}[b]{0.24\textwidth}
        \centering
        \includegraphics[width=\textwidth]{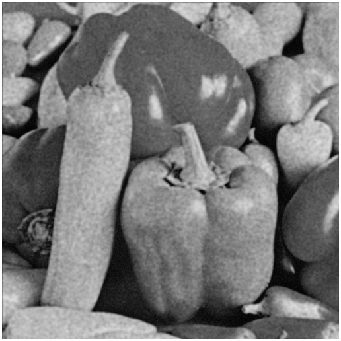}
        \caption{TDFM}
    \end{subfigure}
    \begin{subfigure}[b]{0.24\textwidth}
        \centering
        \includegraphics[width=\textwidth]{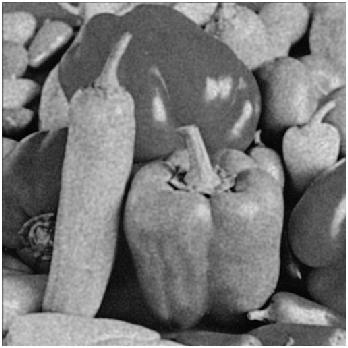}
        \caption{Proposed}
    \end{subfigure}

    \caption{
        The first column contains noisy pepper gray images with noise level Look = 1, 3, 5, 10. 
        Subsequent columns: Restored pepper gray images using different models (HPCPDE, TDFM, Proposed Model).
    }
    \label{fig:gamma_noise_results_gray_1}
\end{figure}

\begin{figure}[H]
    \centering
    \begin{subfigure}[b]{0.24\textwidth}
        \centering
        \includegraphics[width=\textwidth]{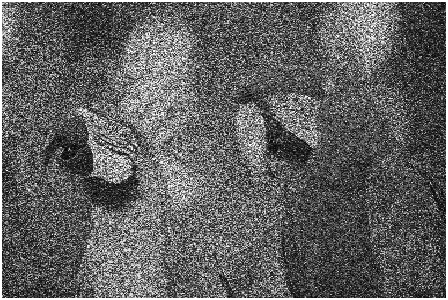}
        \caption{Look=1}
    \end{subfigure}
    \begin{subfigure}[b]{0.24\textwidth}
        \centering
        \includegraphics[width=\textwidth]{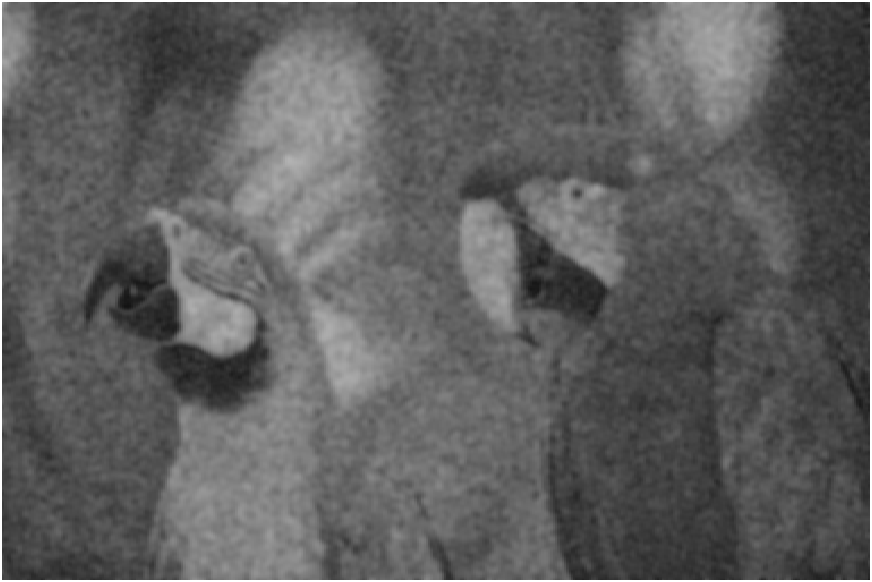}
        \caption{HPCPDE}
    \end{subfigure}
    \begin{subfigure}[b]{0.24\textwidth}
        \centering
        \includegraphics[width=\textwidth]{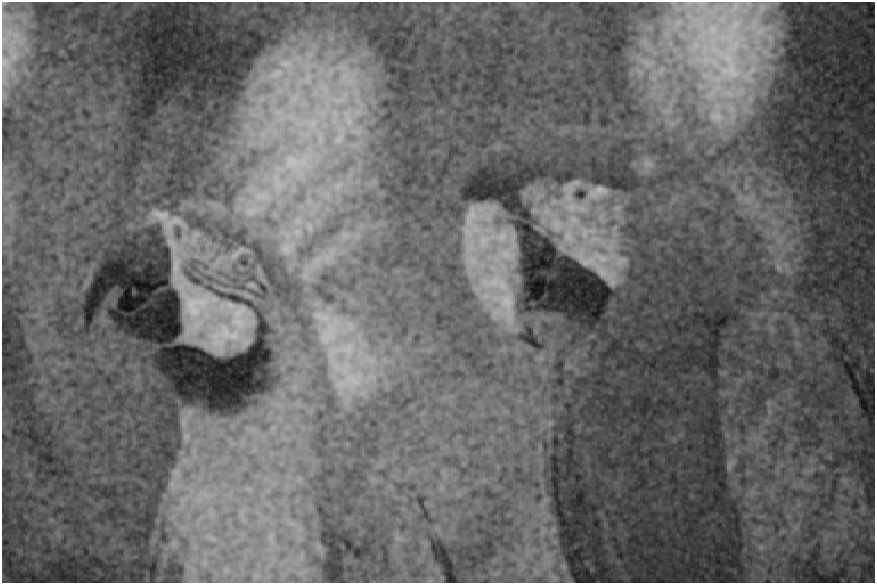}
        \caption{TDFM}
    \end{subfigure}
    \begin{subfigure}[b]{0.24\textwidth}
        \centering
        \includegraphics[width=\textwidth]{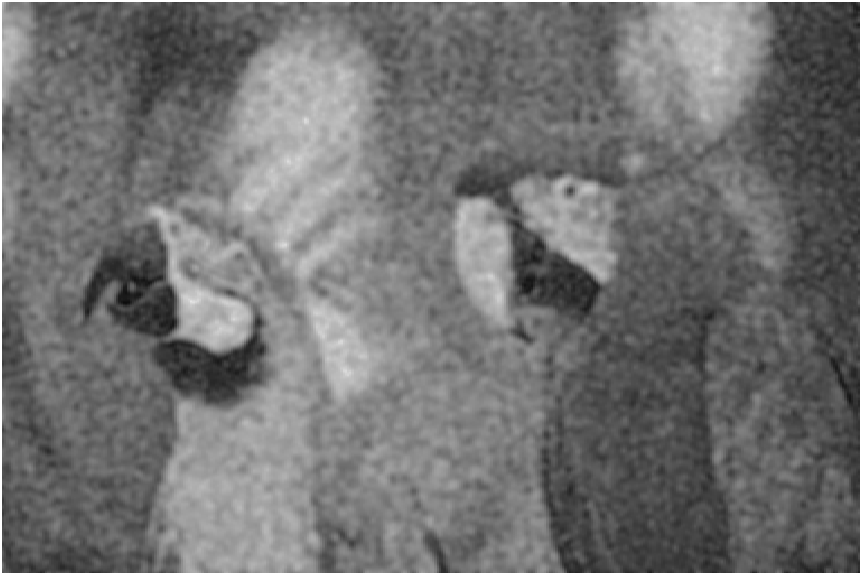}
        \caption{Proposed}
    \end{subfigure}

    \begin{subfigure}[b]{0.24\textwidth}
        \centering
        \includegraphics[width=\textwidth]{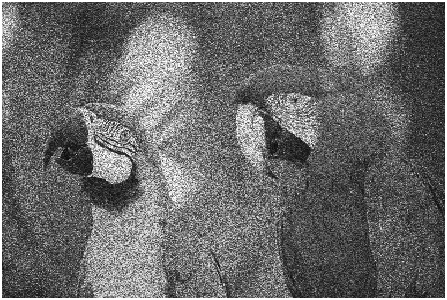}
        \caption{Look=3}
    \end{subfigure}
    \begin{subfigure}[b]{0.24\textwidth}
        \centering
        \includegraphics[width=\textwidth]{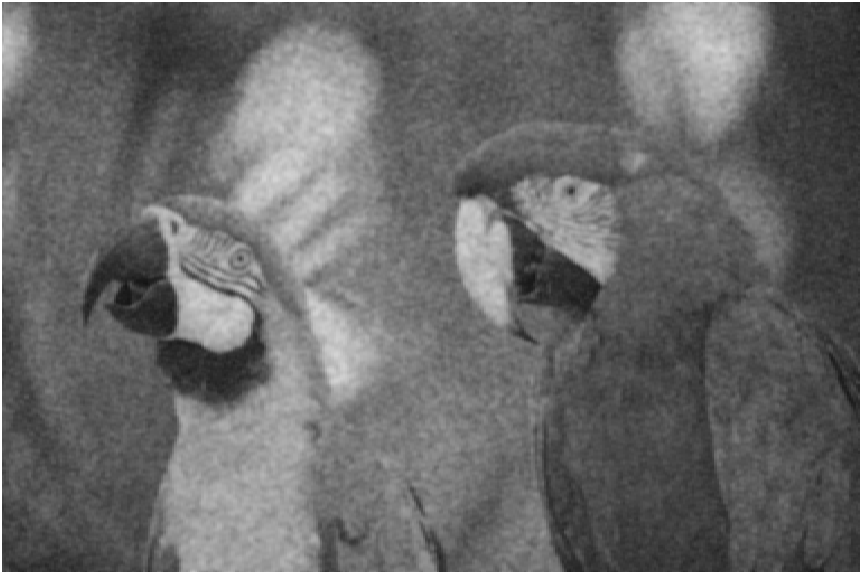}
        \caption{HPCPDE}
    \end{subfigure}
    \begin{subfigure}[b]{0.24\textwidth}
        \centering
        \includegraphics[width=\textwidth]{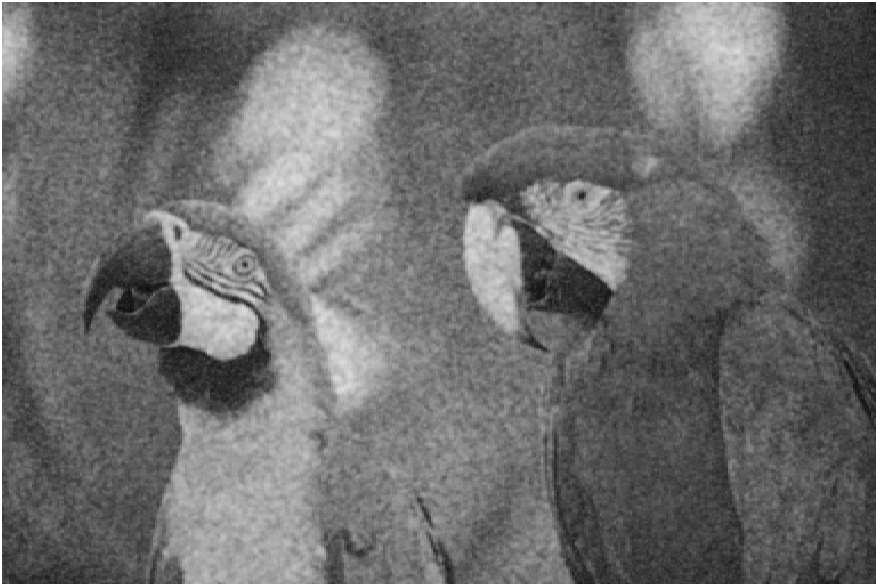}
        \caption{TDFM}
    \end{subfigure}
    \begin{subfigure}[b]{0.24\textwidth}
        \centering
        \includegraphics[width=\textwidth]{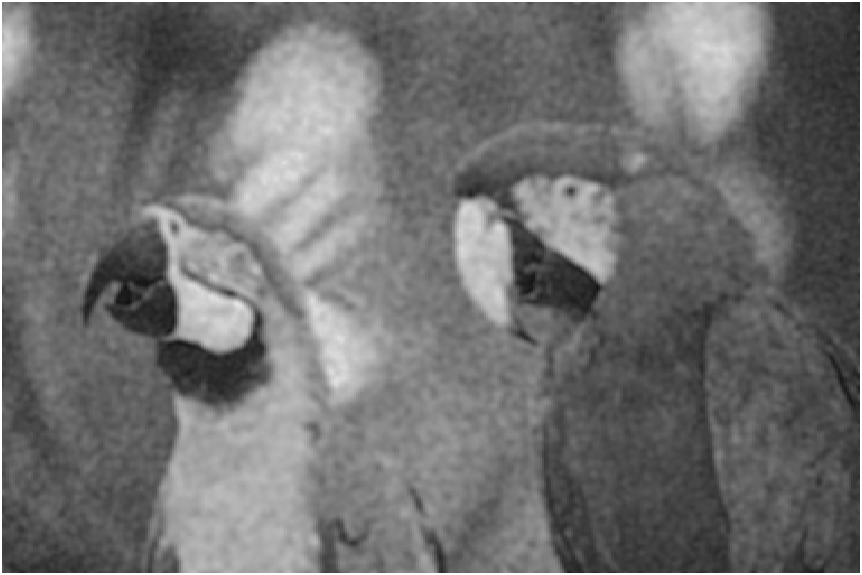}
        \caption{Proposed}
    \end{subfigure}

    \begin{subfigure}[b]{0.24\textwidth}
        \centering
        \includegraphics[width=\textwidth]{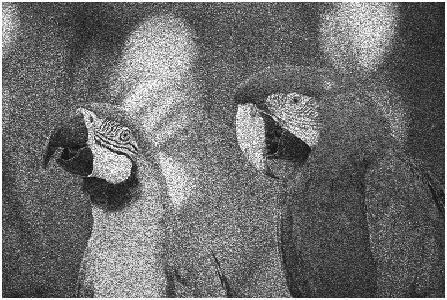}
        \caption{Look=5}
    \end{subfigure}
    \begin{subfigure}[b]{0.24\textwidth}
        \centering
        \includegraphics[width=\textwidth]{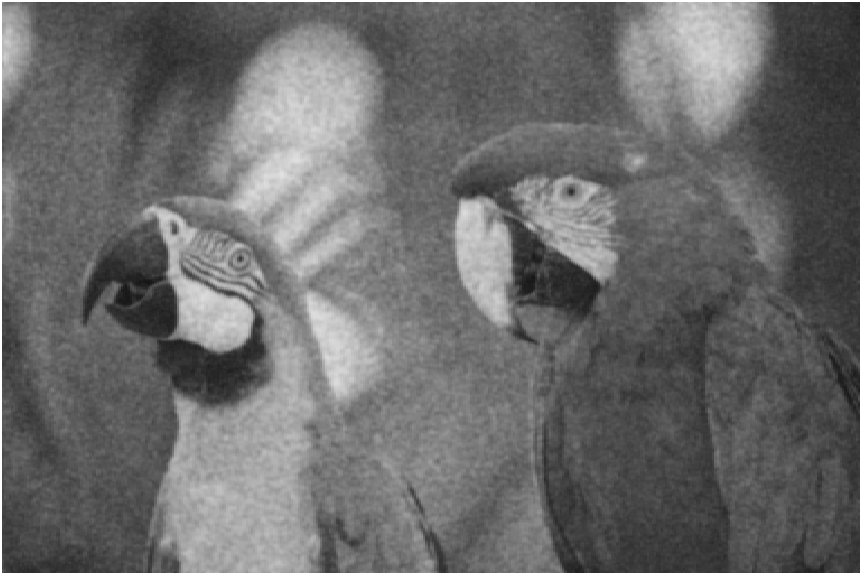}
        \caption{HPCPDE}
    \end{subfigure}
    \begin{subfigure}[b]{0.24\textwidth}
        \centering
        \includegraphics[width=\textwidth]{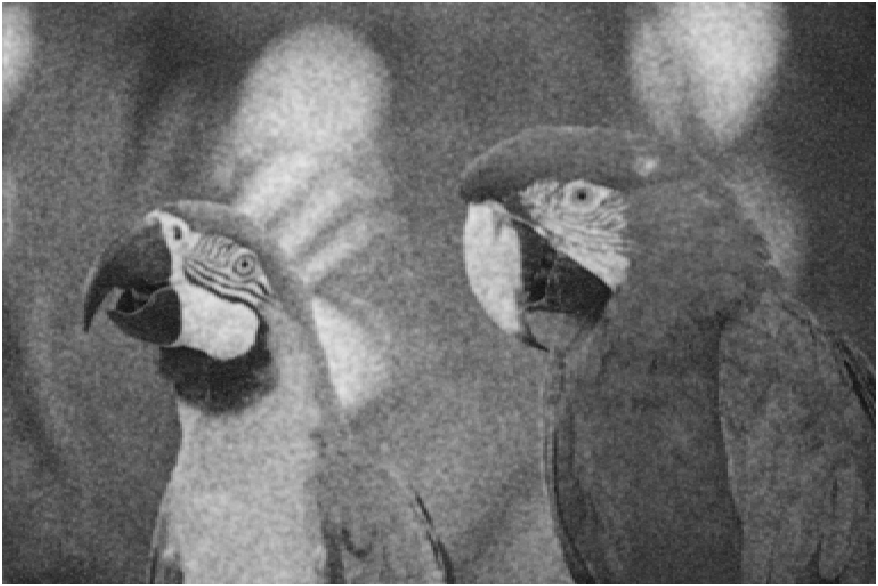}
        \caption{TDFM}
    \end{subfigure}
    \begin{subfigure}[b]{0.24\textwidth}
        \centering
        \includegraphics[width=\textwidth]{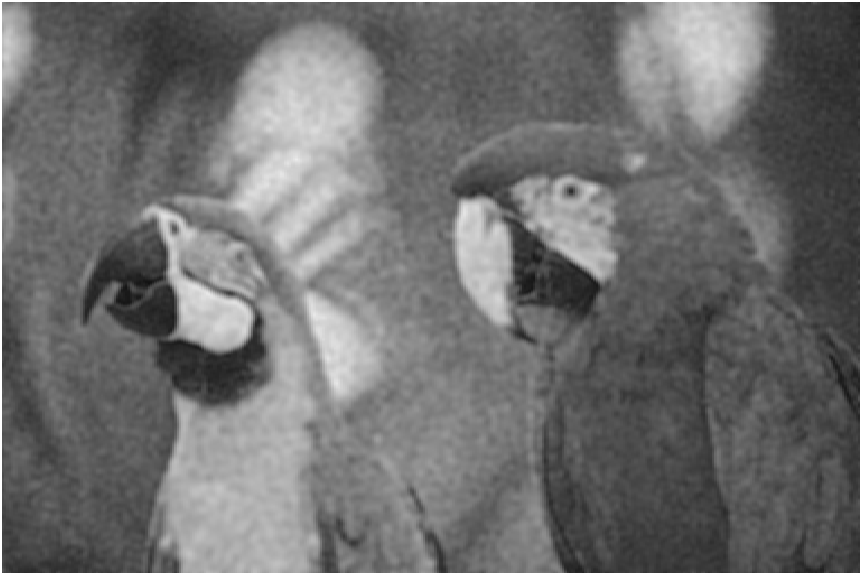}
        \caption{Proposed}
    \end{subfigure}

    \begin{subfigure}[b]{0.24\textwidth}
        \centering
        \includegraphics[width=\textwidth]{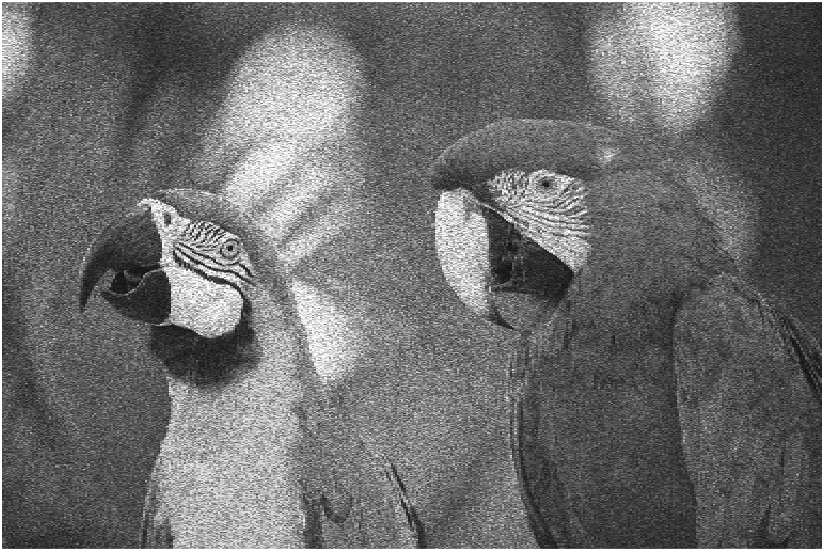}
        \caption{Look=10}
    \end{subfigure}
    \begin{subfigure}[b]{0.24\textwidth}
        \centering
        \includegraphics[width=\textwidth]{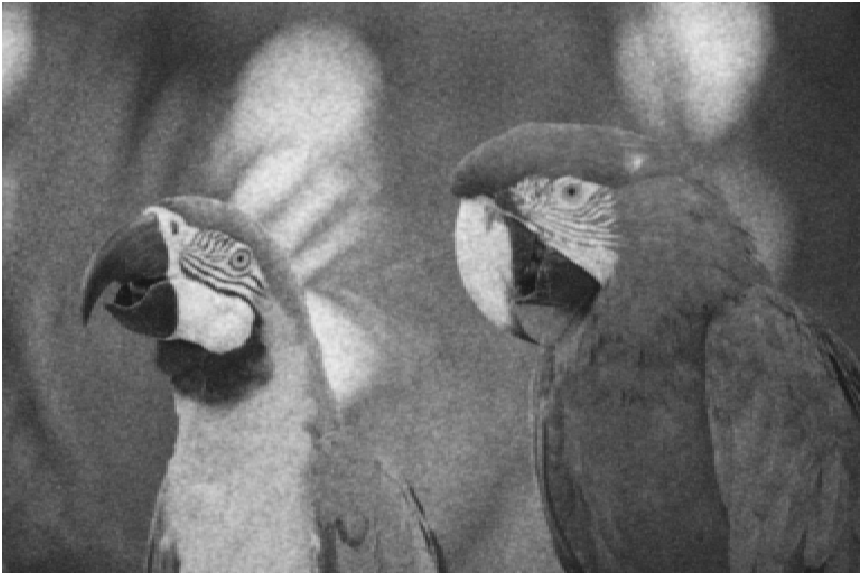}
        \caption{HPCPDE}
    \end{subfigure}
    \begin{subfigure}[b]{0.24\textwidth}
        \centering
        \includegraphics[width=\textwidth]{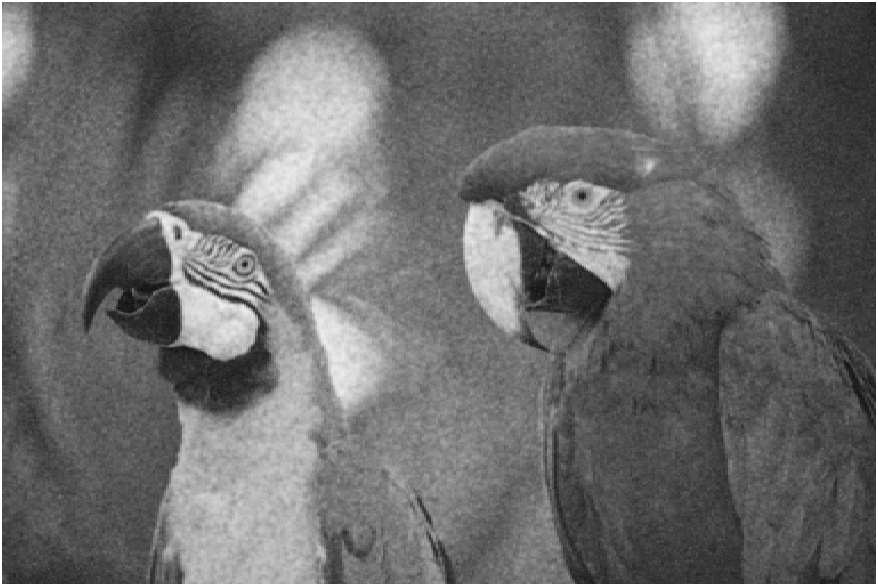}
        \caption{TDFM}
    \end{subfigure}
    \begin{subfigure}[b]{0.24\textwidth}
        \centering
        \includegraphics[width=\textwidth]{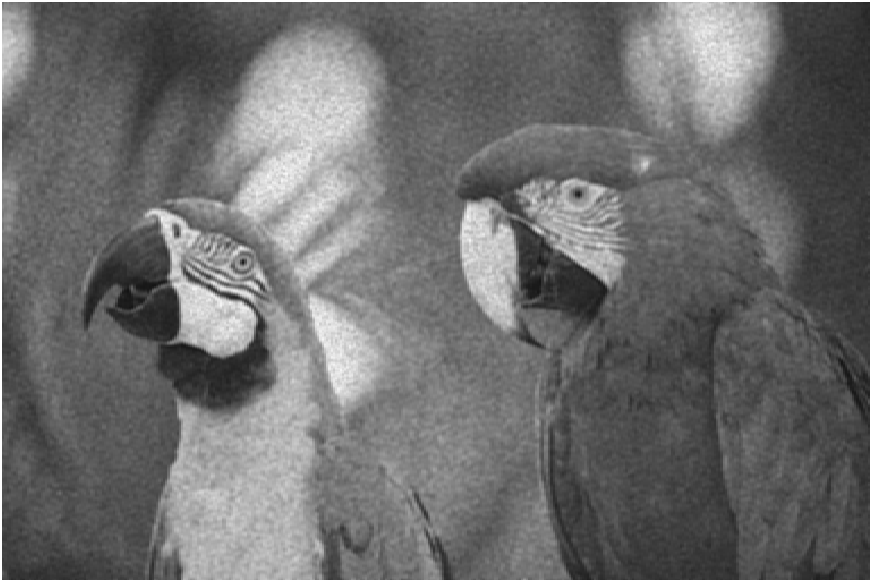}
        \caption{Proposed}
    \end{subfigure}

    \caption{
     The first column contains noisy parrots gray images with noise level Look = 1, 3, 5, 10. 
        Subsequent columns: Restored parrots gray images using different models (HPCPDE, TDFM, Proposed Model).
    }
    \label{fig:gamma_noise_results_gray_2}
\end{figure}

\begin{figure}[H]
\centering
    \begin{subfigure}[b]{0.24\textwidth}
        \centering
        \includegraphics[width=\textwidth]{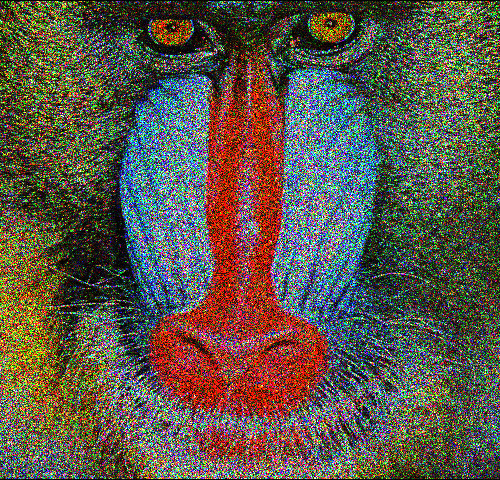}
        \caption{Look=1}
    \end{subfigure}
    \begin{subfigure}[b]{0.24\textwidth}
        \centering
        \includegraphics[width=\textwidth]{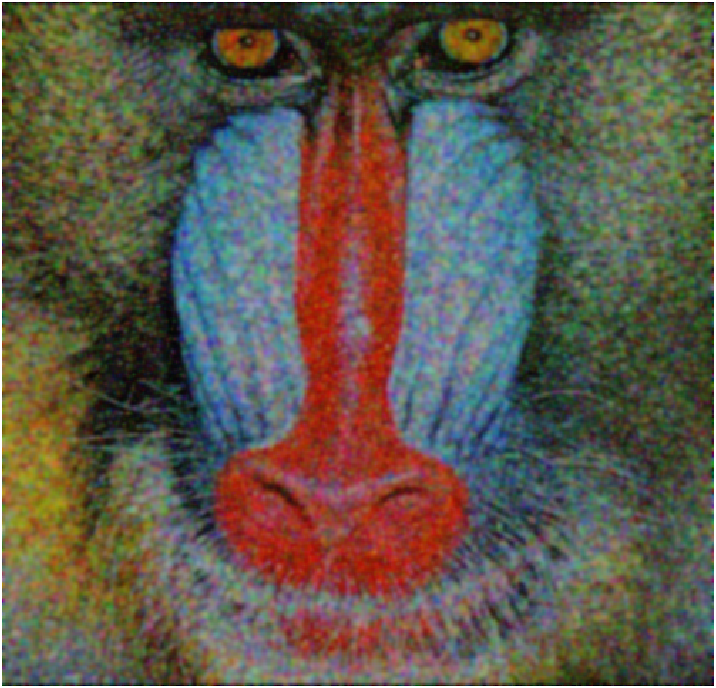}
        \caption{HPCPDE}
    \end{subfigure}
    \begin{subfigure}[b]{0.24\textwidth}
        \centering
        \includegraphics[width=\textwidth]{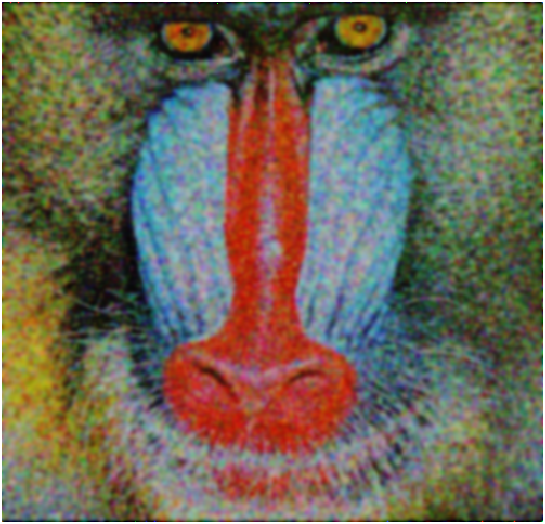}
        \caption{TDFM}
    \end{subfigure}
    \begin{subfigure}[b]{0.24\textwidth}
        \centering
        \includegraphics[width=\textwidth]{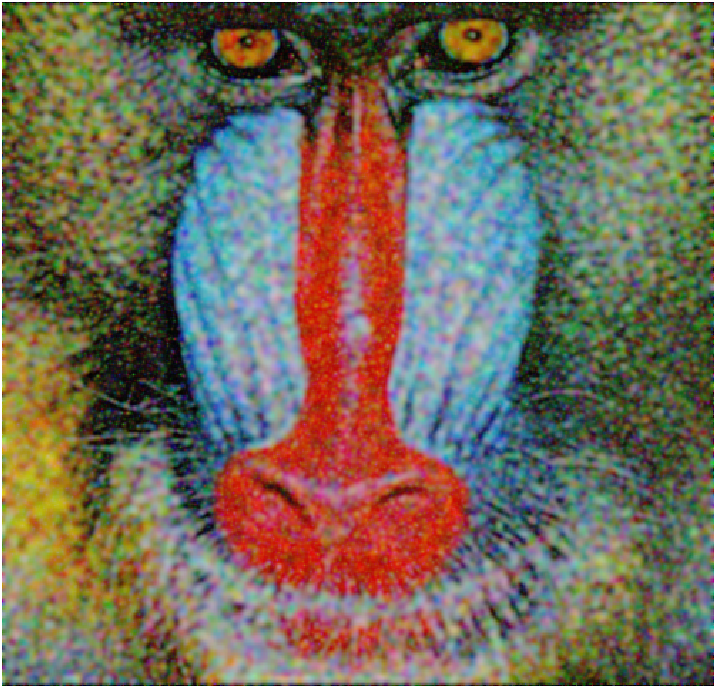}
        \caption{Proposed}
    \end{subfigure}

    \begin{subfigure}[b]{0.24\textwidth}
        \centering
        \includegraphics[width=\textwidth]{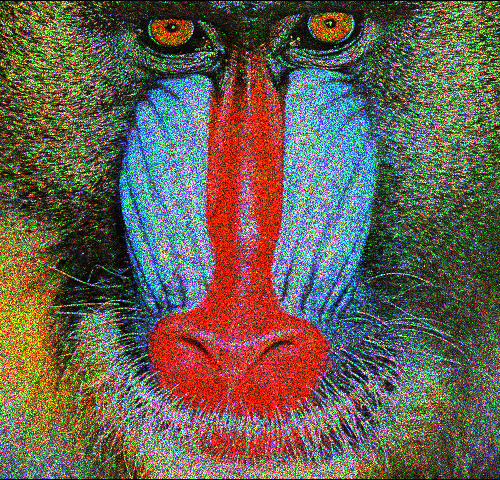}
        \caption{Look=3}
    \end{subfigure}
    \begin{subfigure}[b]{0.24\textwidth}
        \centering
        \includegraphics[width=\textwidth]{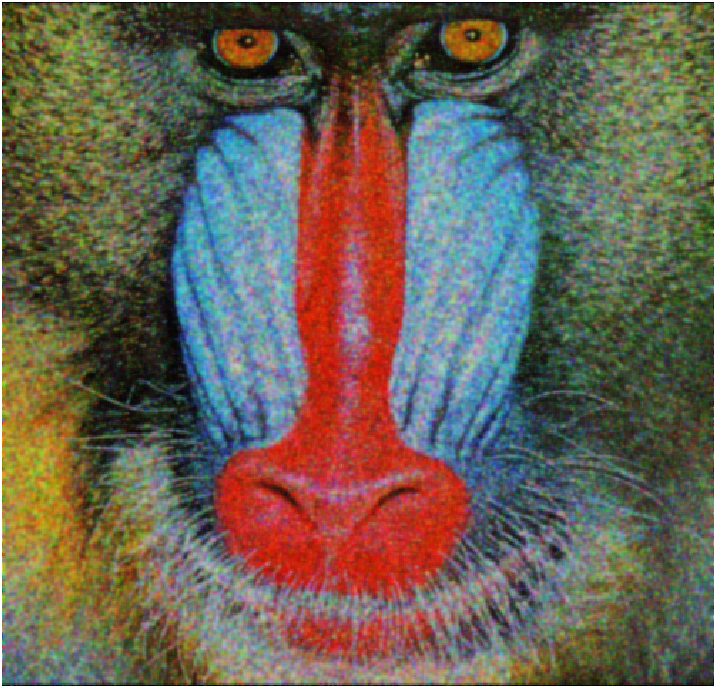}
        \caption{HPCPDE}
    \end{subfigure}
    \begin{subfigure}[b]{0.24\textwidth}
        \centering
        \includegraphics[width=\textwidth]{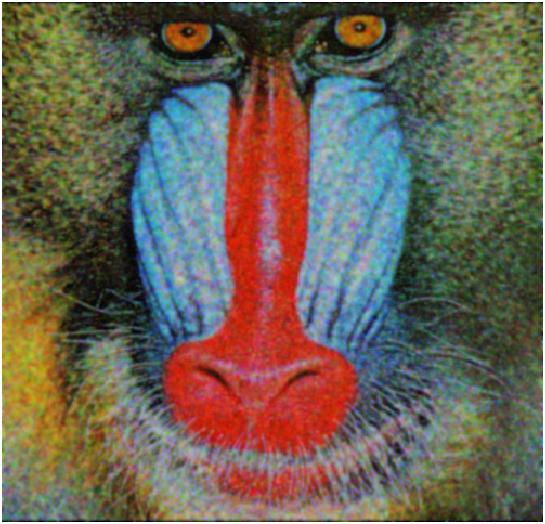}
        \caption{TDFM}
    \end{subfigure}
    \begin{subfigure}[b]{0.24\textwidth}
        \centering
        \includegraphics[width=\textwidth]{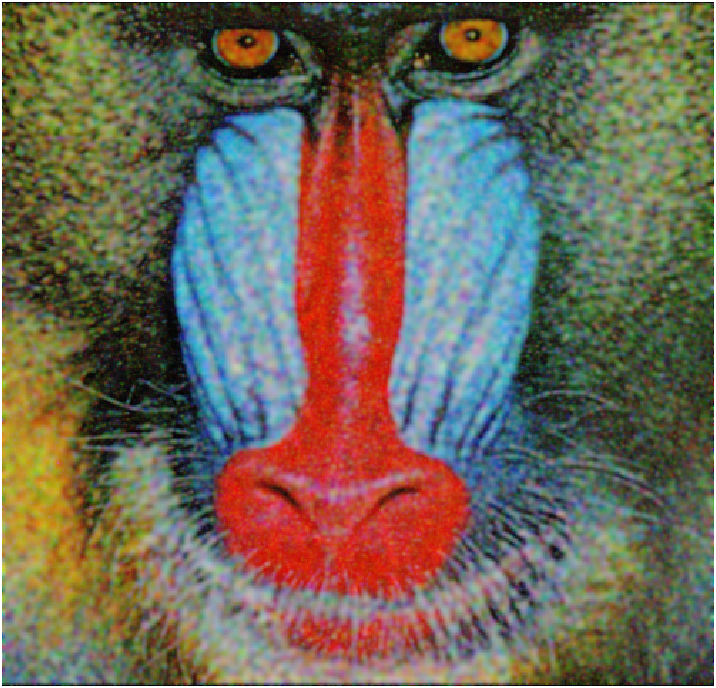}
        \caption{Proposed}
    \end{subfigure}

    \begin{subfigure}[b]{0.24\textwidth}
        \centering
        \includegraphics[width=\textwidth]{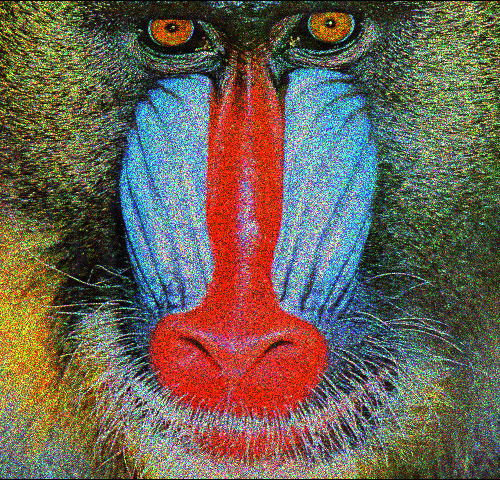}
        \caption{Look=5}
    \end{subfigure}
    \begin{subfigure}[b]{0.24\textwidth}
        \centering
        \includegraphics[width=\textwidth]{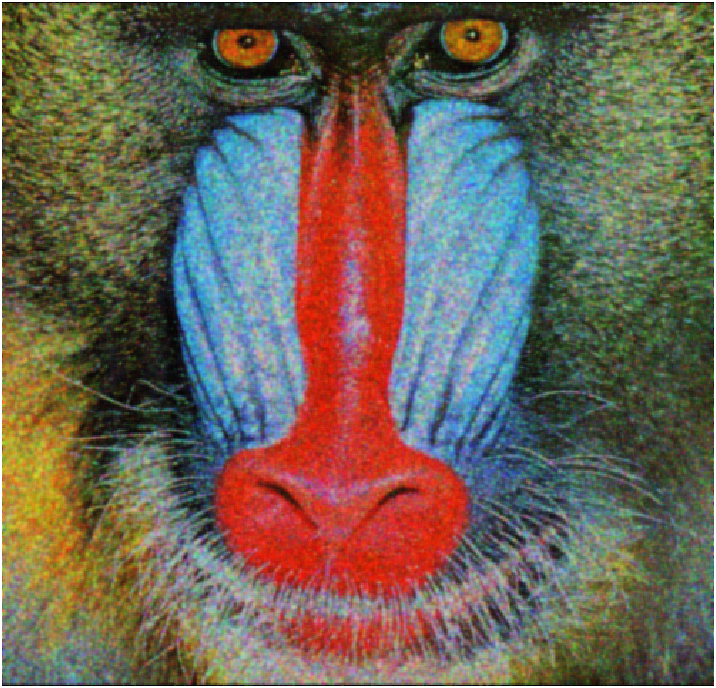}
        \caption{HPCPDE}
    \end{subfigure}
    \begin{subfigure}[b]{0.24\textwidth}
        \centering
        \includegraphics[width=\textwidth]{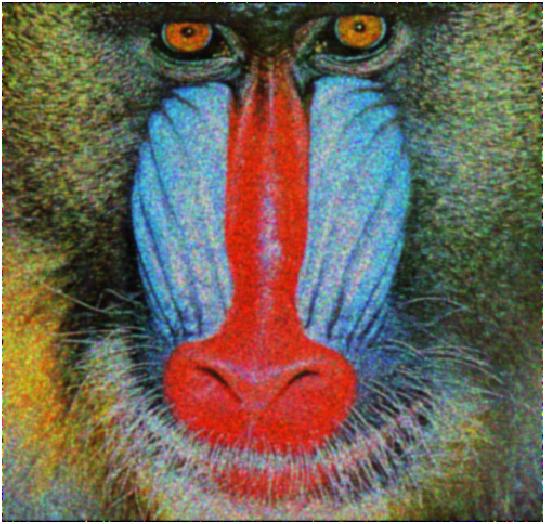}
        \caption{TDFM}
    \end{subfigure}
    \begin{subfigure}[b]{0.24\textwidth}
        \centering
        \includegraphics[width=\textwidth]{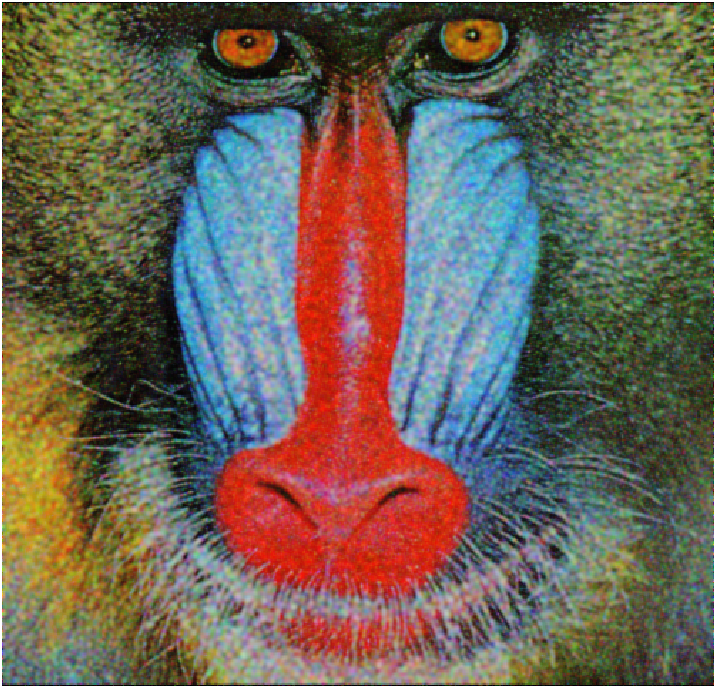}
        \caption{Proposed}
    \end{subfigure}

    \begin{subfigure}[b]{0.24\textwidth}
        \centering
        \includegraphics[width=\textwidth]{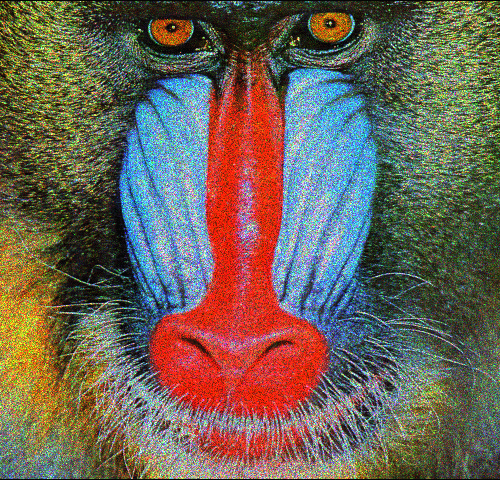}
        \caption{Look=10}
    \end{subfigure}
    \begin{subfigure}[b]{0.24\textwidth}
        \centering
        \includegraphics[width=\textwidth]{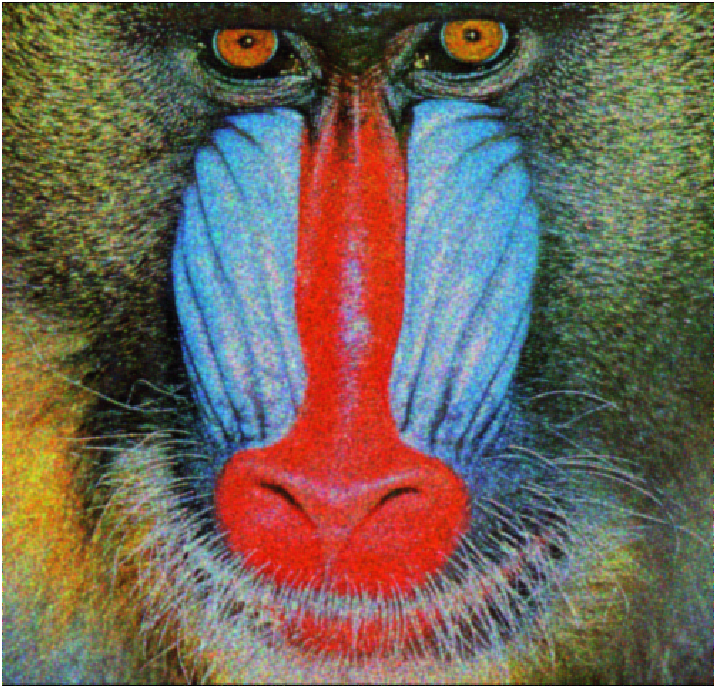}
        \caption{HPCPDE}
    \end{subfigure}
    \begin{subfigure}[b]{0.24\textwidth}
        \centering
        \includegraphics[width=\textwidth]{Colour_coupled_TDM_baboon_restored_look_10.png}
        \caption{TDFM}
    \end{subfigure}
    \begin{subfigure}[b]{0.24\textwidth}
        \centering
        \includegraphics[width=\textwidth]{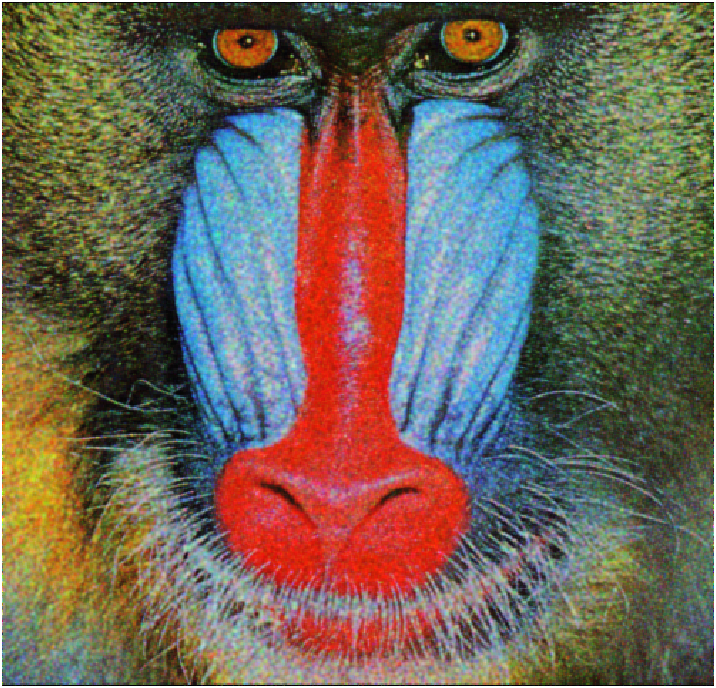}
        \caption{Proposed}
    \end{subfigure}

    \caption{
     The first column contains noisy baboon color images with noise level Look = 1, 3, 5, 10. 
        Subsequent columns: Restored baboon color images using different models (HPCPDE, TDFM, Proposed Model).
    }
    \label{fig:gamma_noise_results_color_1}
\end{figure}

\begin{figure}[H]
\centering
    \begin{subfigure}[b]{0.24\textwidth}
        \centering
        \includegraphics[width=\textwidth]{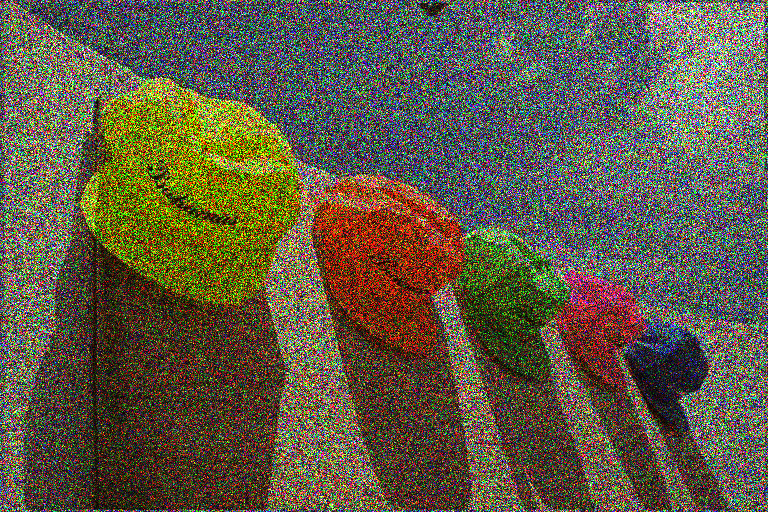}
        \caption{Look=1}
    \end{subfigure}
    \begin{subfigure}[b]{0.24\textwidth}
        \centering
        \includegraphics[width=\textwidth]{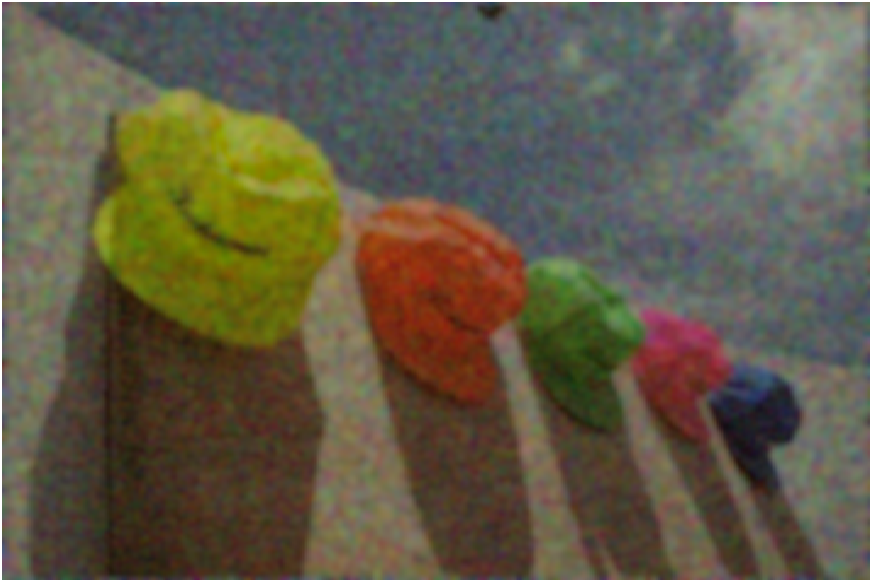}
        \caption{HPCPDE}
    \end{subfigure}
    \begin{subfigure}[b]{0.24\textwidth}
        \centering
        \includegraphics[width=\textwidth]{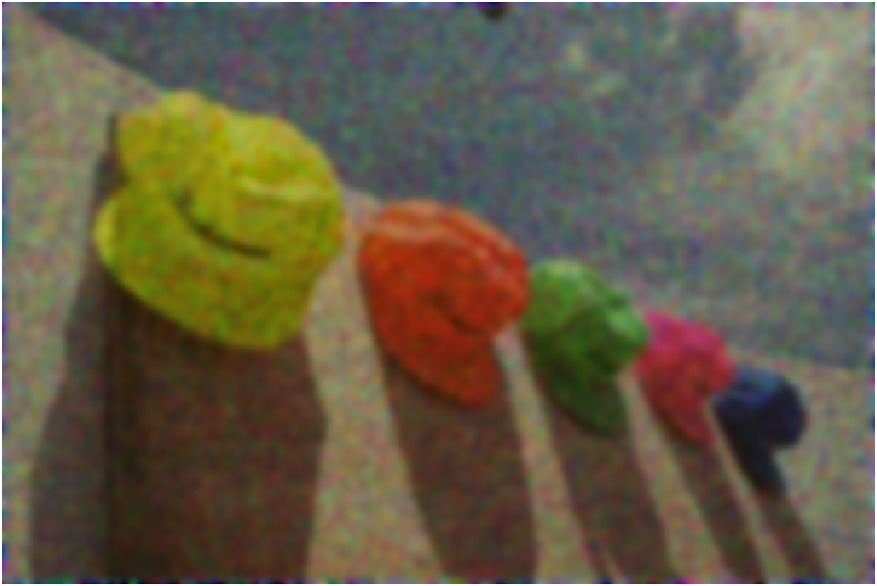}
        \caption{TDFM}
    \end{subfigure}
    \begin{subfigure}[b]{0.24\textwidth}
        \centering
        \includegraphics[width=\textwidth]{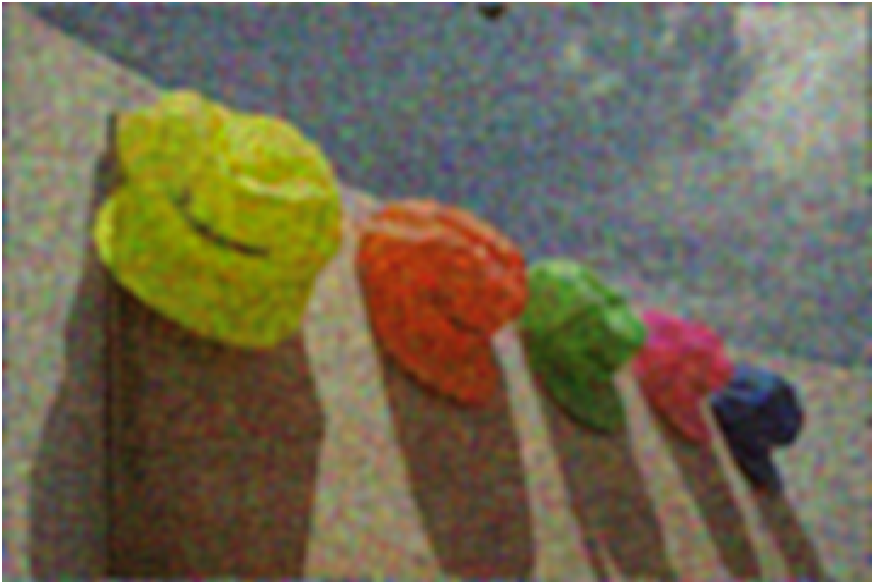}
        \caption{Proposed}
    \end{subfigure}

    \begin{subfigure}[b]{0.24\textwidth}
        \centering
        \includegraphics[width=\textwidth]{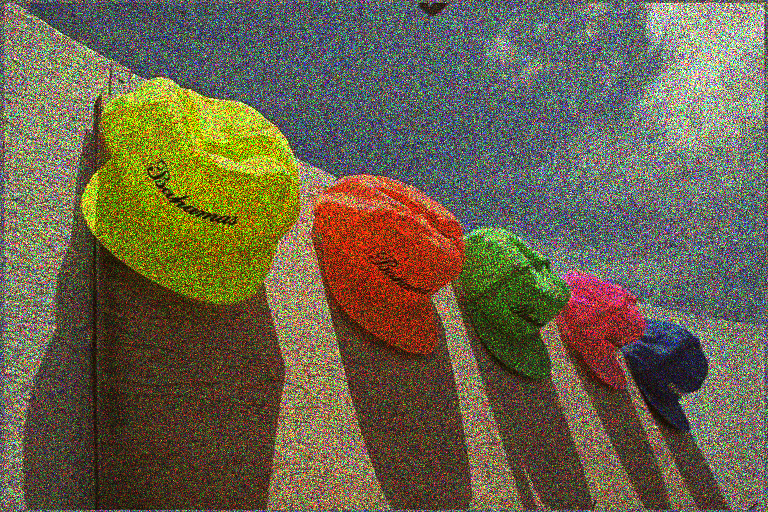}
        \caption{Look=3}
    \end{subfigure}
    \begin{subfigure}[b]{0.24\textwidth}
        \centering
        \includegraphics[width=\textwidth]{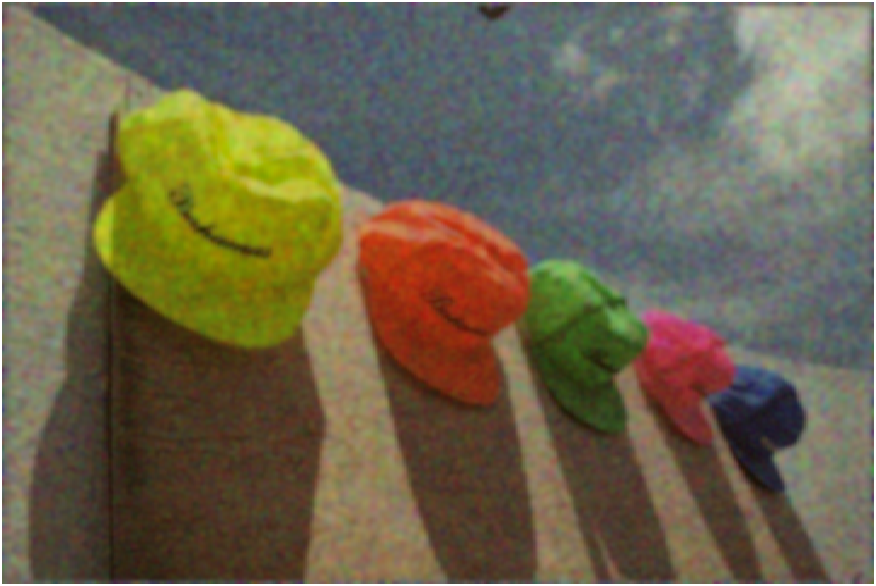}
        \caption{HPCPDE}
    \end{subfigure}
    \begin{subfigure}[b]{0.24\textwidth}
        \centering
        \includegraphics[width=\textwidth]{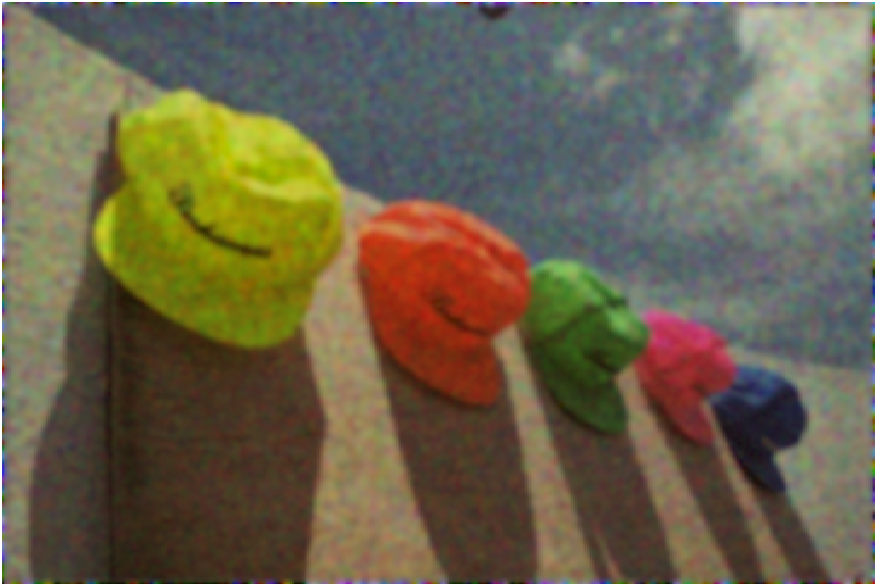}
        \caption{TDFM}
    \end{subfigure}
    \begin{subfigure}[b]{0.24\textwidth}
        \centering
        \includegraphics[width=\textwidth]{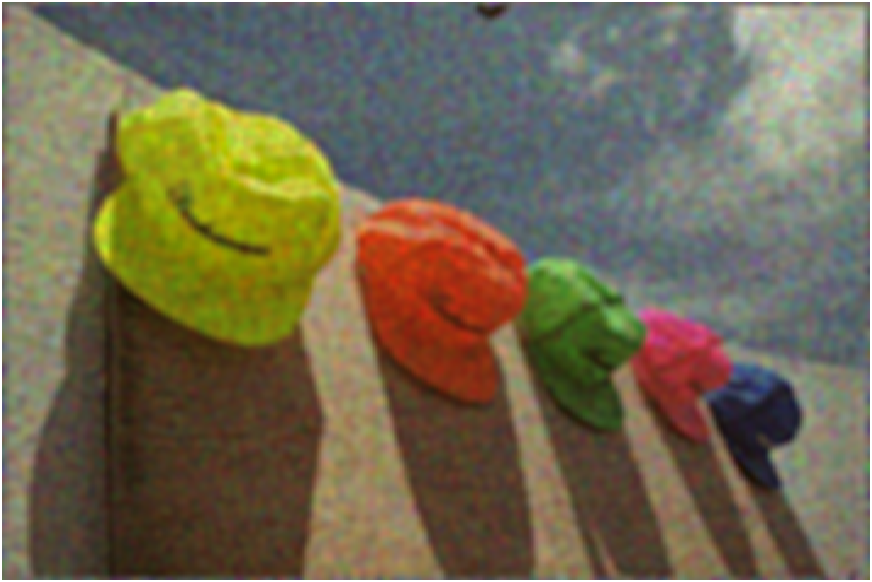}
        \caption{Proposed}
    \end{subfigure}

    \begin{subfigure}[b]{0.24\textwidth}
        \centering
        \includegraphics[width=\textwidth]{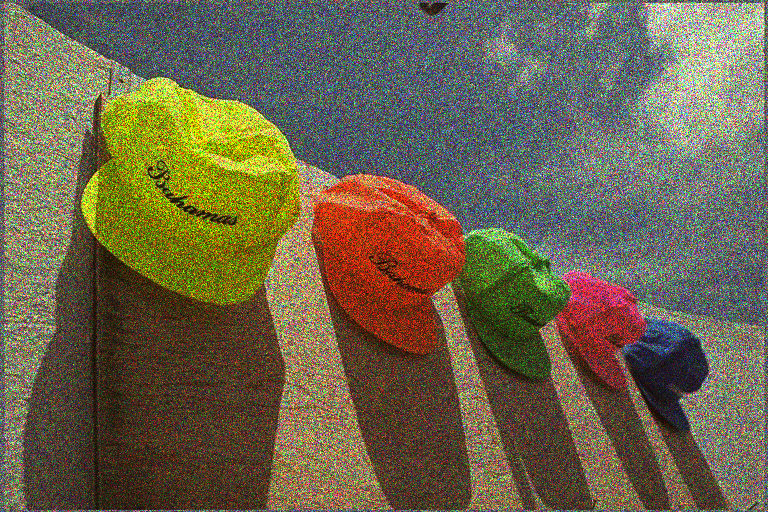}
        \caption{Look=5}
    \end{subfigure}
    \begin{subfigure}[b]{0.24\textwidth}
        \centering
        \includegraphics[width=\textwidth]{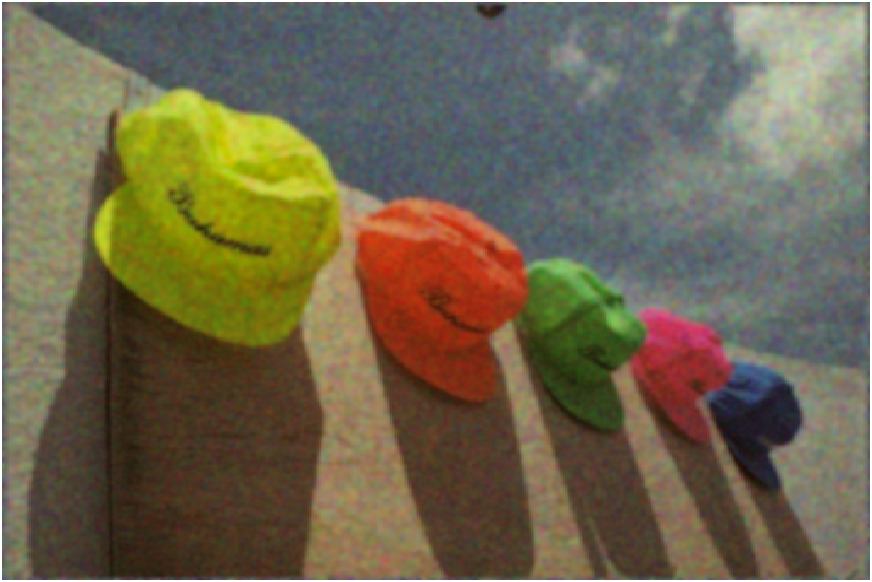}
        \caption{HPCPDE}
    \end{subfigure}
    \begin{subfigure}[b]{0.24\textwidth}
        \centering
        \includegraphics[width=\textwidth]{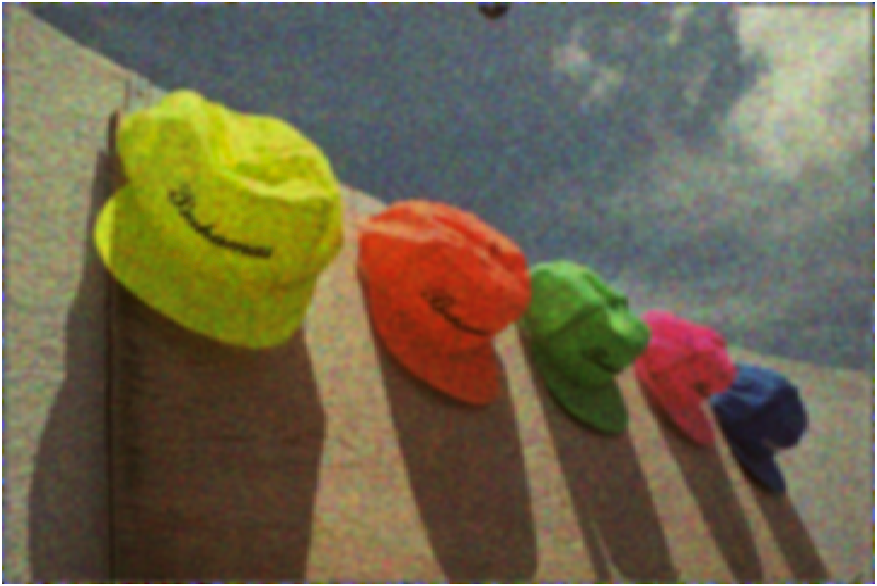}
        \caption{TDFM}
    \end{subfigure}
    \begin{subfigure}[b]{0.24\textwidth}
        \centering
        \includegraphics[width=\textwidth]{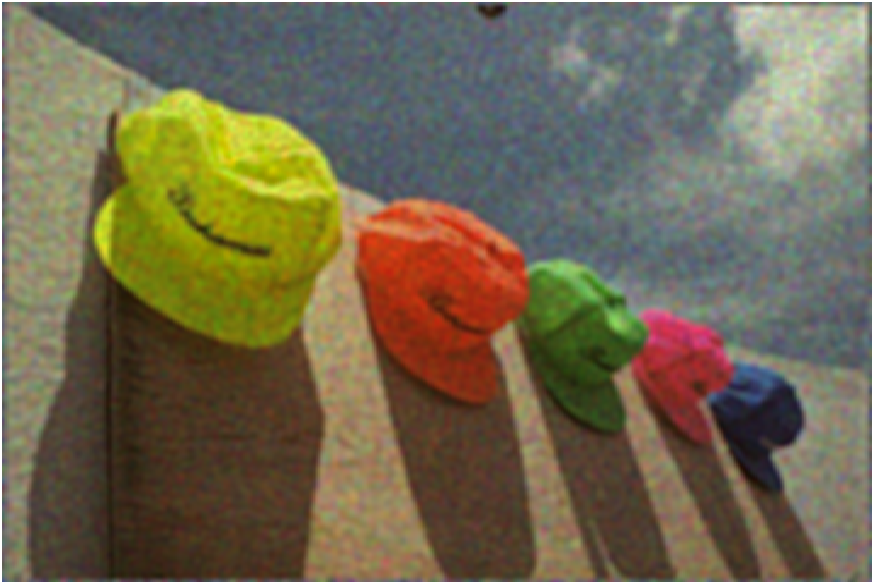}
        \caption{Proposed}
    \end{subfigure}

    \begin{subfigure}[b]{0.24\textwidth}
        \centering
        \includegraphics[width=\textwidth]{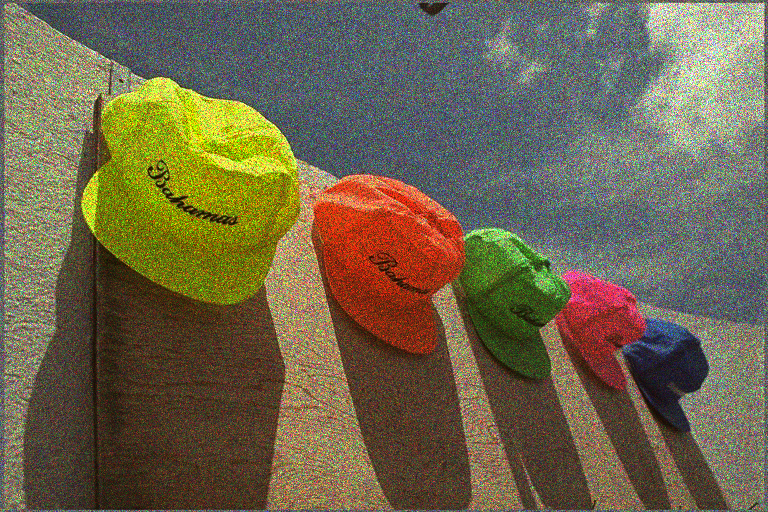}
        \caption{Look=10}
    \end{subfigure}
    \begin{subfigure}[b]{0.24\textwidth}
        \centering
        \includegraphics[width=\textwidth]{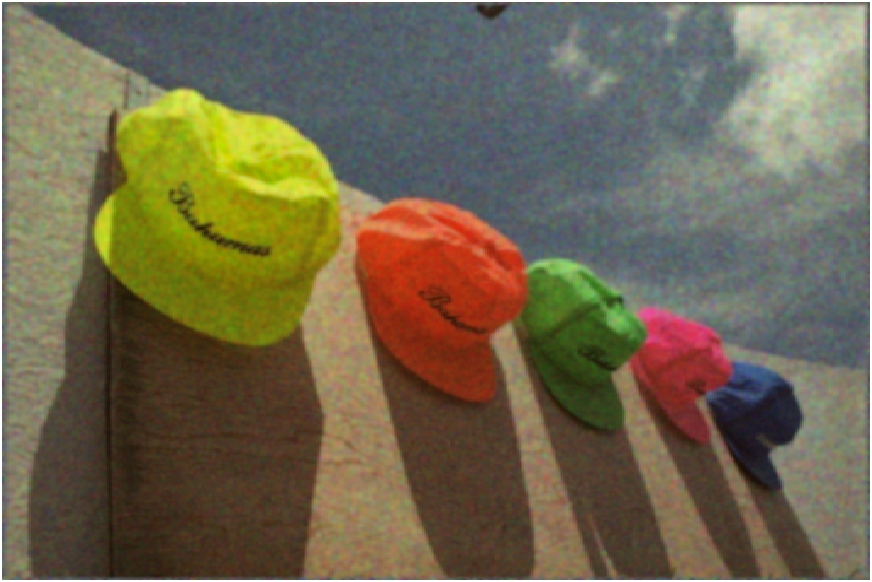}
        \caption{HPCPDE}
    \end{subfigure}
    \begin{subfigure}[b]{0.24\textwidth}
        \centering
        \includegraphics[width=\textwidth]{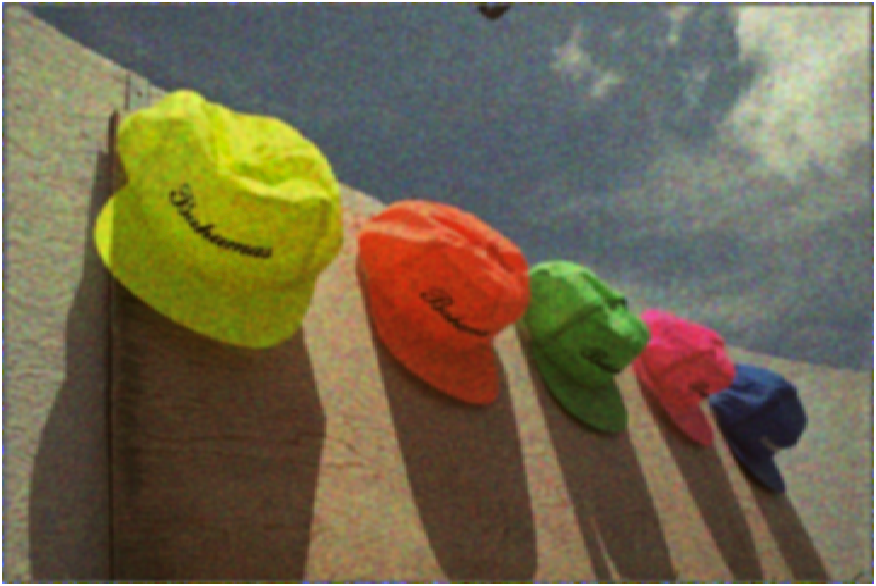}
        \caption{TDFM}
    \end{subfigure}
    \begin{subfigure}[b]{0.24\textwidth}
        \centering
        \includegraphics[width=\textwidth]{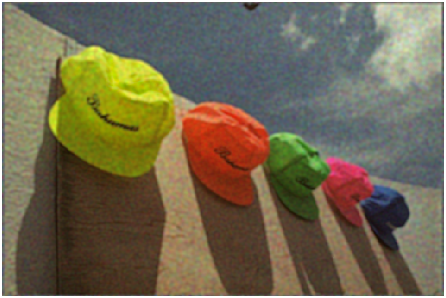}
        \caption{Proposed}
    \end{subfigure}

    \caption{
      The first column contains images of noisy caps with noise levels of Look = 1, 3, 5, and 10. 
        Subsequent columns: Restored color images using different models (HPCPDE, TDFM, Proposed Model)..
    }
    \label{fig:gamma_noise_resultscolor_2}
\end{figure}

Figures~\ref{fig:images_0} and~\ref{fig:images_0c} show the original grayscale and color images, respectively. Speckle multiplicative noise with looks \( 1, 3, 5, 10\) is added to both grayscale and color images, with the corresponding noisy results presented in Figures ~\ref{fig:gamma_noise_results_gray_1}~\ref{fig:gamma_noise_results_gray_2}~\ref{fig:gamma_noise_results_color_1}~\ref{fig:gamma_noise_resultscolor_2}. These figures demonstrate the performance of the proposed denoising model under varying speckle noise levels. The test images used include widely recognized benchmarks such as peppers, parrots, baboons, and caps, offering a comprehensive evaluation of the model’s robustness across diverse textures and structural features. Visual inspection allows us to qualitatively assess the denoising effect by comparing the noisy and denoised images; however, visually comparing the performance of two different models is challenging. Therefore, we quantitatively evaluate the models using Peak Signal-to-Noise Ratio (PSNR) and Mean Structural Similarity Index (MSSIM) metrics for both gray and color images. Tables ~\ref{tab:comparison_gray} and ~\ref{tab:comparison_color} summarize these results, demonstrating that the proposed model outperforms existing state-of-the-art methods by a significant margin. Higher PSNR and MSSIM values indicate superior denoising performance and better preservation of image details, confirming the effectiveness of our approach.

All models were evaluated at the same timestep with fixed parameters; however, some models use adjustable parameters. These parameters are not fixed because even small changes can have a significant impact on model performance. To achieve optimal denoising results and maximize PSNR and MSSIM values, we carefully tuned the parameters for different image types. Tables ~\ref{tab:parameter_gray} and ~\ref{tab:parameter_color} detail the parameter settings used for grayscale and color images, respectively. This detailed parameter configuration ensures that the experiments are reproducible and that the results are reliable.

\begin{figure}[H]
    \centering
    \begin{subfigure}[b]{0.25\textwidth}
        \centering
        \includegraphics[width=\textwidth, keepaspectratio]{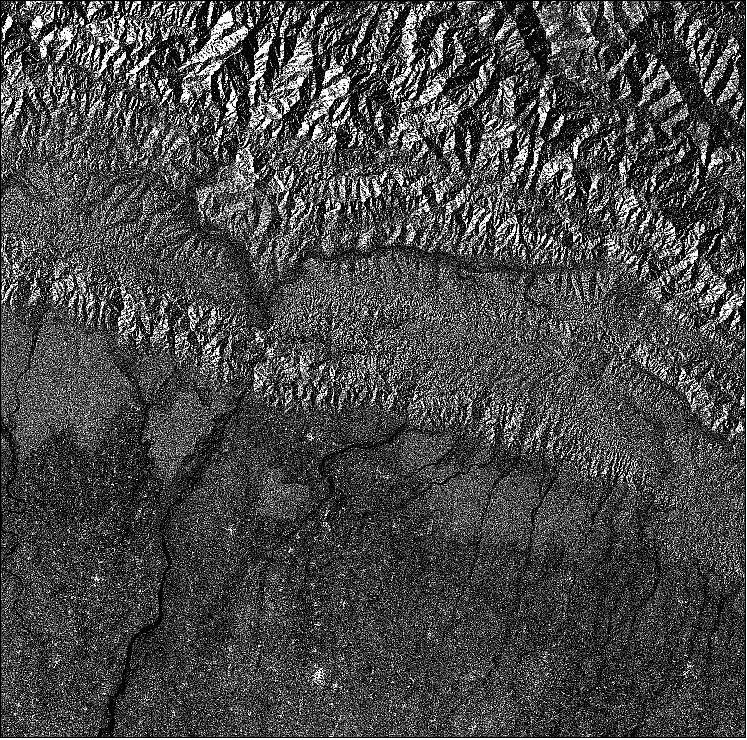}
        \caption{Image 1}
        \label{fig:image1_noisy1}
    \end{subfigure}
    \begin{subfigure}[b]{0.25\textwidth}
        \centering
        \includegraphics[width=\textwidth, keepaspectratio]{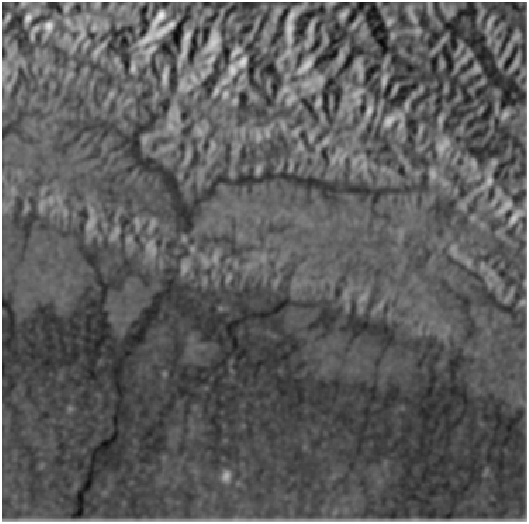}
        \caption{TDFM}
        \label{fig:image1_forth}
    \end{subfigure}
    \begin{subfigure}[b]{0.25\textwidth}
        \centering
        \includegraphics[width=\textwidth, keepaspectratio]{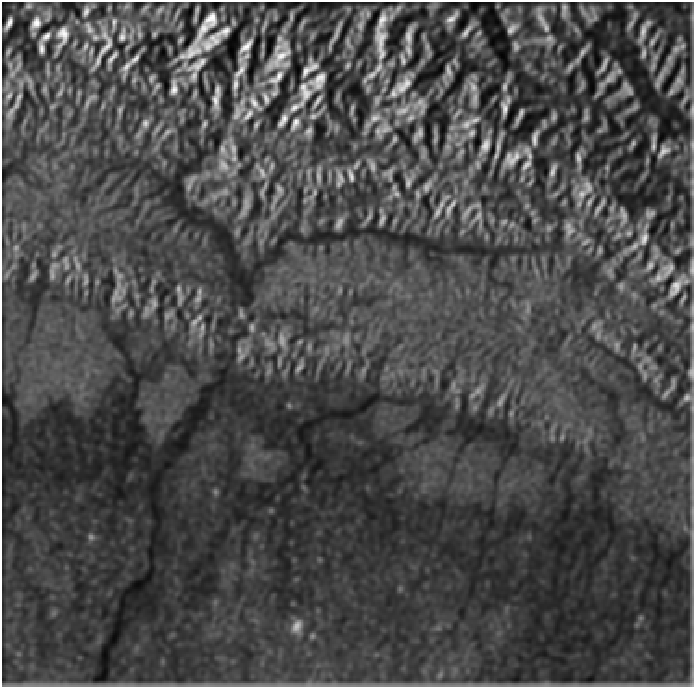}
        \caption{Proposed}
        \label{fig:image1_proposed}
    \end{subfigure}

    \begin{subfigure}[b]{0.25\textwidth}
        \centering
        \includegraphics[width=\textwidth, keepaspectratio]{ULTRASOUND3.png}
        \caption{Image 2}
        \label{fig:image2_noisy2}
    \end{subfigure}
    \begin{subfigure}[b]{0.25\textwidth}
        \centering
        \includegraphics[width=\textwidth, keepaspectratio]{Forth_Hyperbolic_ultrasound3_Restored.png}
        \caption{TDFM}
        \label{fig:image2_forth}
    \end{subfigure}
    \begin{subfigure}[b]{0.25\textwidth}
        \centering
        \includegraphics[width=\textwidth, keepaspectratio]{coupled_ultrasound3_FORTH_restored.png}
        \caption{Proposed}
        \label{fig:image2_proposed}
    \end{subfigure}
    
    \caption{Comparison of SAR and Ultrasound Image Restoration. Each row presents: (1) Noisy Image, (2) TDFM Restored Image, and (3) Proposed Model Restored Image.}
    \label{fig:sar_results}
\end{figure}

\begin{figure}[H]
    \centering

    \begin{subfigure}[b]{0.40\textwidth}
        \includegraphics[width=\linewidth]{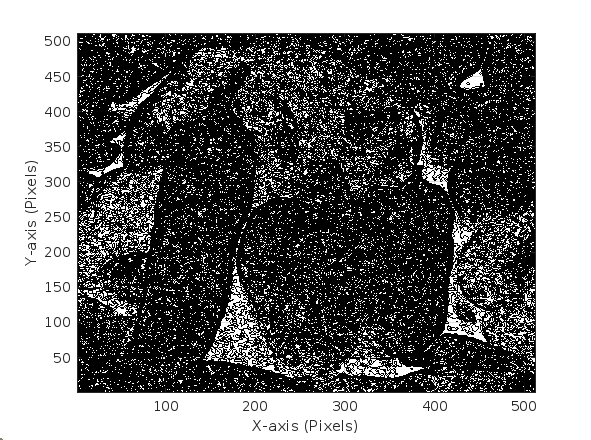}
        \caption{Noisy Image}
        \label{fig:noisy}
    \end{subfigure}
    \hspace{1em}
    \begin{subfigure}[b]{0.40\textwidth}
        \includegraphics[width=\linewidth]{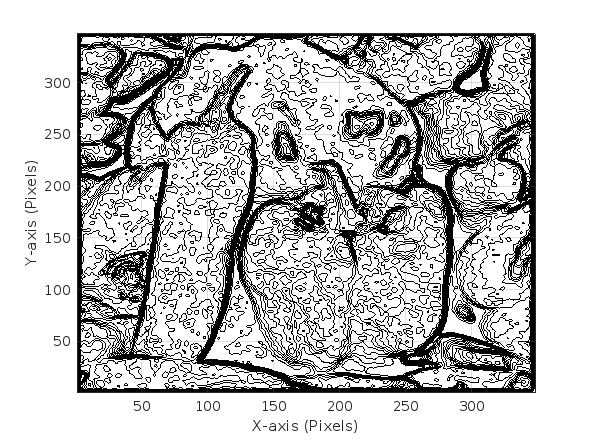}
        \caption{HPCPDE Restored}
        \label{fig:shan}
    \end{subfigure}
    
    \vspace{1em}
    \begin{subfigure}[b]{0.40\textwidth}
        \includegraphics[width=\linewidth]{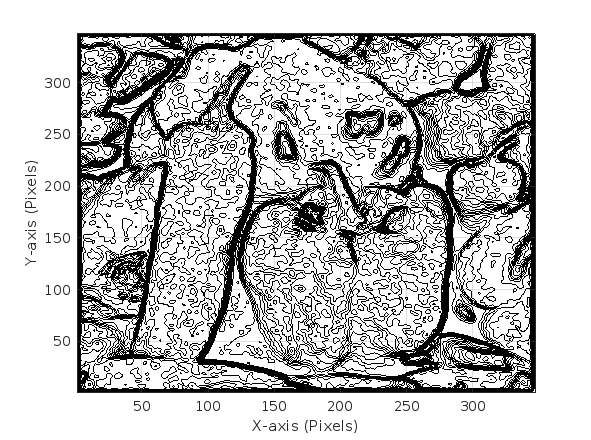}
        \caption{TDFM Restored}
        \label{fig:tde}
    \end{subfigure}
    \hspace{1em}
    \begin{subfigure}[b]{0.40\textwidth}
        \includegraphics[width=\linewidth]{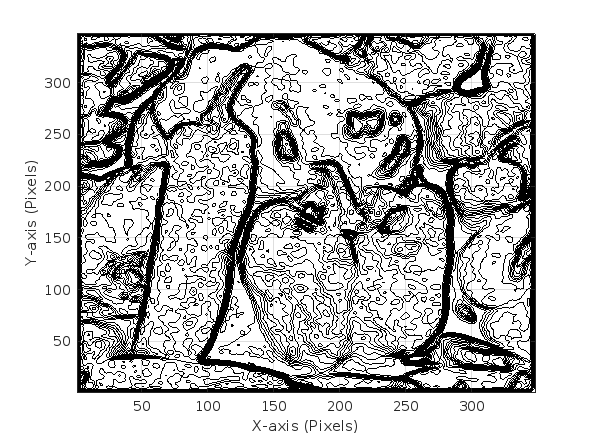}
        \caption{Proposed Restored}
        \label{fig:proposed}
    \end{subfigure}

    \caption{2D contour plots of the Peppers grayscale image at Look = 3. 
The comparison illustrates (a) noisy input, (b) restoration using the HPCPDE model, 
(c) restoration using the TDFM, and (d) restoration using the proposed model.}

    \label{fig:2d_comparisons}
\end{figure}

\begin{figure}[H]
    \centering
    \begin{minipage}{0.40\textwidth}
        \centering
        \includegraphics[width=\linewidth]{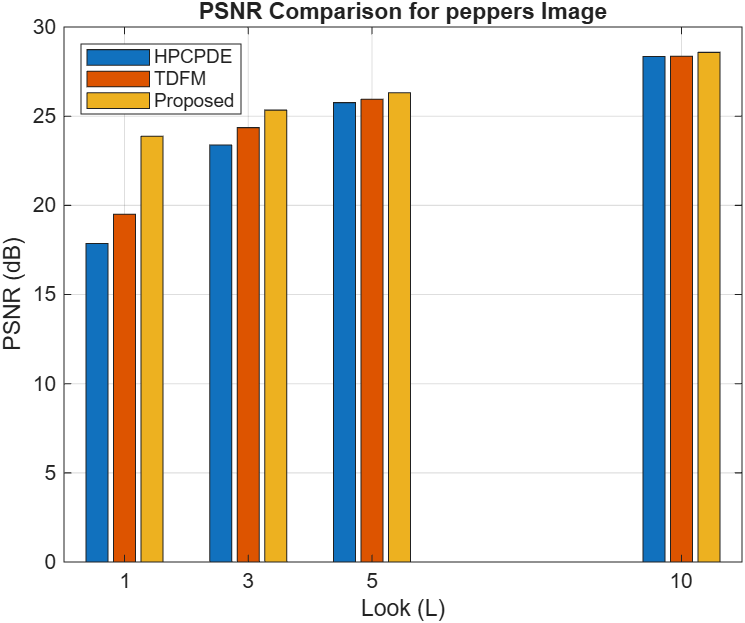}
        
    \end{minipage}
    \hfill
    \begin{minipage}{0.40\textwidth}
        \centering
        \includegraphics[width=\linewidth]{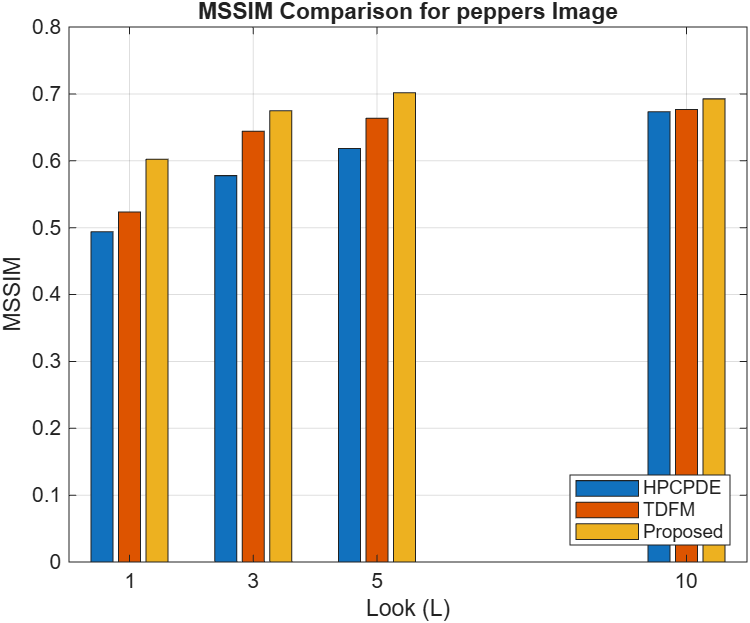}
       
    \end{minipage}
    
   \caption{PSNR and MSSIM performance comparison of the proposed model against state-of-the-art approaches for the Peppers grayscale image.}

    \label{tab:PSNR1}
\end{figure}

\begin{figure}[H]
    \centering
    \begin{minipage}{0.40\textwidth}
        \centering
        \includegraphics[width=\linewidth]{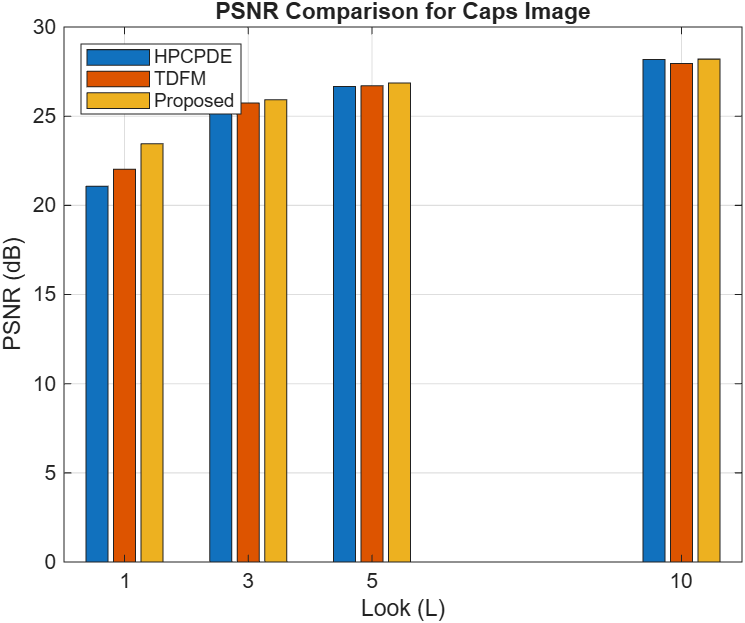}
        
    \end{minipage}
    \hfill
    \begin{minipage}{0.40\textwidth}
        \centering
        \includegraphics[width=\linewidth]{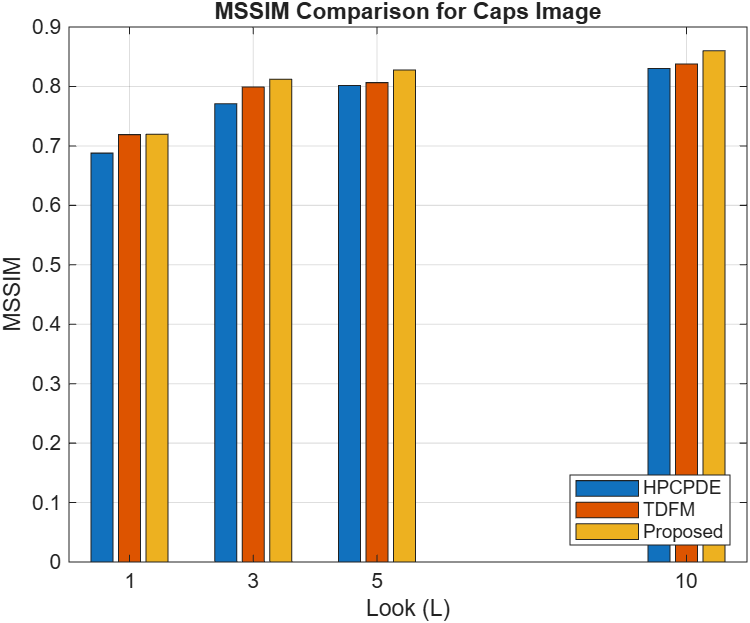}
       
    \end{minipage}
    
    \caption{PSNR and MSSIM performance comparison of the proposed model against state-of-the-art approaches for the caps color image.}

    \label{tab:psnr2}
\end{figure}

In this study, we evaluate our proposed denoising model on two real-world images: a Synthetic Aperture Radar (SAR) image, designated as Image ~\ref{fig:image1_noisy1}, and an ultrasound image, designated as Image ~\ref{fig:image2_noisy2}. Both images naturally contain speckle noise, a granular, multiplicative noise pattern that originates from the physical imaging process and environmental factors, rather than being artificially introduced. Because ground truth images are unavailable for these real datasets, quantitative evaluation is performed using the Speckle Index (S.I.), a common metric that measures the level of speckle noise. A lower value of S.I. indicates more effective noise removal and better image restoration. Figure~\ref {fig:sar_results} illustrates a comparative visual and quantitative evaluation between our proposed method and a conventional fourth-order PDE-based denoising model, highlighting the superior performance of our approach. Table~\ref {tab:si_sar_images1} lists the parameter settings used during these experiments, alongside the Speckle Index values computed for both the original noisy images and the corresponding denoised outputs. This thorough analysis confirms that our method effectively suppresses speckle noise while preserving important structural information, making it highly suitable for practical applications involving SAR and medical ultrasound images. To illustrate the denoising performance of the proposed model,  we present a 2D contour plot analysis of the pepper image with speckle noise at look 3 in figure ~\ref{fig:2d_comparisons}. The noisy image predominantly contains significant fluctuations, which make preserving important image details challenging. Lower fluctuations in the contour plot indicate reduced variation and better preservation of essential image features. Upon visual inspection, it is evident that the proposed model significantly outperforms other methods, as it produces contours with markedly fewer fluctuations. This reduction in noise-induced variability highlights the superior capability of the proposed approach to minimize speckle noise while maintaining the integrity of the underlying image structures. To comprehensively assess the effectiveness of the proposed denoising model, Figures ~\ref{tab:PSNR1} and ~\ref{tab:psnr2} present the PSNR and MSSIM bar graphs for grayscale and color images, respectively. The proposed model consistently outperforms competing methods across all tested speckle noise levels (looks). The consistent improvement across different noise intensities highlights the robustness and effectiveness of the proposed approach in both grayscale and color image denoising scenarios.

\begin{figure}[H]
    \centering
    
    \begin{subfigure}[b]{0.23\textwidth}
        \centering
        \includegraphics[width=\linewidth]{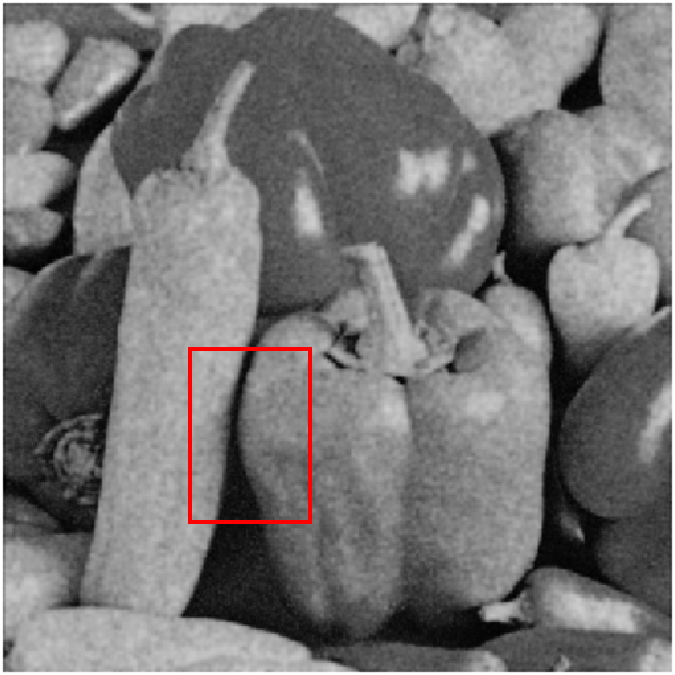}
        \caption{HPCPDE}
    \end{subfigure}
    \begin{subfigure}[b]{0.23\textwidth}
        \centering
        \includegraphics[width=\linewidth]{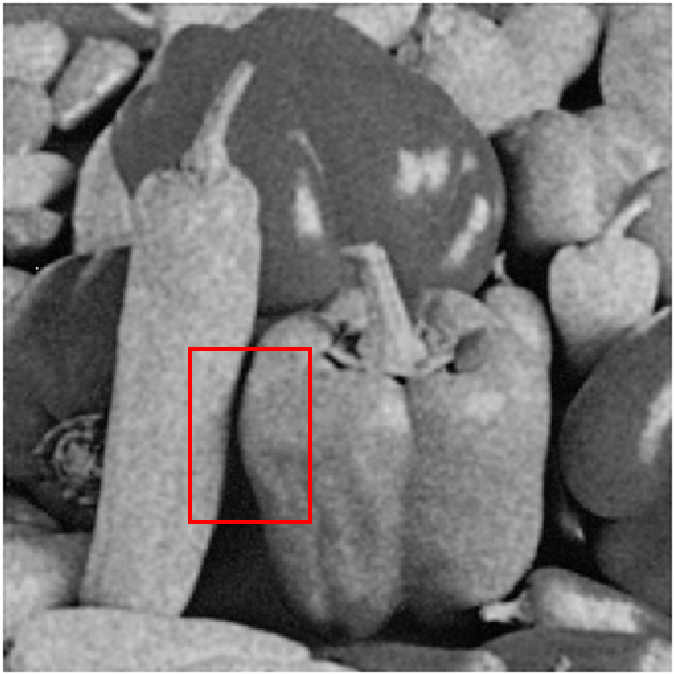}
        \caption{TDFM}
    \end{subfigure}
    \begin{subfigure}[b]{0.23\textwidth}
        \centering
        \includegraphics[width=\linewidth]{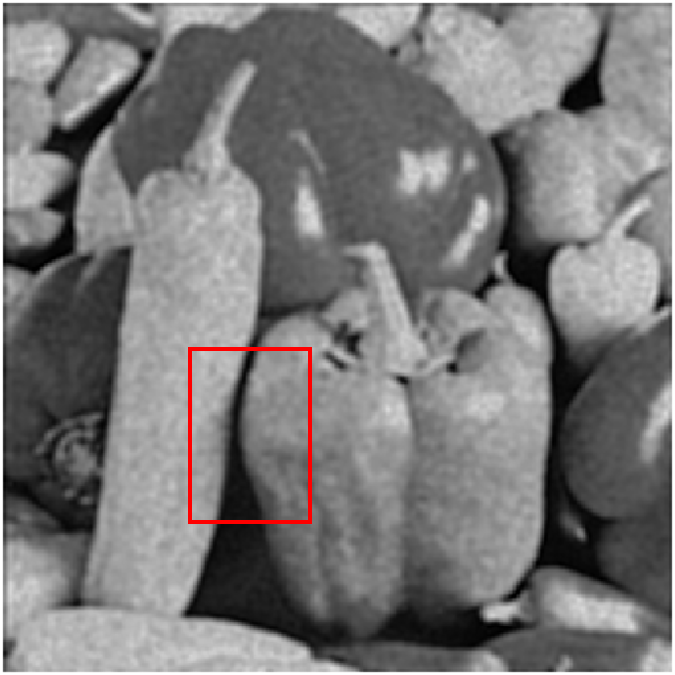}
        \caption{Proposed}
    \end{subfigure}

    \vspace{0.5cm} 

    \begin{subfigure}[b]{0.23\textwidth}
        \centering
        \includegraphics[width=\linewidth]{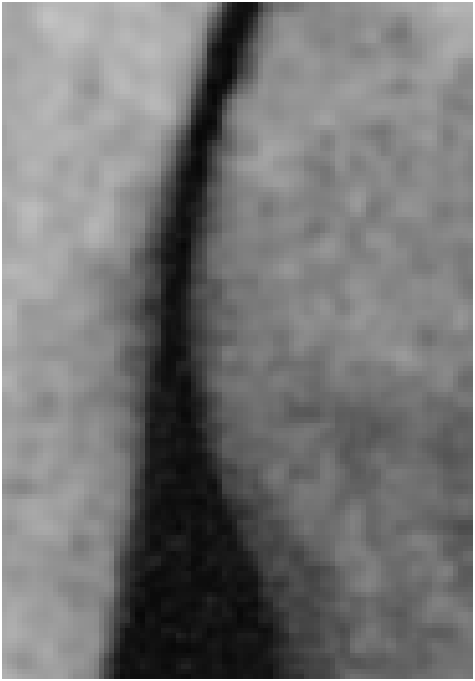}
        \caption{HPCPDE}
    \end{subfigure}
    \begin{subfigure}[b]{0.23\textwidth}
        \centering
        \includegraphics[width=\linewidth]{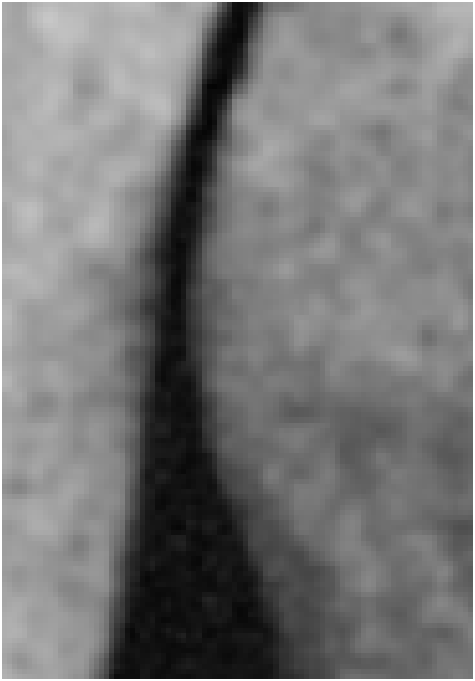}
        \caption{TDFM}
    \end{subfigure}
    \begin{subfigure}[b]{0.23\textwidth}
        \centering
        \includegraphics[width=\linewidth]{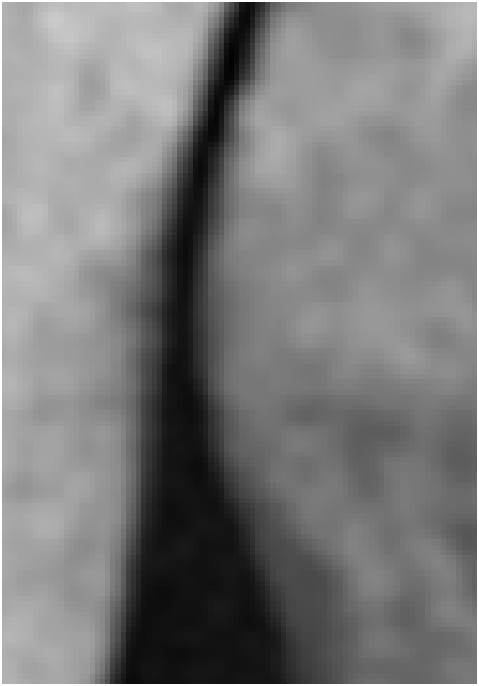}
        \caption{Proposed }
    \end{subfigure}

    \caption{The first row shows full Pepper gray images (Look = 5) with red boxes marking ROIs, while the second row zooms into these ROIs to compare noise removal across models.}
    \label{fig:box_gray}
    \end{figure}

\begin{figure}[H]
    \centering
    
    \begin{subfigure}[b]{0.23\textwidth}
        \centering
        \includegraphics[width=\linewidth]{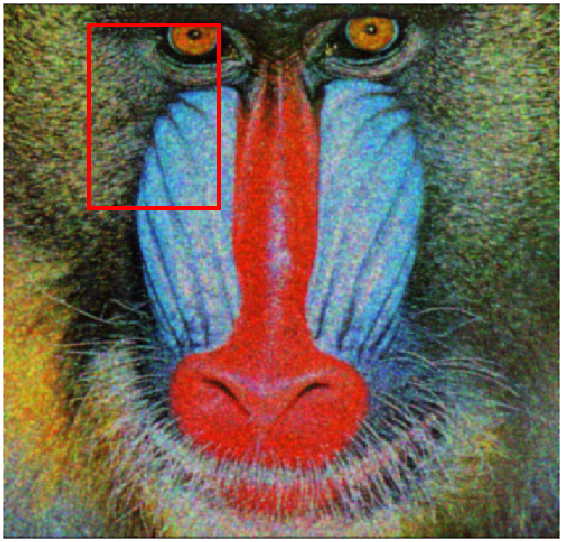}
        \caption{HPCPDE}
    \end{subfigure}
    \begin{subfigure}[b]{0.23\textwidth}
        \centering
        \includegraphics[width=\linewidth]{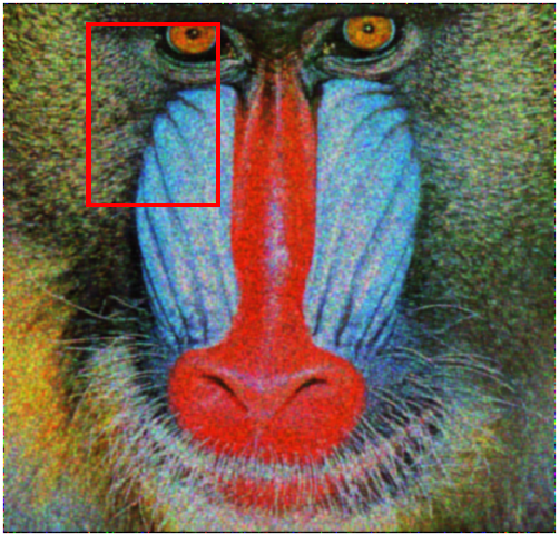}
        \caption{Forth Order Model}
    \end{subfigure}
    \begin{subfigure}[b]{0.23\textwidth}
        \centering
        \includegraphics[width=\linewidth]{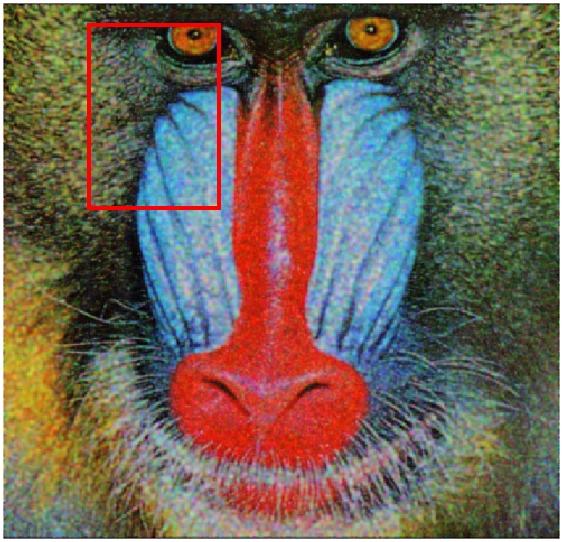}
        \caption{Proposed Model}
    \end{subfigure}

    \vspace{0.5cm} 

    \begin{subfigure}[b]{0.23\textwidth}
        \centering
        \includegraphics[width=\linewidth]{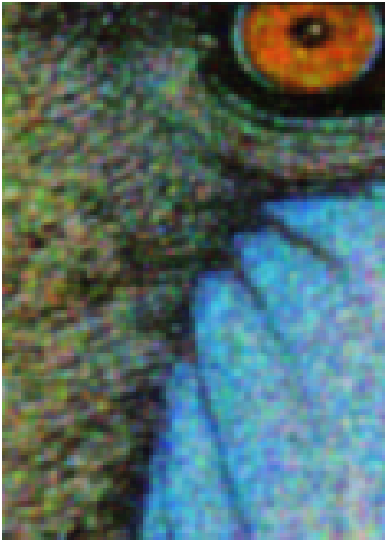}
        \caption{HPCPDE}
    \end{subfigure}
    \begin{subfigure}[b]{0.23\textwidth}
        \centering
        \includegraphics[width=\linewidth]{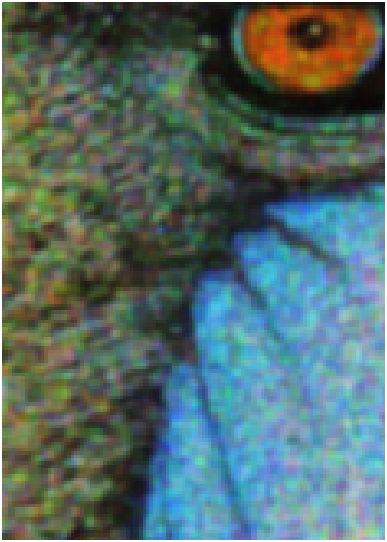}
        \caption{Forth Order Model}
    \end{subfigure}
    \begin{subfigure}[b]{0.23\textwidth}
        \centering
        \includegraphics[width=\linewidth]{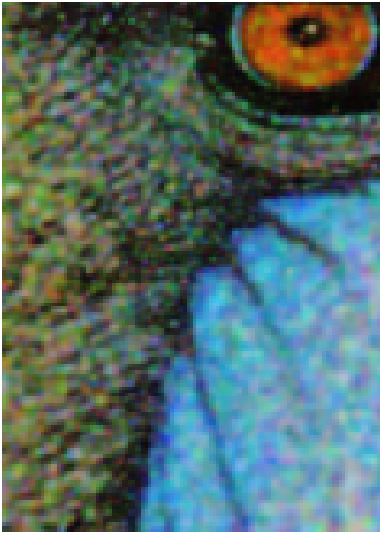}
        \caption{Proposed}
    \end{subfigure}

    \caption{The first row shows full Baboon color images (Look = 5) with red ROI boxes; the second row zooms in for detailed noise removal comparison.}
    \label{fig:box_color}
    \end{figure}

Lastly, Figures ~\ref{fig:box_gray} and ~\ref{fig:box_color} focus on selected Regions of Interest (ROIs) extracted from the Pepper grayscale image and the Baboon color image, both taken with look 5 speckle noise conditions.Upon close examination of the Figures~\ref{fig:box_gray} and ~\ref{fig:box_color} , it is evident that the proposed denoising model produces noticeably smoother results while effectively preserving important edges and fine details within the zoomed-in region. The model not only reduces noise but also captures and maintains the essential structural features of the image, which is critical for maintaining visual fidelity. These qualitative visualizations provide an in-depth comparison of despeckling performance, reinforcing the advantage of the proposed approach in balancing noise suppression with detail preservation. The results underscore that our method surpasses existing techniques by offering superior edge retention and noise removal, thus enabling better image clarity and interpretability in practical applications.


\begin{table}[H]
\centering
\caption{PSNR and MSSIM comparison for grayscale images (Peppers and Parrots) at different looks ($L=1,3,5,10$). Results are shown for the HPCPDE, the TDFM, and the proposed model.}
\label{tab:comparison_gray}
\begin{tabular}{|l|l|cc|cc|cc|}
\hline
\multicolumn{1}{|c|}{Image} & \multicolumn{1}{c|}{Looks} & \multicolumn{2}{c|}{HPCPDE} & \multicolumn{2}{c|}{TDFM} & \multicolumn{2}{c|}{Proposed} \\
\hline
 &  & MSSIM & PSNR & MSSIM & PSNR & \textbf{MSSIM} & \textbf{PSNR} \\
\hline
Peppers gray &
\begin{tabular}[c]{@{}l@{}}1\\3\\5\\10\end{tabular} &
\begin{tabular}[c]{@{}l@{}}0.4938\\0.5778\\0.6184\\0.6734\end{tabular} &
\begin{tabular}[c]{@{}l@{}}17.86\\23.39\\25.76\\28.35\end{tabular} &
\begin{tabular}[c]{@{}l@{}}0.5233\\0.6443\\0.6636\\0.6766\end{tabular} &
\begin{tabular}[c]{@{}l@{}}19.50\\24.36\\25.95\\28.36\end{tabular} &
\textbf{\begin{tabular}[c]{@{}l@{}}0.6023\\0.6749\\0.7018\\0.6926\end{tabular}} &
\textbf{\begin{tabular}[c]{@{}l@{}}23.88\\25.34\\26.31\\28.58\end{tabular}} \\
\hline
Parrots gray &
\begin{tabular}[c]{@{}l@{}}1\\3\\5\\10\end{tabular} &
\begin{tabular}[c]{@{}l@{}}0.5670\\0.5910\\0.6176\\0.6658\end{tabular} &
\begin{tabular}[c]{@{}l@{}}19.10\\23.93\\25.87\\28.24\end{tabular} &
\begin{tabular}[c]{@{}l@{}}0.5811\\0.5926\\0.6673\\0.6934\end{tabular} &
\begin{tabular}[c]{@{}l@{}}20.63\\24.34\\26.15\\28.18\end{tabular} &
\textbf{\begin{tabular}[c]{@{}l@{}}0.6100\\0.7044\\0.7372\\0.7191\end{tabular}} &
\textbf{\begin{tabular}[c]{@{}l@{}}22.16\\25.09\\26.30\\28.39\end{tabular}} \\
\hline
\end{tabular}
\end{table}

\begin{table}[H]
\centering
\caption{Optimum parameter values for different models on grayscale images.}
\label{tab:parameter_gray}
\begin{tabular}{|l|l|ccc|cccc|cccc|}
\hline
Image  & Look  & \multicolumn{3}{c|}{HPCPDE} & \multicolumn{4}{c|}{TDFM} & \multicolumn{4}{c|}{Proposed} \\ \hline
       & L     & \(\alpha\) & \(\beta\) & \(\gamma\) & \(\alpha\) & \(\beta\) & \(\gamma\) & \(\lambda\) & \(\alpha\) & \(\beta\) & \(\gamma\) & \(\lambda\) \\ \hline
pepper gray   & \begin{tabular}[c]{@{}l@{}}1\\ 3\\ 5\\ 10\end{tabular} 
       & \begin{tabular}[c]{@{}l@{}}1\\ 2\\ 2\\ 2\end{tabular}    
       & \begin{tabular}[c]{@{}l@{}}3\\ 3\\ 3\\ 3\end{tabular}                
       & \begin{tabular}[c]{@{}l@{}}5\\ 5\\ 5\\ 5\end{tabular} 

       & \begin{tabular}[c]{@{}l@{}}1\\ 1\\ 1\\ 1\end{tabular}       
       & \begin{tabular}[c]{@{}l@{}}2\\ 2\\ 2\\ 2\end{tabular}
       & \begin{tabular}[c]{@{}l@{}}5\\ 5\\ 5\\ 5\end{tabular} 
       & \begin{tabular}[c]{@{}l@{}}0.05\\ 0.05\\ 0.05\\ 0.1\end{tabular} 

       & \begin{tabular}[c]{@{}l@{}}1\\ 1\\ 1\\ 2\end{tabular}         
       & \begin{tabular}[c]{@{}l@{}}10\\ 10\\ 10\\ 10\end{tabular} 
       & \begin{tabular}[c]{@{}l@{}}1\\ 1\\ 1\\ 1\end{tabular}    
       & \begin{tabular}[c]{@{}l@{}}0.1\\ 0.1\\ 0.1\\ 0.5\end{tabular} \\ \hline

parrots gray  & \begin{tabular}[c]{@{}l@{}}1\\ 3\\ 5\\ 10\end{tabular} 
       & \begin{tabular}[c]{@{}l@{}}1\\ 1\\ 1\\ 1\end{tabular}            
       & \begin{tabular}[c]{@{}l@{}}3\\ 3\\ 3\\ 3\end{tabular}                
       & \begin{tabular}[c]{@{}l@{}}1\\ 1\\ 1\\ 1\end{tabular} 

       & \begin{tabular}[c]{@{}l@{}}1\\ 1\\ 1\\ 1\end{tabular}     
       & \begin{tabular}[c]{@{}l@{}}2\\ 2\\ 2\\ 2\end{tabular} 
       & \begin{tabular}[c]{@{}l@{}}5\\ 5\\ 5\\ 5\end{tabular} 
       & \begin{tabular}[c]{@{}l@{}}0.05\\ 0.1\\ 0.1\\ 0.2\end{tabular} 

       & \begin{tabular}[c]{@{}l@{}}2\\ 2\\ 2\\ 2\end{tabular}         
       & \begin{tabular}[c]{@{}l@{}}10\\ 10\\ 10\\ 10\end{tabular} 
       & \begin{tabular}[c]{@{}l@{}}1\\ 1\\ 1\\ 2\end{tabular}    
       & \begin{tabular}[c]{@{}l@{}}0.05\\ 0.05\\ 0.05\\ 0.05\end{tabular} \\ \hline

\end{tabular}
\end{table}

\begin{table}[H]
    \centering
    \caption{PSNR and MSSIM comparison for color images (Baboon and Caps) at different looks ($L=1,3,5,10$). Results are shown for the HPCPDE, the TDFM, and the proposed model.}
    \label{tab:comparison_color}
    \begin{tabular}{|c|c|cc|cc|cc|}
        \hline
        \multirow{2}{*}{Image} & \multirow{2}{*}{Look} 
        & \multicolumn{2}{c|}{HPCPDE} 
        & \multicolumn{2}{c|}{TDFM} 
        & \multicolumn{2}{c|}{Proposed Model} \\ 
        \cline{3-8}
        &  & MSSIM & PSNR & MSSIM & PSNR & MSSIM & PSNR \\ 
        \hline
        \multirow{4}{*}{Baboon}  
        & 1  & 0.5489 & 16.72 & 0.5687 & 17.91 & \textbf{0.5706} & \textbf{18.08} \\
        & 3  & 0.6904 & 19.60 & 0.6821 & 19.51 & \textbf{0.6924} & \textbf{19.77} \\
        & 5  & 0.7558 & 20.23 & 0.7366 & 20.32 & \textbf{0.7579} & \textbf{20.67} \\
        & 10 & 0.8197 & 22.18 & 0.7998 & 21.60 & \textbf{0.8254} & \textbf{22.20} \\
        \hline
        \multirow{4}{*}{Caps}  
        & 1  & 0.6879 & 21.07 & 0.7190 & 22.03 & \textbf{0.7195} & \textbf{23.45} \\
        & 3  & 0.7709 & 25.41 & 0.7992 & 25.74 & \textbf{0.8122} & \textbf{25.92} \\
        & 5  & 0.8016 & 26.67 & 0.8065 & 26.71 & \textbf{0.8276} & \textbf{26.86} \\
        & 10 & 0.8303 & 28.18 & 0.8377 & 27.96 & \textbf{0.8601} & \textbf{28.20} \\
        \hline
    \end{tabular}
\end{table}

\begin{table}[H]
\centering
\caption{Optimum parameter values for different models on color images.}
\label{tab:parameter_color}
\begin{tabular}{|l|l|ccc|cccc|cccc|}
\hline
Image  & Look  & \multicolumn{3}{c|}{HPCPDE} & \multicolumn{4}{c|}{TDFM} & \multicolumn{4}{c|}{Proposed} \\ \hline
       & L     & \(\alpha\) & \(\beta\) & \(\gamma\) & \(\alpha\) & \(\beta\) & \(\gamma\) & \(\lambda\) & \(\alpha\) & \(\beta\) & \(\gamma\) & \(\lambda\) \\ \hline
baboon   & \begin{tabular}[c]{@{}l@{}}1\\ 3\\ 5\\ 10\end{tabular} 
       & \begin{tabular}[c]{@{}l@{}}3\\ 3\\ 3\\ 3\end{tabular}    
       & \begin{tabular}[c]{@{}l@{}}3\\ 3\\ 3\\ 3\end{tabular}                
       & \begin{tabular}[c]{@{}l@{}}3\\ 3\\ 3\\ 3\end{tabular} 
       & \begin{tabular}[c]{@{}l@{}}1\\ 1.1\\ 1.2\\ 1.2\end{tabular}       
       & \begin{tabular}[c]{@{}l@{}}2\\ 2\\ 2\\ 2\end{tabular}
       & \begin{tabular}[c]{@{}l@{}}5\\ 5\\ 5\\ 5\end{tabular} 
       & \begin{tabular}[c]{@{}l@{}}0.1\\ 0.1\\ 0.1\\ 0.1\end{tabular} 
       & \begin{tabular}[c]{@{}l@{}}1\\ 1\\ 1\\ 2\end{tabular}         
       & \begin{tabular}[c]{@{}l@{}}10\\ 10\\ 10\\ 10\end{tabular} 
       & \begin{tabular}[c]{@{}l@{}}3\\ 3\\ 3\\ 3\end{tabular}    
       & \begin{tabular}[c]{@{}l@{}}0.07\\ 0.07\\ 0.07\\ 0.07\end{tabular} \\ \hline

caps  & \begin{tabular}[c]{@{}l@{}}1\\ 3\\ 5\\ 10\end{tabular} 
       & \begin{tabular}[c]{@{}l@{}}1\\ 1\\ 1\\ 2\end{tabular}            
       & \begin{tabular}[c]{@{}l@{}}3\\ 3\\ 3\\ 3\end{tabular}                
       & \begin{tabular}[c]{@{}l@{}}3\\ 3\\ 3\\ 3\end{tabular} 
       & \begin{tabular}[c]{@{}l@{}}1\\ 1\\ 1\\ 1\end{tabular}     
       & \begin{tabular}[c]{@{}l@{}}2\\ 2\\ 2\\ 2\end{tabular} 
       & \begin{tabular}[c]{@{}l@{}}5\\ 5\\ 5\\ 5\end{tabular} 
       & \begin{tabular}[c]{@{}l@{}}0.5\\ 0.5\\ 0.5\\ 0.3\end{tabular} 
       & \begin{tabular}[c]{@{}l@{}}1\\ 1\\ 1\\ 1.8\end{tabular}         
       & \begin{tabular}[c]{@{}l@{}}10\\ 10\\ 10\\ 10\end{tabular} 
       & \begin{tabular}[c]{@{}l@{}}2.5\\ 2.5\\ 2.5\\ 2\end{tabular}    
       & \begin{tabular}[c]{@{}l@{}}0.01\\ 0.01\\ 0.02\\ 0.05\end{tabular} \\ \hline

\end{tabular}
\end{table}

\begin{table}[H]
\centering
\caption{Speckle Index (S.I.) comparison on SAR images for various parameter settings \((\gamma, \alpha, \lambda, \beta)\) using the proposed method.}

\begin{tabular}{cccc}
\hline
\textbf{Image} & \textbf{Noisy} & \textbf{Parameters} & \textbf{Proposed Method} \\ \hline
               &                & \(\boldsymbol{\gamma}  \quad\boldsymbol{\alpha} \quad \quad\boldsymbol{\lambda}\quad\quad \boldsymbol{\beta}\) &                \\ \hline
Image\_1       & 1.02           & \(5 \quad 0.1 \quad 0.007 \quad 4\) & 0.2911           \\
Image\_2       & 0.92           & \(2 \quad 0.1 \quad 0.001 \quad 4\) & 0.3275           \\ \hline
\end{tabular}
\label{tab:si_sar_images1}
\end{table}



\section{Conclusion}
This work presents a novel coupled fourth-order partial differential equation (PDE) model for image despeckling, particularly in SAR and medical ultrasound images. The model integrates a nonlinear hyperbolic-parabolic PDE system, combining wave-like propagation with high-order diffusion to achieve effective noise suppression while preserving essential image structures. A key advancement is the coupling of image intensity evolution from edge detection through an auxiliary variable, enabling precise diffusion control for enhanced edge and texture preservation. The diffusion coefficient dynamically adapts to local image features by combining intensity and curvature information, improving robustness against multiplicative speckle noise. Extensive experiments on synthetic and real grayscale, color, SAR, and ultrasound images show consistently superior performance compared to existing second- and fourth-order PDE-based methods. Quantitative metrics (PSNR, MSSIM, Speckle Index), along with visual and ROI analyses, confirm the model’s efficacy across different noise levels and image types. The flexible framework supports both grayscale and color restoration, with stable convergence ensured by norm- or PSNR-based criteria, making it a promising tool for practical use. Future work may focus on optimizing computational efficiency to reduce processing time. Integration with deep learning-based priors may also be investigated to combine model-driven and data-driven despeckling strategies for improved results. These advancements aim to further enhance despeckling performance, ensure robustness across diverse imaging scenarios, and make the framework more practical for real-world applications.

\vspace{11pt}
\noindent\textbf{Author Declaration} \\
The authors declare that there are no conflicts of interest.

\vspace{11pt}
\noindent\textbf{Data Availability} \\
The data that support the findings of this study are available from the corresponding
author upon reasonable request.

\end{document}